\documentclass[twocolumn,iop]{emulateapj}
\usepackage{color}

\begin{document}


\title{A Gemini/GMOS Study of Intermediate Luminosity Early-Type Virgo Cluster Galaxies. I. Globular Cluster and Stellar Kinematics}


\author{Biao Li\altaffilmark{1,2}}
\author{Eric W. Peng\altaffilmark{1,2}}
\author{Hong-xin Zhang\altaffilmark{1,2,3,4,5,7}}
\author{John P. Blakeslee\altaffilmark{6}}
\author{Patrick C\^ot\'e\altaffilmark{6}}
\author{Laura Ferrarese\altaffilmark{6}}
\author{Andr\'es Jord\'an\altaffilmark{7,8}}
\author{Chengze Liu\altaffilmark{9,10}}
\author{Simona Mei\altaffilmark{11,12,13}}
\author{Thomas H. Puzia\altaffilmark{7}}
\author{Marianne Takamiya\altaffilmark{14}}
\author{Gelys Trancho\altaffilmark{15}}
\author{Michael J. West\altaffilmark{16}}

\altaffiltext{1}{Department of Astronomy, Peking University, Beijing  100871, China; biaolee@pku.edu.cn; peng@pku.edu.cn}
\altaffiltext{2}{Kavli Institute for Astronomy and Astrophysics, Peking University, China}
\altaffiltext{3}{Chinese Academy of Sciences South America Center for Astronomy, Camino EI Observatorio \#1515, Las Condes, Santiago, Chile}
\altaffiltext{4}{National Astronomical Observatories, Chinese Academy of Sciences, Beijing 100012, China}
\altaffiltext{5}{CAS-CONICYT Fellow}
\altaffiltext{6}{National Research Council of Canada, Herzberg Astronomy and Astrophysics Program, 5071 West Saanich Road, Victoria, BC V9E 2E7, Canada} 
\altaffiltext{7}{Instituto de Astrof\'isica, Facultad de F\'isica,
    Pontificia Universidad Cat\'olica de Chile, Av.\ Vicu\~na Mackenna
    4860, 7820436 Macul, Santiago, Chile}
\altaffiltext{8}{Millennium Institute of Astrophysics, Av.\ Vicu\~na Mackenna
    4860, 7820436 Macul, Santiago, Chile}
\altaffiltext{9}{Department of Physics, Shanghai Jiao Tong University, 800 Dongchuan Road, Shangai 200240, China}
\altaffiltext{10}{Shanghai Key Lab for Particle Physics and Cosmology, Shanghai Jiao Tong University, Shanghai 200240, China}
\altaffiltext{11}{GEPI, Observatoire de Paris, PSL Research University, CNRS, University of Paris Diderot, 61, Avenue de l'Observatoire 75014, Paris France}
\altaffiltext{12}{Universit\'{e} Paris Denis Diderot, Universit\'e Paris Sorbonne Cit\'e, 75205 Paris Cedex 13, France}
\altaffiltext{13}{California Institute of Technology, Pasadena, CA 91125, USA}
\altaffiltext{14}{Institute for Astronomy, University of Hawaii, 2680 Woodlawn Dr., Honolulu, HI 96822, USA}
\altaffiltext{15}{Giant Magellan Telescope Organization, 251 South Lake Avenue, Pasadena, CA 91101, USA}
\altaffiltext{16}{Maria Mitchell Observatory, 4 Vestal Street, Nantucket, MA, 02554, USA}



\begin{abstract}
We present a kinematic analysis of the globular cluster systems and diffuse stellar light of four intermediate luminosity (sub-$L^{\ast}$) early-type galaxies in the Virgo cluster based on Gemini/GMOS data. Our galaxy sample is fainter ($-23.8<M_K<-22.7$) than most previous studies, nearly doubling the number of galaxies in this magnitude range that now have GC kinematics. The data for the diffuse light extends to $4R_e$, and the data for the globular clusters reaches 8--$12R_e$. We find that the kinematics in these outer regions are all different despite the fact that these four galaxies have similar photometric properties, and are uniformly classified as ``fast rotators'' from their stellar kinematics within $1R_e$. The globular cluster systems exhibit a wide range of kinematic morphology. The rotation axis and amplitude can change between the inner and outer regions, including a case of counter-rotation. This difference shows the importance of wide-field kinematic studies, and shows that stellar and GC kinematics can change significantly as one moves beyond the inner regions of galaxies. Moreover, the kinematics of the globular cluster systems can differ from that of the stars, suggesting that the formation of the two populations are also distinct.
\end{abstract}



\section{Introduction}
Early-type galaxies (ETGs) are thought to be the final products of mergers and accretion processes in a hierarchical universe. This history of collapse and merging is preserved in the motions of their stars. Stellar kinematics can therefore be used to constrain galactic formation and evolution. Absorption line profiles measured from integrated spectroscopy can measure these bulk stellar motions. However, these observations are usually restricted to within 1--2 effective radii (R $\leq$ 1--2$R_{e}$) of galaxy centers \citep{ge01,pin03,pro05,spo08} due to the faintness of the stellar halo. With the development of integral field units (IFUs), this technique has given us a new two-dimensional view of the line-of-sight velocity distribution (LOSVD) in galaxies \citep{sauron01, ca11,califa12, muse14,manga15,guerou15}. \cite{ca11} observed 260 ETGs in the nearby universe with the Spectrographic Areal Unit for Research on Optical Nebulae (SAURON) IFU in the $\text{ATLAS}^{\text{3D}}$ project. They were able to measure the projected angular momentum per unit mass, and used this parameter to classify 260 ETGs as either ``fast'' or ``slow'' rotators, finding that nearly all but the most massive ETGs are fast rotators \citep{em11}. They argue for a shift in the existing paradigm for ETGs, where galaxies generally classified morphologically into disc-like S0 galaxies and spheroidal-like E systems should instead be classified kinematically.

However, like earlier studies, the $\text{ATLAS}^{\text{3D}}$ project is limited to $1R_{e}$, where relaxation processes make inferring formation histories increasingly difficult, and where feedback processes complicate things even further. The fraction of the total mass in this region may only be $\sim$15--30\% \citep{wu14}. In addition, the outer regions may present more obvious signatures of merging or accretion events. There are only a few studies that extend out to $\sim$2--4$R_{e}$ \citep{me00, pro09,wei09,for10,ar13}. \cite{pro09} developed a new technique for extracting integrated spectra of the field stars, or galaxy diffuse light (GDL), contained in multi-slit observations, and found a variety of rotation profiles beyond $1R_{e}$ for five ETGs. They found that the stellar kinematics in their galaxies' inner regions were not necessarily predictive of those in the outer regions. One possibility is that dissipation processes in gas-rich major mergers produce dramatic kinematic differences between the inner and outer regions \citep{ho10}. Alternatively, the galaxies could have formed in two separate phases, with the inner region forming early by dissipation, and the outer region forming later by accretion of smaller galaxies \citep{os10}. Further studying stellar kinematics at large radius is clearly an important direction to pursue.

Compared to stellar kinematics, which are limited by low surface brightness, observations of globular clusters (GCs) can extend out to much larger distances ($\sim$10$R_{e}$) from the galaxy center. Because GCs are compact and luminous, they are easily observed in the halos of galaxies. GCs are old, coeval collections of stars that can survive a Hubble time, and they preserve information on the assembly history of their host galaxies. 

In recent years, large samples of GC velocities around ETGs have been produced (e.g., M87, \citealt{str11}; NGC 1399, \citealt{sc10}; NGC~5128, \citealt{pe04}, \citealt{wo10}). For practical reasons, much of the work in this field has focused on massive early-type galaxies and their rich GC systems. ETGs at intermediate mass, however, are much more common, and perhaps less complicated than their more massive counterparts. For these ETGs, with luminosities at or fainter than the knee of the galaxy luminosity function ($M_{K}=-24.2$~mag, \citealt[$h=0.7$]{cole01}){, there are only a few studies (NGC 3115, \citealt{Ar11}; NGC 3379, \citealt{pu04}, \citealt{pi06}, \citealt{be06}; NGC 4494, \citealt{fo11}; NGC 821, 3377, 1400, 4278, 7454, \citealt{po13}). Due to lower surface brightnesses and sparser GC systems, these galaxies generally lack kinematic data at large radii. 

\cite{ro03} and \cite{do07} used planetary nebulae as kinematic tracers and found some intermediate luminosity ETGs (NGC 3379, NGC 821 and NGC 4494) with low mass-to-light ratios , which appeared to conflict with the dark matter fractions predicted in cosmological simulations. Compared to massive ETGs, intermediate luminosity ETGs may also have more sporadic merger histories with different mass ratios, potentially preserving the wide range of assembly processes at work, and making them important and interesting targets of study. 

In this paper, we study the GC and stellar kinematics of four intermediate luminosity ETGs in the Virgo cluster out to large radii. We compare GC and stellar kinematic trends (rotation amplitude and position angle) with each other, and at different radii. In our next paper (Li et al., 2015b, in prep), we will study the mass distribution and dark matter content of these intermediate luminosity ETGs.

\section{Sample, Observations, and Data}

\subsection{Sample}

We selected four intermediate-luminosity ETGs from the ACS Virgo Cluster Survey (ACSVCS; \citealt{co04}), a homogeneous Hubble Space Telescope survey of 100 ETGs in the nearby Virgo cluster of galaxies using the Advanced Camera for Surveys (ACS; \citealt{fo98}). These deep images, with their high spatial resolution, allow for excellent selection of globular clusters \citep{jo09}. This is crucial for our sample galaxies, as they do not have very rich GC systems, and contamination from foreground and background sources can significantly reduce the spectroscopic yield. 

The four galaxies in our study are VCC 1231, 2000, 1062, 685 (NGC 4473, 4660, 4442, and 4350, respectively, see Table~\ref{tbl-1}). They span the interesting $L^{*}$ to sub-$L^{*}$ luminosity range ($-20.5< M_{B}<-19$), making them fainter than galaxies targeted in typical GC kinematic studies. For its luminosity range, this sample now doubles the number of galaxies with GC kinematics. According to NED, two have morphological type E5, and the other two are classified as S0. All galaxies are classified as ``fast rotators'' by $\text{ATLAS}^{\text{3D}}$. \citep{em11}

\begin{deluxetable*}{cccccccccc}
\tabletypesize{\scriptsize}
\centering 
\tablecaption{Sample properties\label{tbl-1}}
\tablewidth{0pt}
\tablehead{
\colhead{ID} & \colhead{Type} & \colhead{R.A.(J2000)} & \colhead{Decl.(J2000)} & 
\colhead{$V_{\rm sys}$} & \colhead{$R_{e}$} & \colhead{D} &
\colhead{$\rm M_{K}$} & \colhead{$\rm PA_{K}$} & \colhead{$(\rm b/a)_{K}$}  \\
\colhead{} & \colhead{} & \colhead{(deg)} & \colhead{(deg)} & 
\colhead{($\rm km\ s^{-1}$)} & \colhead{(\rm kpc)} & \colhead{(\rm Mpc)} &
\colhead{(mag)} & \colhead{(degree)} & \colhead{}  \\
\colhead{(1)} & \colhead{(2)} & \colhead{(3)} & \colhead{(4)} & \colhead{(5)} &
\colhead{(6)} & \colhead{(7)} & \colhead{(8)} &
\colhead{(9)} & \colhead{(10)} 
}  
\startdata
VCC1231 (NGC4473)&E5&187.453659& +13.429320&2260 &1.2 &15.2 & $-23.77$& $275$ & 0.54\\

VCC2000 (NGC4660) & E5 & 191.133209 & +11.190533 & 1087 & 0.8 &  15.0 & $-22.69$ & $280$ & 0.5\\

VCC1062 (NGC4442) & SB0 & 187.016220 & +09.803620 & 547 & 1.3&  15.3 & $-23.63$ & $82$ & 0.51 \\

VCC0685 (NGC4350) & SA0 & 185.990891 & +16.693356  & 1210    & 0.9&  15.4 & $-23.13$ & $30$ & 0.3
\enddata
\tablecomments{Hubble types (2) are from the NED database. The position of galaxy center, RA (3) and DEC (4), the galaxy systemic velocity (5) and the K band absolute magnitude (8) are all from $\text{ATLAS}^{\text{3D}}$ project \citep{ca11}, where (8) are derived from the 2MASS extended source catalog \citep{ja00}. The effective radii (6) are from \cite{Fe06}. The distances (7) are from \cite{bl09}, except VCC 685 which is from the GC luminosity function estimates by \cite{jo07}. The photometric position angle (9) and axis ratio (10) are from 2MASS \citep{sk06}.}
\end{deluxetable*}

\subsection{Observation}

We observed these galaxies with the Gemini Multi-Object Spectrographs (GMOS, \citealt{ho04}), twin instruments on the Gemini North and Gemini South telescopes. Our target galaxies have sizes ($R_{e}\sim10$--$18\arcsec$) that fit well within the GMOS field-of-view (5.5 $\rm arcmin^{2}$), providing coverage out to 10--$16R_{e}$. Each galaxy contained $\sim50$ targetable GCs with $V<23$~mag. VCC 1231, VCC 1062 and VCC 2000 data were taken with GMOS-South, whereas data for VCC 685 was taken with GMOS-North. We observed three masks for each galaxy to overcome the issue of slit-crowding at the galaxy centers. VCC 685 was only observed with one mask due to it being lower priority in the queue. Table~\ref{tbl:ob} summarizes the observing information. Because there are two gaps in the GMOS CCD focal plane, we observed with two wavelength centers, 5080\AA\ and 5120\AA, dithering exposures between the two. For each central wavelength, we took two exposures. We used the B600\_G5323(GMOS North) and B600\_5303(GMOS South) grisms with $1\farcs0$ wide slits, giving a dispersion of 0.5 and 0.45 $\rm \AA\ pixel^{-1}$, respectively. Both grisms give a spectral resolution of $R=1688$. \citep{pu13} 

\begin{deluxetable*}{cccccc}
\tabletypesize{\scriptsize} 
\centering 
\tablecaption{Summary of Gemini/MOS Observation \label{tbl:ob}}
\tablewidth{0pt}
\tablehead{
\colhead{Mask} & \colhead{R.A.(J2000)} & \colhead{Decl.(J2000)} & 
\colhead{Exp.Time} & \colhead{Slits} & \colhead{Night} \\
\colhead{} & \colhead{(deg)} & \colhead{(deg)} & 
\colhead{(sec)} & \colhead{} & \colhead{} 
}
\startdata
GS2008AQ008-01&191.134109&+11.176211&3040&22&2008-04-06\\

GS2008AQ008-02&191.134109&+11.176211&3040&23&2008-04-07\\

GS2008AQ008-03&191.134109&+11.176211&3040&20&2008-04-08\\

GS2008AQ008-04&187.453537&+13.427367&3040&30&2008-03-14\\

GS2008AQ008-05&187.453537&+13.427367&3040&25&2008-04-01\\

GS2008AQ008-06&187.453537&+13.427367&3040&27&2008-04-05\\

GS2008AQ008-07&187.024216&+09.806433&2700&21&2008-03-02\\

GS2008AQ008-08&187.024216&+09.806433&2700&21&2008-03-09\\

GS2008AQ008-09&187.024216&+09.806433&2700&21&2008-03-14\\

GN2008AQ073-01&186.002594&+16.686339&2820&24&2008-03-31

\enddata
\end{deluxetable*}

\subsection{Data Reduction}
The data reduction was performed by using the Gemini/GMOS IRAF package (Version 1.11). (1) The cosmic ray rejection was done using the GSCRSPEC script, which was originally written by Bryan Mlller.  GSCRSPEC is an MEF wrapper for the spectroscopic version of the Laplacian Cosmic Ray Identification routine \citep{va01}; (2) The bias subtraction and flat-fielding were done by the GSDEDUCE; (3) We used GSWAVELENGTH to establish the wavelength calibration from the CuAr arc frames. Typical wavelength solution residuals were $\sim0.1 \rm\AA$--$0.3\rm\AA$; (4) The wavelength calibration was applied to the object frames by using GSTRANSFORM; (5) We used GSSKYSUB to do the sky-subtracted interactively and GSEXTRACT to extract 1D spectra from each object frame; (6) Finally, we used the SCOMBINE to median combine extracted spectra from the same slits.

After extracting the target spectrum from each of the mask slitlets, we get a ``background" spectrum for each slitlets. For slits closer to the galaxy center, this ``background" spectrum contains not only sky but also useful amounts of host GDL. If the slit is far from the galaxy center, the spectrum should be dominated by ``pure sky''. In order to recover information from the slits with significant amounts of galaxy light, we carefully choose slits at the greatest distance ($\sim$$7R_{e}$) from the galaxy center and combine their spectra to produce a pure sky spectrum for non-local sky subtraction. We do not use slits located near the edges of the mask where the vignetting effects are serious. After subtracting the sky spectrum from the remaining ``background'' spectra we are left with the GDL spectra. 
Figure~\ref{spec} shows GC and GDL spectra with different signal-to-noise ratio. Our spectra typically cover a wavelength range from 3500\AA\ to 6500\AA, but the wavelength range depends on the slit position on the mask and some spectra can reach wavelengths as red as 7200\AA.

\begin{figure}
\centering
\includegraphics[width=1.0\columnwidth]{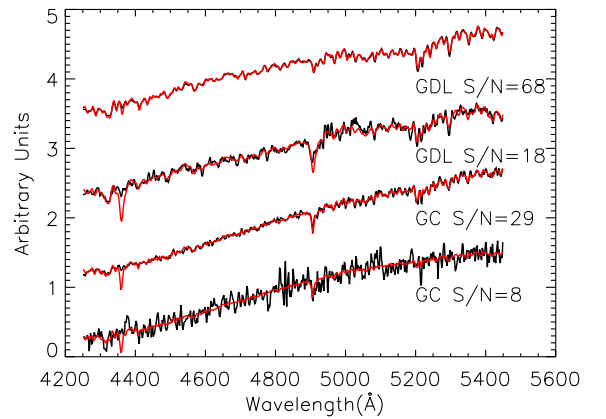}
\caption{Reduced GC and GDL spectra. The red lines are the best-fitt PPXF templates. The signal-to-noise (S/N) is calculated from 4500\AA\  to 4750\AA.}
\label{spec}
\end{figure}

\subsection{Radial Velocity Determination}
We use the FXCOR task in IRAF to measure the GC and GDL velocities. The template we use for FXCOR is a K-type M31 globular cluster spectrum from a list of Smithsonian Astrophysical Observatory (SAO) spectral templates. For the GDL, we also use the penalized pixel-fitting method (pPXF) as implemented by \cite{cap04} to measure radial velocities and velocity dispersions. For pPXF, we use the MILES library of stellar spectra as templates \citep{miles}. The radial velocities measured using FXCOR and pPXF are very consistent, so we choose to use the FXCOR velocities. The velocity uncertainties are estimated by FXCOR and scaled to those determined with repeated measurements of science targets. For velocity dispersion, we use the region of 4850$\rm \AA$  to 5350$\rm \AA$  which contains the $\rm H\beta$, Mg b and Fe5270 spectroscopic features. To be conservative, we only measure the velocity dispersion within 3$R_e$. The signal-to-noise of the spectra are typically too low to measure accurately the velocity dispersion in the outer regions.

\begin{figure}
\includegraphics[scale=.50]{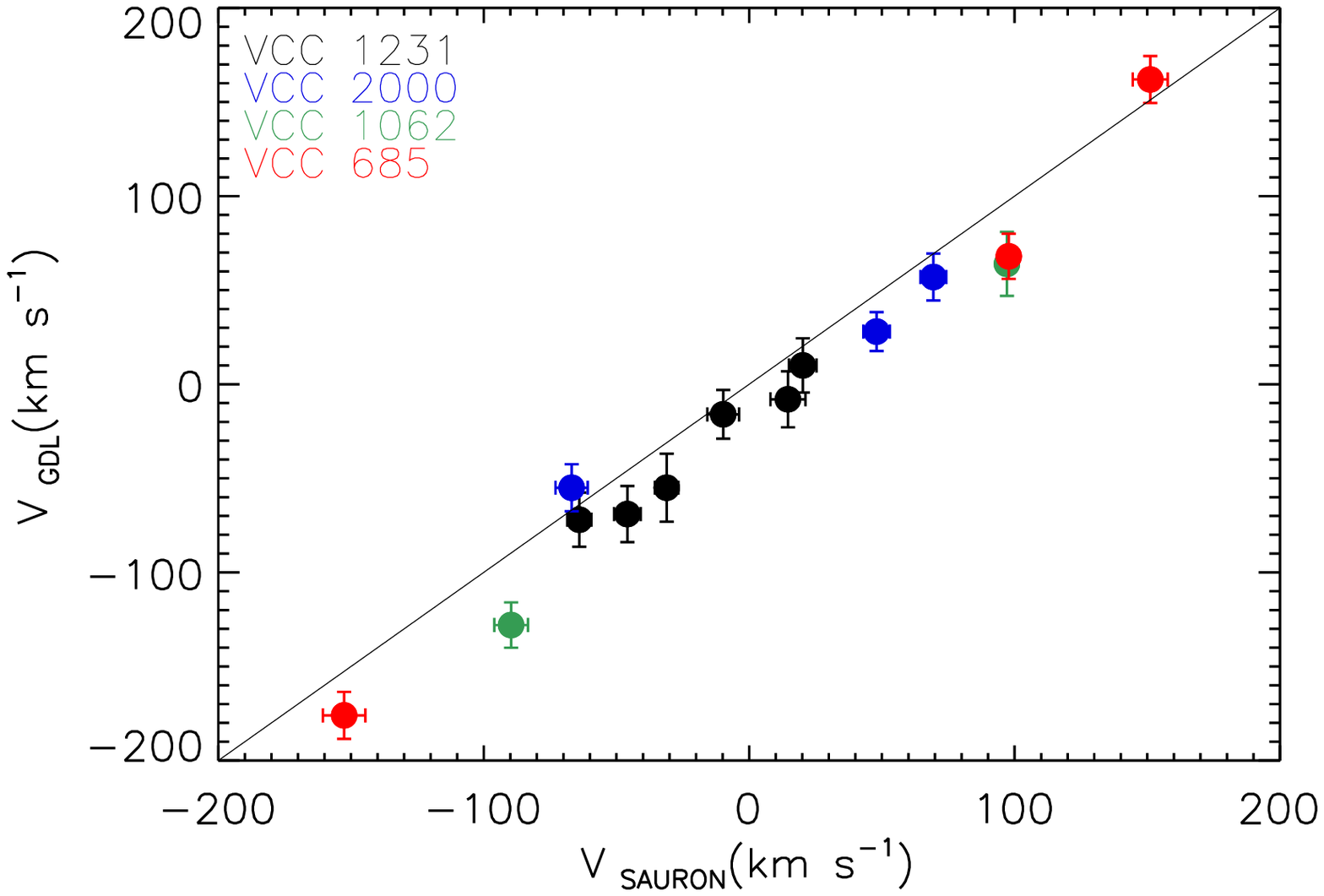}
\includegraphics[scale=.50]{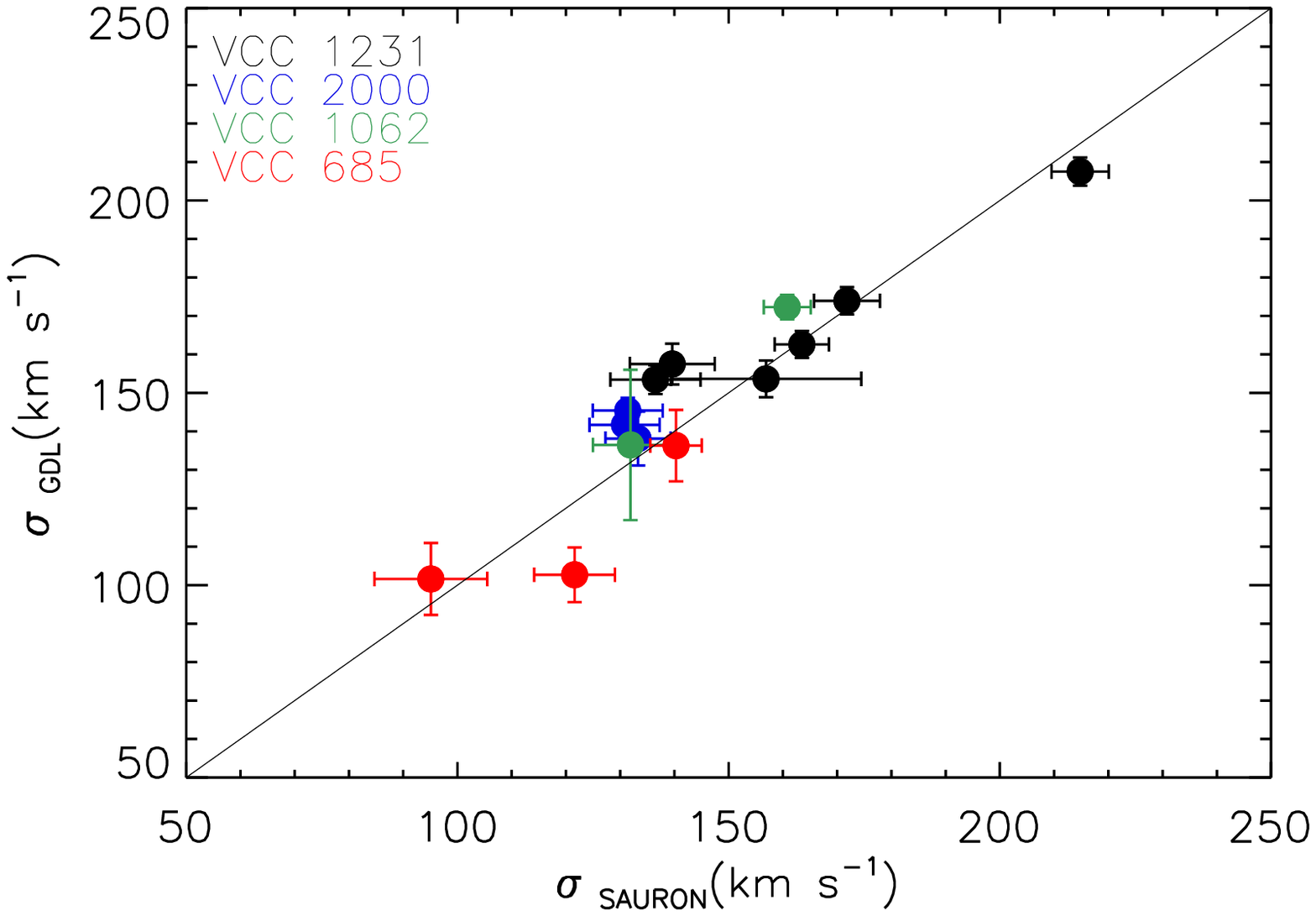}
\caption{A comparison of our GDL radial velocity and velocity dispersion measurements with those from the SAURON survey.  The solid line is a 1:1 relation. The mean offset is 16 km/s and 4 km/s for the velocity and velocity dispersion, respectively. The two surveys show good agreement.}
\label{diff}
\end{figure}

We divide our GCs into red and blue populations according to the $(g-z)_{0}$ colour distributions in \cite{pe06}. The colors of GCs are from the ACSVCS and the Next Generation Virgo Survey (NGVS, \citealt{fe12}). The NGVS data are used for objects that fall outside the ACS field of view. The transformation that we use between the ACSVCS and NGVS filters are:
\begin{equation}
g_{ACS}-g_{NGVS}=0.12\pm0.06
\end{equation}
\begin{equation}
z_{ACS}-z_{NGVS}=-0.06\pm0.10
\label{eq}
\end{equation}
The number of spectroscopically confirmed blue and red GCs in our galaxy sample is listed in Table~\ref{tbl-2}. Most of the GCs belonging to VCC 685 and VCC 2000 are blue. 

In addition to the Gemini/GMOS spectra of the GCs and GDL, all four of our sample galaxies have published IFU spectroscopy from SAURON \citep{em04, ca11}. Since we have some GDL slits in the SAURON regions, we compare our measured radial velocity and velocity dispersion with those obtained with SAURON (Figure~\ref{diff}). The velocities and velocity dispersions of SAURON data are from a 2D interpolation of their radial velocity and velocity dispersion field. The solid line is a 1:1 relation. We found generally good agreement between these two data sets. SAURON velocities are a little bit larger than our GDL velocities. The offset is 16 km/s and 4 km/s for the velocity and velocity dispersion, respectively, between our GDL data and SAURON data. To facilitate comparison to the SAURON data, we add a 16 km/s velocity offset to our measured velocities. We do not change our velocity dispersion measurements, as 4 km/s is well within the estimated uncertainties.

One possible problem of obtaining stellar kinematics with the GDL technique is that the ``background'' spectrum may be contaminated by residual GC light. To check whether this is a problem, we compare the velocity of the GC and GDL from the same VCC 1231 slits. In Figure~\ref{corr}, we show that there is no correlation between these two velocities, suggesting that the effect of the GC light on the GDL is negligible.

\begin{figure}
\includegraphics[scale=.50]{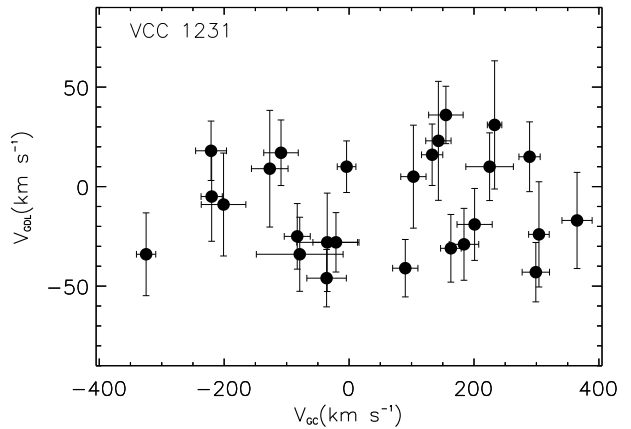}
\caption{Velocity comparison between the GDL and the target observed in the same slits. We see the data point is random, which suggest there is no correlation between the velocity measured for the target and the GDL. }
\label{corr}
\end{figure}

We present the full table of GC and GDL positions, velocities, velocity dispersions (GDL only), and uncertainties in Tables~\ref{gc_table} and \ref{gdl_table}. The radial velocity distribution is shown in Figure~\ref{v_r}.

 \begin{figure*}
\centering 
\plottwo{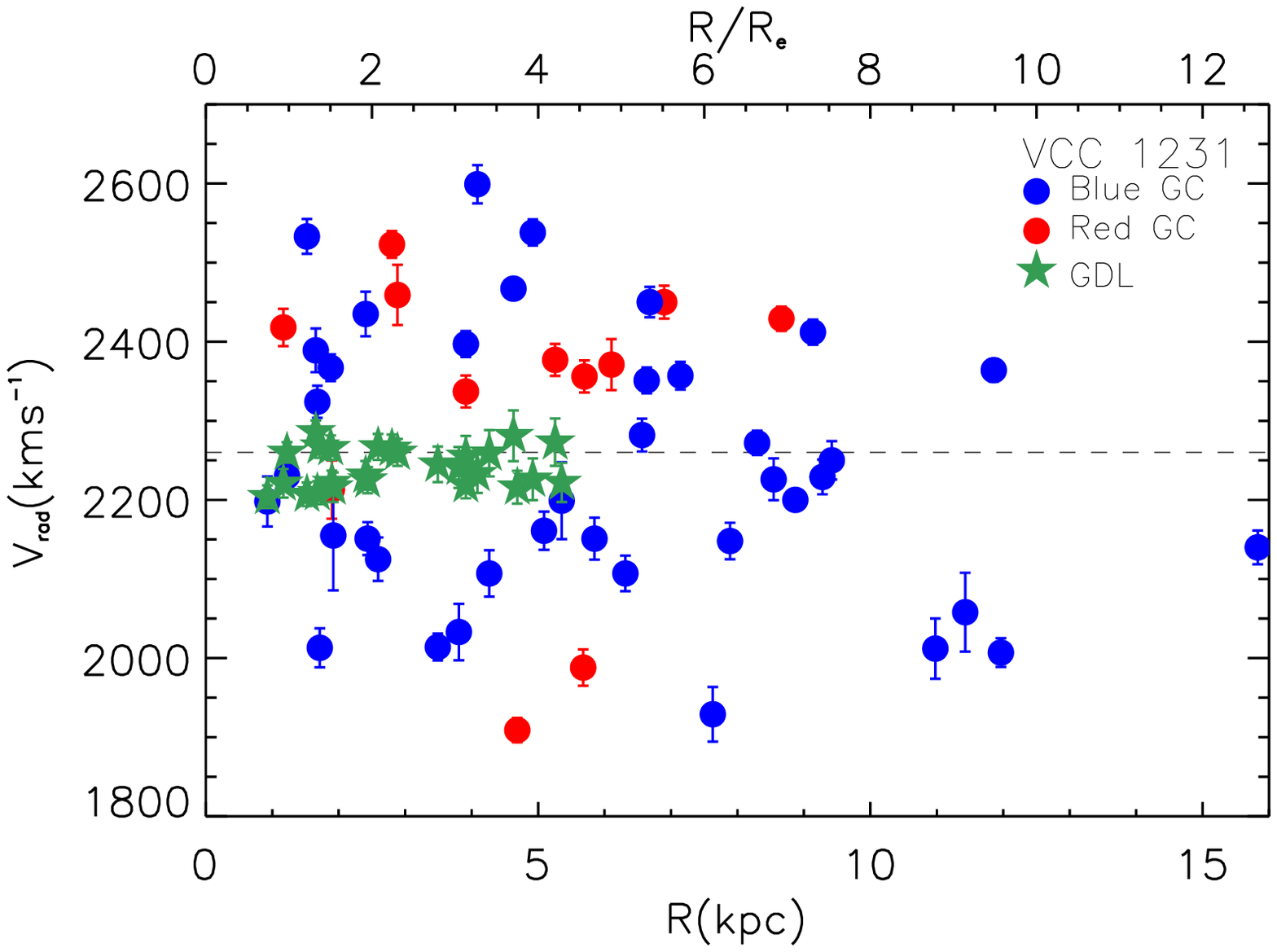}{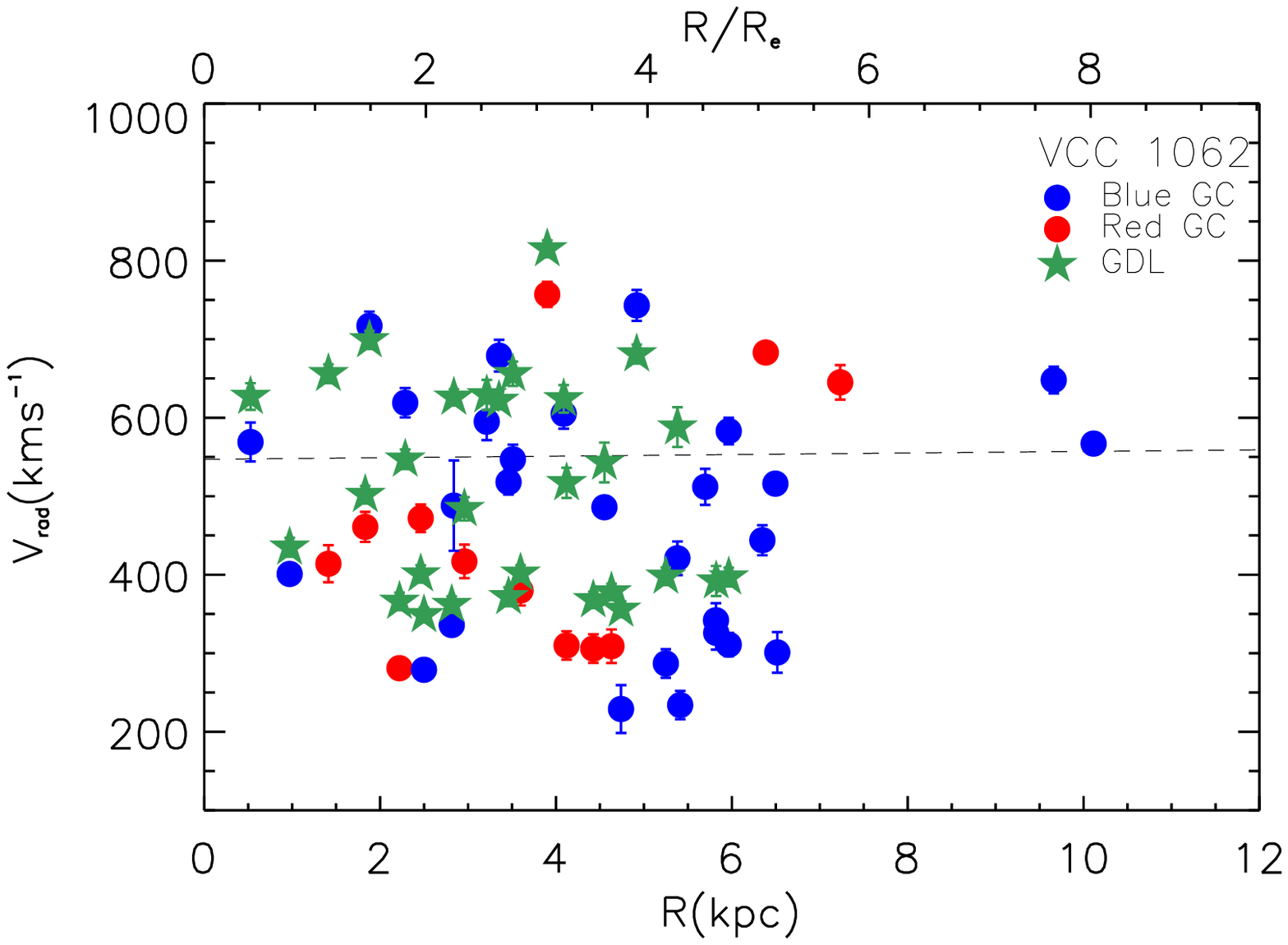}
\plottwo{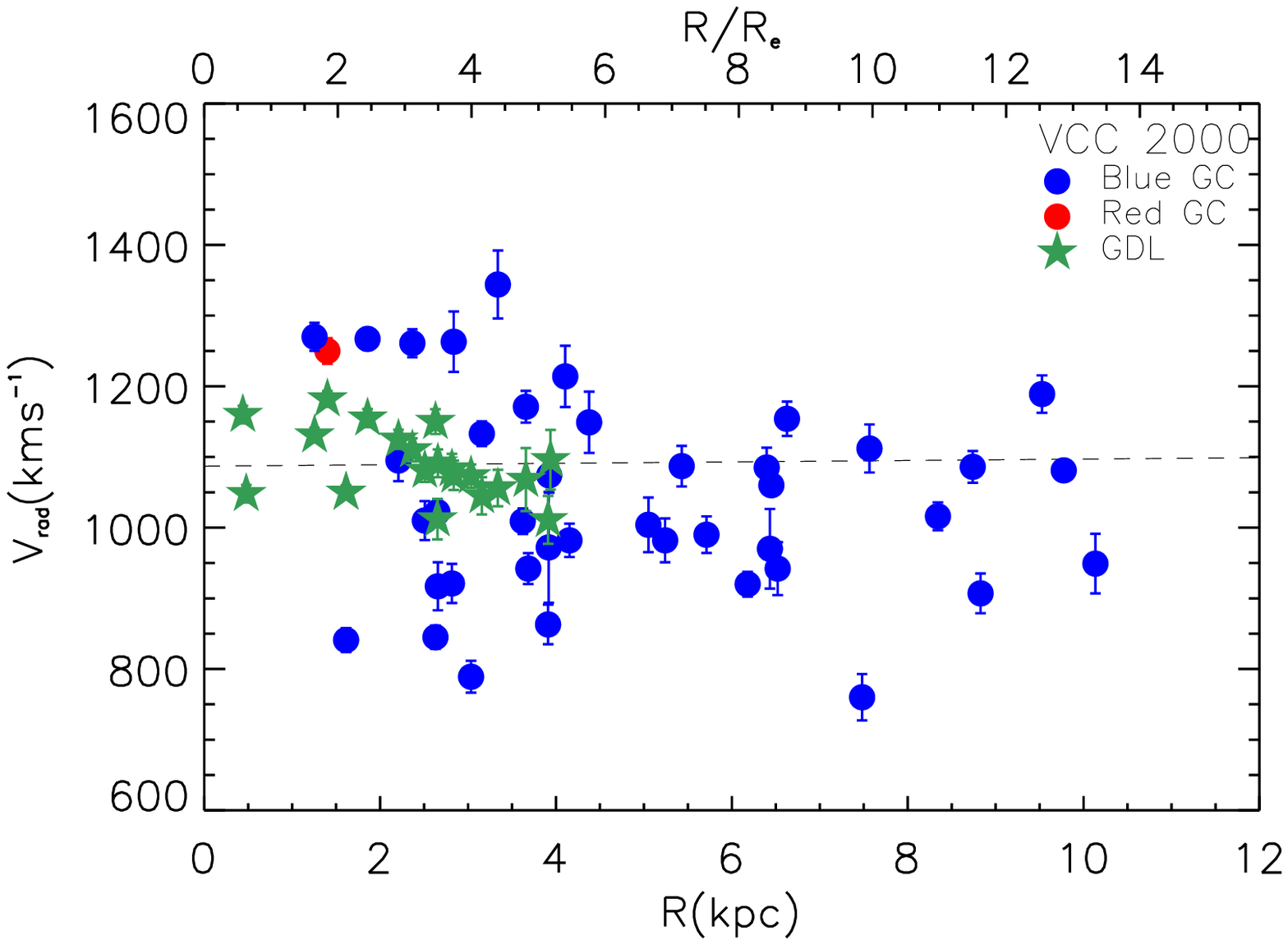}{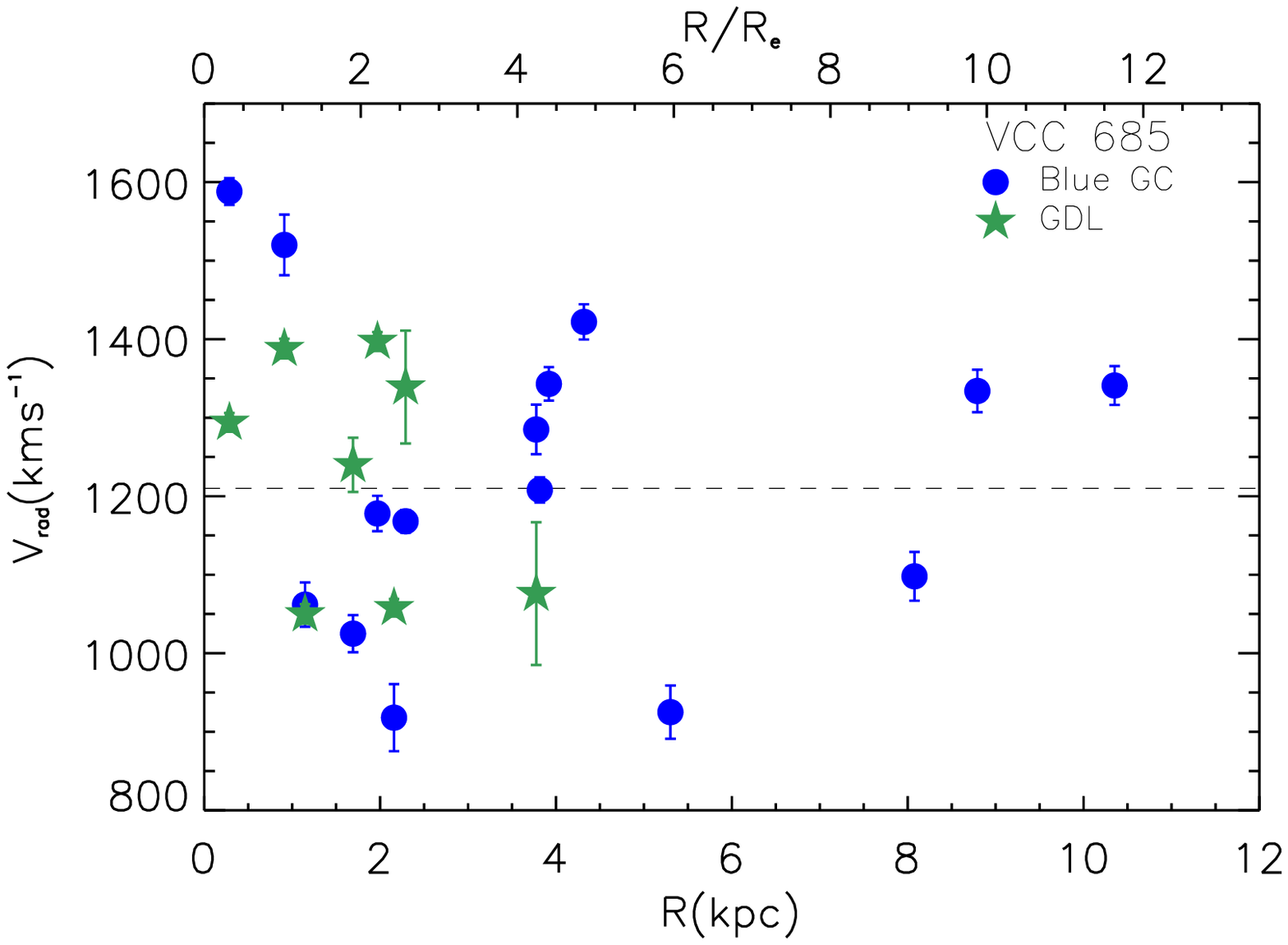}
\caption{The GC and GDL velocity distribution as function of radius. The blue, red circle, and green stars represent blue GCs, red GCs, and GDL, respectively. The galaxy systemic velocity is plotted as a dashed line.} 
\label{v_r}
\end{figure*}

\section{Kinematics Analysis}

\subsection{Two-dimensional Velocity Field} 

For each galaxy, we show the two-dimensional velocity fields as derived from GCs (e.g., Left panel of Figure~\ref{fig:vfields-1231}). We use a Gaussian kernel function to smooth the velocity field.  At each point, the value of the velocity field is estimated by: 

\begin{equation}
V^{'}(x^{'},y^{'})=\frac{\sum_{j} v_{j}\omega_{j}}{\sum_{j}\omega_{j}}
\label{eq1}
\end{equation}
\begin{equation}
\omega_{j}=\frac{1}{\sqrt{2\pi}\sigma}exp\frac{-D^{2}}{2\sigma^{2}}
\label{eq2}
\end{equation}
where $\omega_{j}$ depends on D, which is the distance from j-th point to ($x^{'},y^{'}$), and $\sigma$ which is the width of the Gaussian kernel. The Gaussian kernel can have a fixed bandwidth \citep{pe04} or vary as a function of local density \citep{co09}. The Gaussian kernel we used has a fixed width of $\sigma = 1$ kpc, which is close to $1R_{e}$ for our galaxy sample. We note that our final result is not strongly sensitive to the exact kernel width, so we choose 1~kpc as a reasonable compromise. We also found that the fixed kernel did not lead to information loss in the outer region of galaxy. In the very inner regions, where it might matter, we typically have few data points due to slit crowding.

To reduce the noise due to the finite spatial sampling of our GCs and GDL, we assume the galaxy is point-symmetric in phase space, a valid assumption for triaxial potentials. Each point ($x, y, v$) thus has a mirror counterpart ($-x, -y, -v$), where the $x$ and $y$ is the position of the GC/GDL on the sky, $v$ is the radial velocity of GC/GDL minus the galaxy's systemic velocity. This technique has been adopted in \cite{ar98,pe04,co09}. For the GCs, we plot the smoothed point-symmetric two-dimensional velocity fields for our galaxy sample (e.g., left panel of Figure~\ref{fig:vfields-1231}). For the GDL, we only plot the point-symmetric two-dimensional velocity fields. In order to compare our GDL data to the SAURON data \citep{em04, ca11}, we plot the SAURON data in the central regions (e.g., right panel of Figure~\ref{fig:vfields-1231}). The raw radial velocity fields without smoothing, and without the assumption of point-symmetry (and hence without doubling the data) are shown in the Appendix. The point-symmetric velocity fields are only used for illustration purposes; all quantitative results in the following sections use the original, non-symmetrized data.

\subsection{Rotation Amplitude and Rotation Axis}
\label{sec:rotationfit}
One of the main kinematic characteristics we wish to explore for halo systems is the degree of rotational support, if any. To obtain the rotation amplitude and axis for the GC systems and stars, we use methods which have been used previously (\citealt{fo11,str11,po13}). For this kinematic analysis, we use the original data without point-symmetric folding or smoothing. For each bin in radius, if we assume the radial velocity distribution is Gaussian, then the equivalent $\chi^{2}$ is:

\begin{equation}
\chi^{2}=\sum_{i}\frac{(v_{i}-v_{\rm mod})^{2}}{\sigma_{\rm p}^{2}+(\Delta v_{i}^{2})}+{\rm{ln}}(\sigma_{\rm p}^{2}+\Delta v_{i}^{2})
\label{eq3}
\end{equation}
\begin{equation}
v_{\rm mod}=v_{0}\pm\frac{\rm V_{rot}}{\sqrt{1+(\frac{\rm tan(\rm PA_{i}-\rm PA_{kin})}{\rm q_{kin}})^{2}}}
\label{eq4}
\end{equation}

where $v_{i}$ is the radial velocity, $\Delta v$ is the velocity error, $\sigma_{\rm p}$ is the velocity dispersion, $\rm V_{rot}$ is the rotation amplitude, $v_{0}$ is the galaxy systemic velocity. $\rm PA_{i}$, $\rm PA_{kin}$, $\rm q_{kin}$ is the position angle of each data point, position angle of the kinematic major axis, and the kinematic axis ratio, respectively. For the GCs, we minimize the Equation~3. For the GDL, the $\chi^{2}_{0}$ is defined as follows:

\begin{equation}
\chi^{2}_{0}=\sum_{i}\frac{(v_{i}-v_{\rm mod})^{2}}{(\Delta v_{i}^{2})}+{\rm{ln}}(\Delta v_{i}^{2})
\label{eq5}
\end{equation}

For both the GCs and GDL, we assume $\rm q_{kin}$ is equal to photometric axis ratio of the galaxy light since $\rm q_{kin}$ is difficult to constrain with our data. We also fit with $\rm q_{kin}$=1 (round), and the results do not show any significant differences. We use Monte Carlo simulations to estimate the errors of the fitted parameters. At each data point, we generate the velocity with the Gaussian distribution defined by the intrinsic dispersion and observed velocity error. After 1000 Monte Carlo simulations, we used the 68 percent confidence intervals as the 1$\sigma$ uncertainties of the fitted parameters. 

\begin{deluxetable*}{ccccccccc}
\tabletypesize{\scriptsize} 
\centering 
\tablecaption{All GC kinematics fitting results\label{tbl-2}}
\tablewidth{0pt}
\tablehead{
\colhead{ID} & \colhead{$N_{blue}$}& \colhead{$N_{red}$} & \colhead{$\rm V_{m}$} & 
\colhead{$\rm V_{rot}$} & \colhead{$\rm Bias$} & \colhead{$\sigma$} &
\colhead{${\rm V_{rot}}/{\sigma}$} & \colhead{$\rm PA_{kin}$} \\
\colhead{} & \colhead{} & \colhead{} & \colhead{($\rm km\ s^{-1}$)} & \colhead{($\rm km\ s^{-1}$)} & \colhead{($\rm km\ s^{-1}$)} &
\colhead{($\rm km\ s^{-1}$)} & \colhead{} & \colhead{(degree)}  \\
\colhead{(1)} & \colhead{(2)} & \colhead{(3)} & \colhead{(4)} & \colhead{(5)} &
\colhead{(6)} & \colhead{(7)} & \colhead{(8)}  & \colhead{(9)} 
}
\startdata
VCC1231&39&12& 2268&35$\pm$40 &31&171$\pm$24&0.2$\pm$0.2&216$\pm$78 \\

VCC2000&42&1& 1062&100$\pm$42 &5&128$\pm$20&0.8$\pm$0.4&198$\pm$32 \\

VCC1062&28&12& 484  &153$\pm$28&8 &108$\pm$18  &1.4$\pm$0.4&92$\pm$12 \\

VCC685&15&0& 1244&66$\pm$120&106&195$\pm$45&0.3$\pm$0.6&92$\pm$91 
\enddata
\tablecomments{GC kinematics results for our galaxy sample. The number of blue GCs, red GCs, the mean velocity, rotation velocity (amplitude), bias of the rotation velocity, velocity dispersion, dominance parameter ($\rm V_{rot}/\sigma$), and kinematic position angle (PA) are shown in columns (2) to (9), respectively. See Section~\ref{sec:rotationfit} for details on the estimation of these parameters.}
\end{deluxetable*}

\begin{deluxetable*}{ccccccccc}
\tabletypesize{\scriptsize} 
\centering 
\tablecaption{Blue and Red GC kinematics fitting results\label{tbl-33}}
\tablewidth{0pt}
\tablehead{
\colhead{ID} & \colhead{${\rm V_{rot,B}}$}& \colhead{${\rm V_{rot,R}}$} & \colhead{$\rm Bias_{B}$} & 
\colhead{$\rm Bias_{R}$} & \colhead{$\sigma_{B}$} & \colhead{$\sigma_R$} &
\colhead{$\rm PA_{kin,B}$} & \colhead{$\rm PA_{kin,R}$} \\
\colhead{} & \colhead{($\rm km\ s^{-1}$)} & \colhead{($\rm km\ s^{-1}$)} & \colhead{($\rm km\ s^{-1}$)} & \colhead{($\rm km\ s^{-1}$)} & \colhead{($\rm km\ s^{-1}$)} &
\colhead{($\rm km\ s^{-1}$)} & \colhead{(degree)} & \colhead{(degree)}  \\
\colhead{(1)} & \colhead{(2)} & \colhead{(3)} & \colhead{(4)} & \colhead{(5)} &
\colhead{(6)} & \colhead{(7)} & \colhead{(8)}  & \colhead{(9)} 
}
\startdata
VCC1231&73$\pm$43&83$\pm$115&21&80&157$\pm$24&183$\pm$51&235$\pm$48&96$\pm$87 \\

VCC2000&97$\pm$44&-&7&-&127$\pm$20&-&196$\pm$30&-\\

VCC1062&145$\pm$37&183$\pm$57&9&20&108$\pm$21&104$\pm$30&88$\pm$16&100$\pm$24 

\enddata
\tablecomments{Blue (B) and red (R) GC kinematics results for our galaxy sample.}
\end{deluxetable*}

\begin{deluxetable*}{cccccccccc}
\tabletypesize{\scriptsize} 
\centering 
\tablecaption{Bright and Faint GC kinematics fitting results\label{tbl-44}}
\tablewidth{0pt}
\tablehead{
\colhead{ID} & \colhead{${\rm V_{rot,b}}$}& \colhead{${\rm V_{rot,f}}$} & \colhead{$\rm Bias_{\rm b}$} & 
\colhead{$\rm Bias_{\rm f}$} & \colhead{$\sigma_{\rm b}$} & \colhead{$\sigma_{\rm f}$} &
\colhead{$\rm PA_{kin,b}$} & \colhead{$\rm PA_{kin,f}$} & \colhead{$g$}\\
\colhead{} & \colhead{($\rm km\ s^{-1}$)} & \colhead{($\rm km\ s^{-1}$)} & \colhead{($\rm km\ s^{-1}$)} & \colhead{($\rm km\ s^{-1}$)} & \colhead{($\rm km\ s^{-1}$)} &
\colhead{($\rm km\ s^{-1}$)} & \colhead{(degree)} & \colhead{(degree)}& \colhead{(magnitude)}  \\
\colhead{(1)} & \colhead{(2)} & \colhead{(3)} & \colhead{(4)} & \colhead{(5)} &
\colhead{(6)} & \colhead{(7)} & \colhead{(8)}  & \colhead{(9)}& \colhead{(10)} 
}
\startdata
VCC1231&118$\pm$62&28$\pm$56&17&54&171$\pm$31&162$\pm$30&224$\pm$45&94$\pm$98&22.274 \\

VCC2000&119$\pm$64&$99\pm46$&20&16&138$\pm$31&$119\pm27$&204$\pm$45&$182\pm31$&22.250\\

VCC1062&139$\pm$29&170$\pm$50&8&13&77$\pm$17&134$\pm$28&88$\pm$16&94$\pm$20&22.458

\enddata
\tablecomments{Col.(2-9): Bright (b) and faint (f) GC kinematics results for our galaxy sample. Col(10): The divided g band magnitude for the bright and faint group.}
\end{deluxetable*}

As discussed in \cite{str11}, allowing the position angle of the rotation axis to be a free parameter creates a tendency to overestimate the rotation amplitude. We use Monte Carlo simulations, similar to those described in \cite{po13}, to estimate the magnitude of this bias. The bias depends on $(\rm V_{rot}/\sigma)$, bin size and the azimuthal distribution of data points. When $(\rm V_{rot}/\sigma)$ and bin size are small, the bias will increase. Sometimes, the estimated bias is comparable to or larger than the measured $v_{rot}$ (i.e., for VCC 685). We list the estimated bias in Table~\ref{tbl-2}. However, in both cases where the bias is comparable to the rotation amplitude, the uncertainty in the rotation amplitude is also large. Thus, our conclusions are based on the rotation amplitudes without bias correction.

To compare the SAURON data, GDL, and GCs, we carry out rolling fits with the raw data binned in elliptical annuli. The number of GCs in each bin depend on the sampling of the GC system. There are 10--20 GCs and 7--10 GDL slits in one bin. We move the bins one data point at a time. The galaxy systemic velocity is fixed to the value in Table~\ref{tbl-1}. The fitting results for the GC systems are listed in Table~\ref{tbl-2}, \ref{tbl-33}, \ref{tbl-44}.

\section{Results} 

\subsection{VCC 1231}

\begin{figure*}
\plottwo{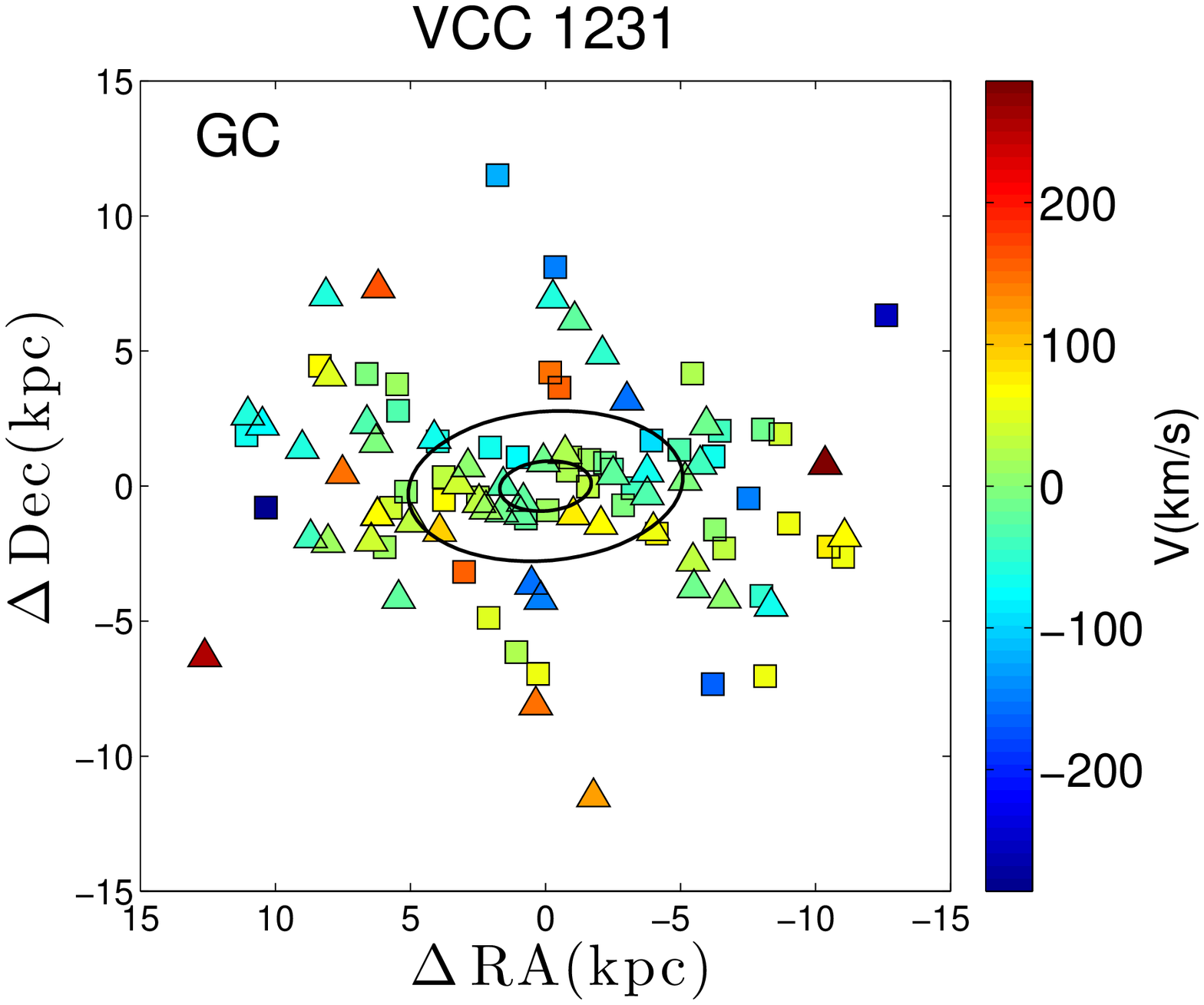}{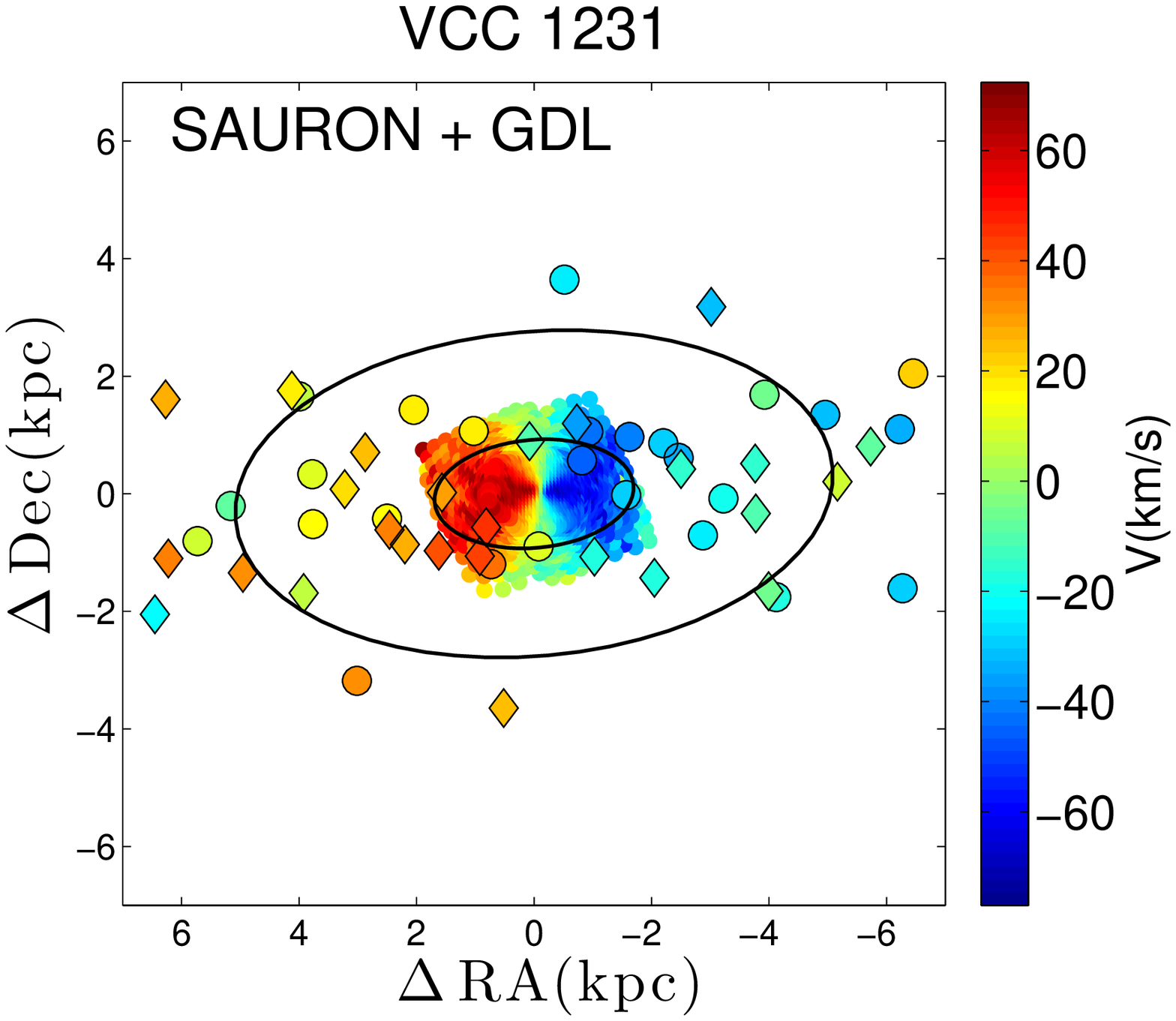}
\caption{The 2D velocity fields of our sample galaxies. {\it Left column}: Smoothed GC velocity fields. Squares represent the location of the GCs, and the triangles represent the point-symmetric counterparts. The color represents the value of the smoothed velocity field at that point, with the color scale on the right hand side of each panel. The ellipses represent 1 and 3 effective radii, the values of which are listed in Table~\ref{tbl-1}.  {\it Right column}: Stellar velocity field as derived from the GDL data without smoothing. The circles represent the location of our slits, and the diamonds represent the point-symmetric counterparts. The SAURON data are plotted in a contiguous rectangle at the galaxy center. Note that the spatial and velocity scales differ in each plot in order to best display the dynamic range. In all plots, North is up, and East is to the left.}
\label{fig:vfields-1231}
\end{figure*}

\begin{figure*}
{\center \includegraphics[scale=.90]{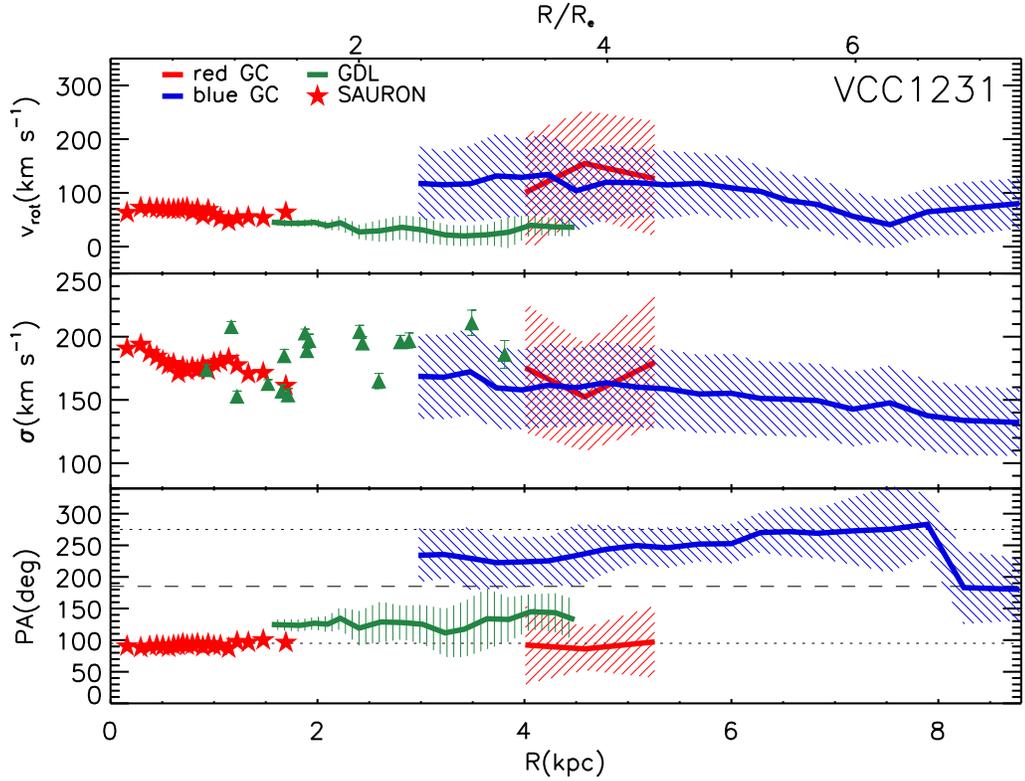}
\caption{The kinematic fitting results as a function of radius in VCC 1231. The top panel shows the rotation amplitude. The middle panel shows the velocity dispersion. The bottom panel shows the kinematic position angle . The red stars represent the fitting results for the SAURON data. The lines represent the fitting results from the multislit galaxy diffuse light data (green line), red and blue GCs (red and blue), with the shaded regions showing the $1\sigma$ uncertainty of the fit. The green triangles represent the velocity dispersion measured in each individual slit within $3R_e$. The photometric major and minor axes are represented with the dotted and dashed line in the bottom panel, respectively. The GDL measurements have a rotation axis similar to that of the SAURON data. However, the rotation amplitude (top panel) is generally lower than for the SAURON data. Red and blue GCs have the same rotation velocity but rotate around the difference axes (although with low significance). The velocity dispersion of the GDL data is consistent with that of the SAURON data. The velocity dispersion profile of blue and red GC stays flat out to large distance.}
\label{1231-1}}
\end{figure*}

\begin{figure}
\centering
\includegraphics[width=1.0\columnwidth]{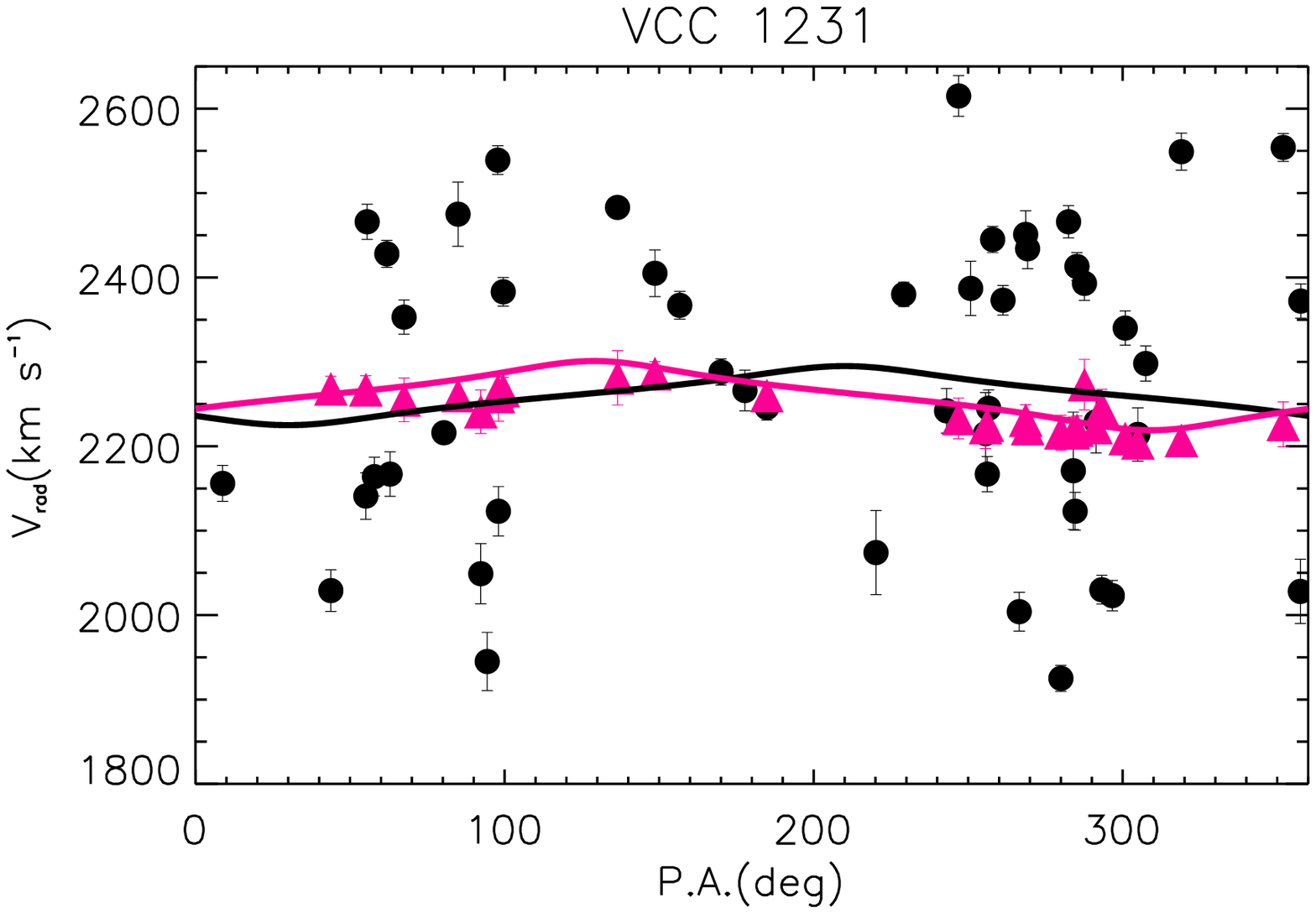}
\caption{Radial velocities versus position angle in VCC 1231. The black circles represent the GC data and pink triangles represent the GDL data. The black and pink solid lines represent the best-fitting rotation curve for the GCs and GDL, respectively. GCs show no rotation or very small rotation. The GDL shows small rotation. The curves are not perfect sinusoids in PA because we include the ellipticity of the system in the fit.}
\label{1231-2}
\end{figure}

The 2D velocity field of VCC 1231 is shown in Figure~\ref{fig:vfields-1231}. The GC and stellar velocity fields are shown in the left and right panels, respectively. The velocity field of the GCs does not show any obvious ordered motion or significant rotation. The central stellar kinematics as shown by the SAURON data, however, show clear rotation, with a peak amplitude of $\rm V_{rot}\sim$ 60 km $\rm s^{-1}$. Our GDL data goes out farther in radius than the SAURON data and is consistent with the SAURON data in the overlap region. In the outer regions, the rotation amplitude declines. Recently, \cite{fo13} also use a similar method to get GDL kinematics in VCC 1231, and their results show both minor and major axis rotation in the outer regions. They argue that it is tell-tale sign of triaxiality (e.g., \citealt{pe04}). Although we do not see this kinematic structure in our GDL data, it possibly due to the fact that most of the slits where we can derive stellar kinematics are along the major axis, so it is difficult to see any major axis rotation. We also lack the data to confirm the minor axis rotation out to 5 kpc. 

In Figure~\ref{1231-1}, we plot the rotation amplitude, $\rm V_{rot}$, velocity dispersion, $\sigma$ and rotation axis, $\rm PA_{kin}$, profiles with radius. We can see that $\sigma$ and the kinematic position angle of the stars in the outer regions (GDL, green solid line or triangle) is generally consistent with that measured by SAURON in the inner regions (red stars). In the region where they overlap, $\rm V_{rot}$ and $\rm PA_{kin}$ of the stars in the outer regions are a little bit offset from the SAURON data. One possibility is our GDL kinematics are not determined from a uniform spatial distribution, which can lead to fitting bias. Moreover, the radial bin size for the SAURON profile is much smaller than for our GDL profile. The $\rm V_{rot}$ of GDL data is in agreement with SAURON data in 2$\sigma$ confidence. 

The GDL rotation amplitude is smaller than that in the inner regions. It decreases with radius and $\rm V_{rot}\sim$ 20 km $\rm s^{-1}$ at 3 kpc. The $\rm PA_{kin}$ is $\sim120^{\rm o}$ and stays the same out to 3 kpc. Similar results were also found in \cite[their Figure 3]{fo13}. We plot the GDL velocity dispersion profile in the middle panel of Figure~\ref{1231-1}. Each green triangle represents the measured velocity dispersion from one slit. We derived the SAURON velocity dispersion profile (red stars) by taking the mean velocity dispersion within an elliptical annulus on the 2D velocity dispersion map. These two measurements are not entirely equivalent, making GDL profile noisier than the SAURON velocity dispersion profile. However, on average, these two datas show a good agreement at the same position (see Figure~\ref{diff}).
 
 The $\rm V_{rot}$ of red and blue GCs are similar and both nominally larger than that of the GDL, but the error is large and the low number of GCs ($10\lesssim N \lesssim20$) in each bin may lead to the  $\rm V_{rot}$ being overestimated. After the bias correction, the rotation velocity of red and blue GC is $3\pm115\rm km\ s^{-1}$ and $52 \pm43\rm km\ s^{-1}$. The red GC rotation is consistent with zero, within the uncertainties.

Although the rotation amplitudes for the two GC subpopulations are similar, the best fit rotation axes are very different, which is intriguing. The red GCs appear to rotate around the photometric minor axis, which is consistent with the stars in the inner regions, while the blue GCs appear to rotate around an axis somewhere between the photometric major and minor axes. These very different PAs lead to the whole GC system showing no significant rotation (see Figure~\ref{1231-2}). The best-fitting rotation amplitude for the full GC system is $35\pm40$ $\rm km\ s^{-1}$ with $\rm PA_{kin}$ = $216\pm78^{\circ}$. To test the significance of this rotation signal, we use the method of ``scrambling'' the properties of the observed GCs and refitting the kinematics. This method  had been applied in testing the rotation of the M87 UCD system \citep{hx15} and the GC systems of low-mass galaxies. We fix the GC position angles and randomly shuffle the velocity and velocity error of each GC. We repeated this exercise 1000 times and each time we apply the same kinematics fitting as we did for our original data. We find that 88\% and 33\% of the simulations for the red and blue GCs, respectively, have rotation amplitudes at least as high as what we fit for our observed data. We conclude that the significance of these rotation amplitudes is not high, but future observations to obtain more GC velocities may shed further light on this issue.

In Figure~\ref{1231-2}, we also plot the GDL best-fitting rotation curve. The GDL best-fitting amplitude is $41\pm6\ \rm km\ s^{-1}$ with $\rm PA_{kin}$ = $129\pm7^{\circ}$, which is smaller than the SAURON data in the inner region.

\subsection{VCC 2000}
\begin{figure*}
\plottwo {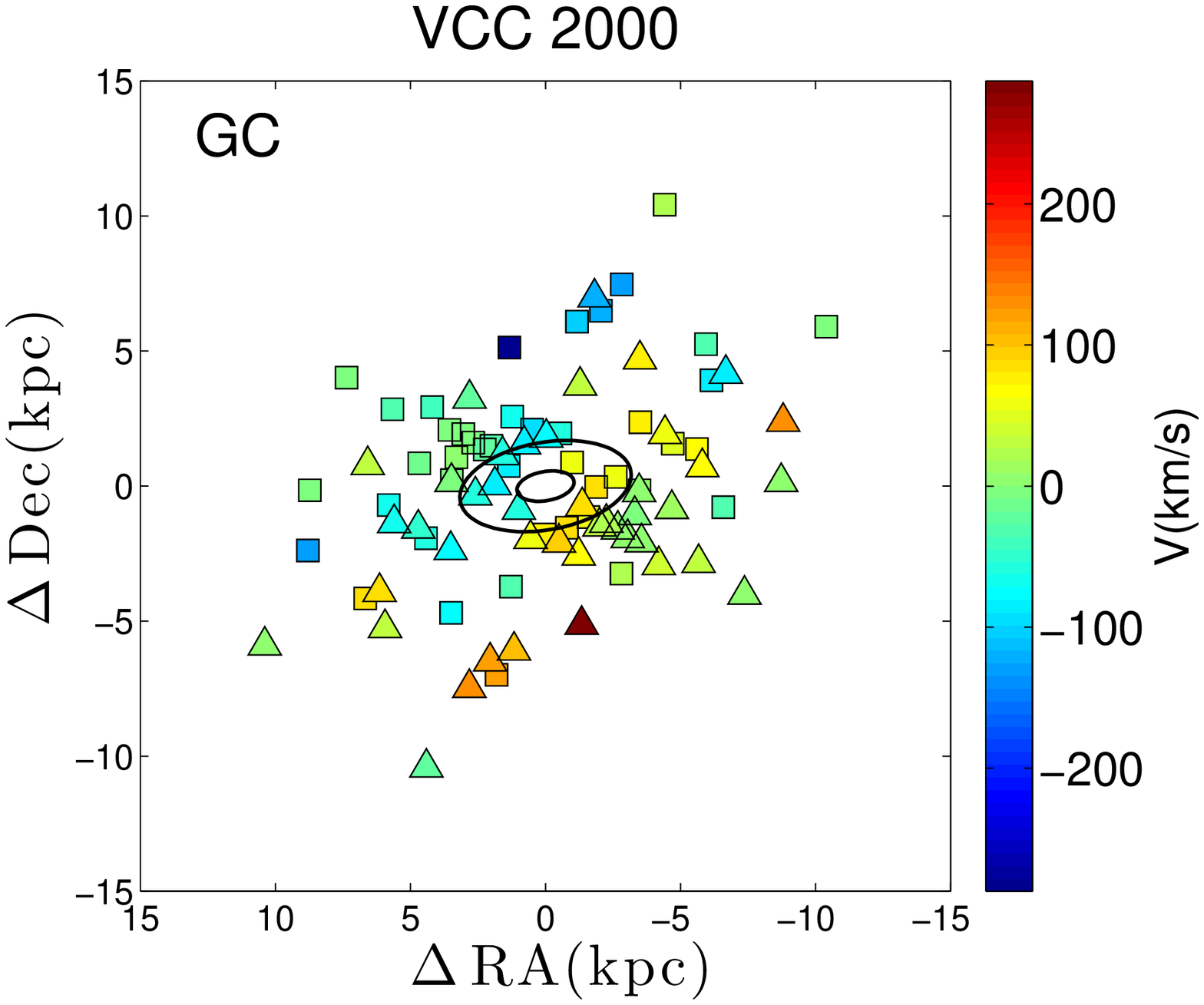}{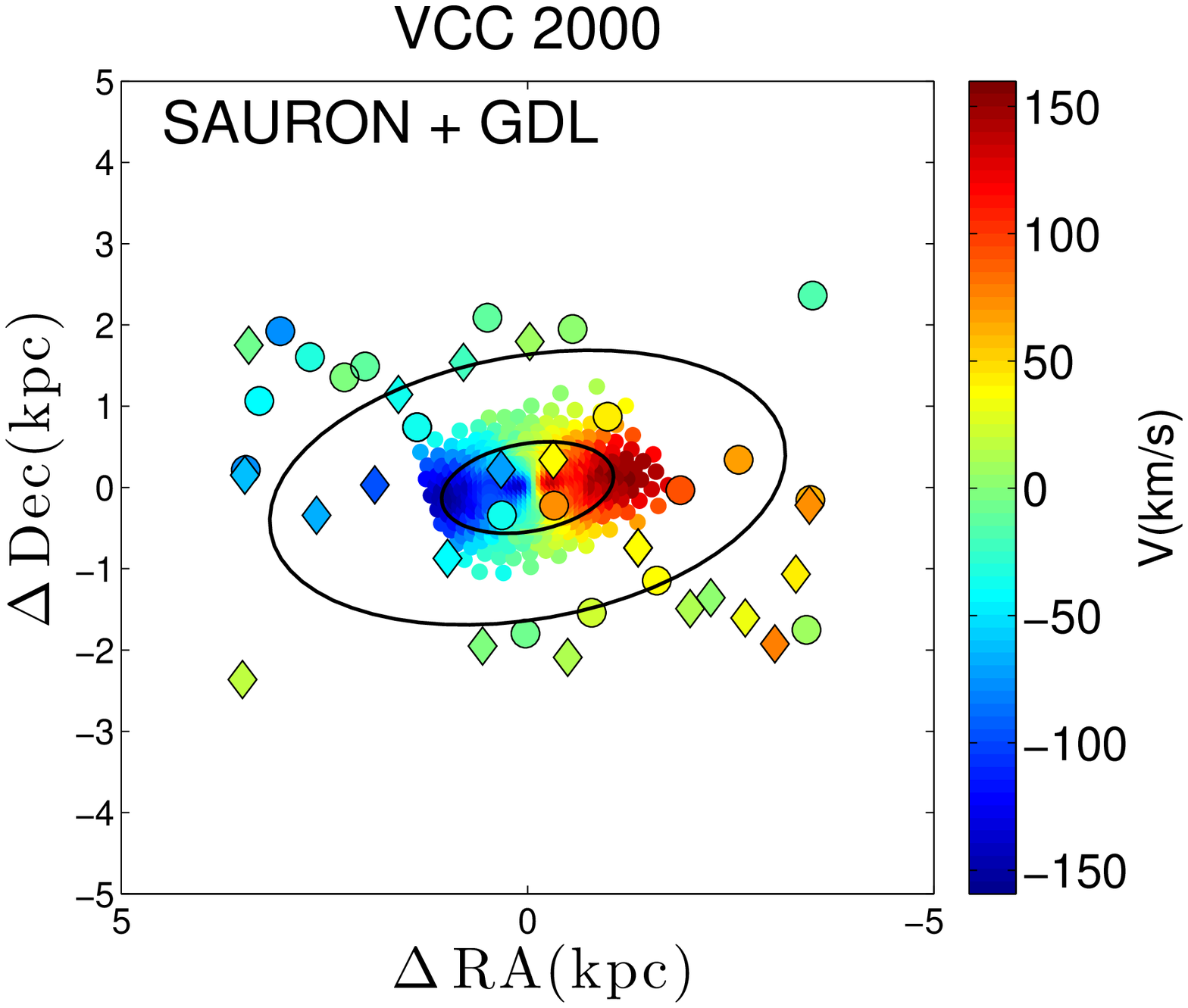}
\caption{Same as in Figure~\ref{fig:vfields-1231}, but for VCC 2000}
\label{fig:vfields-2000}
\end{figure*}

\begin{figure*}
{\center \includegraphics[scale=.90]{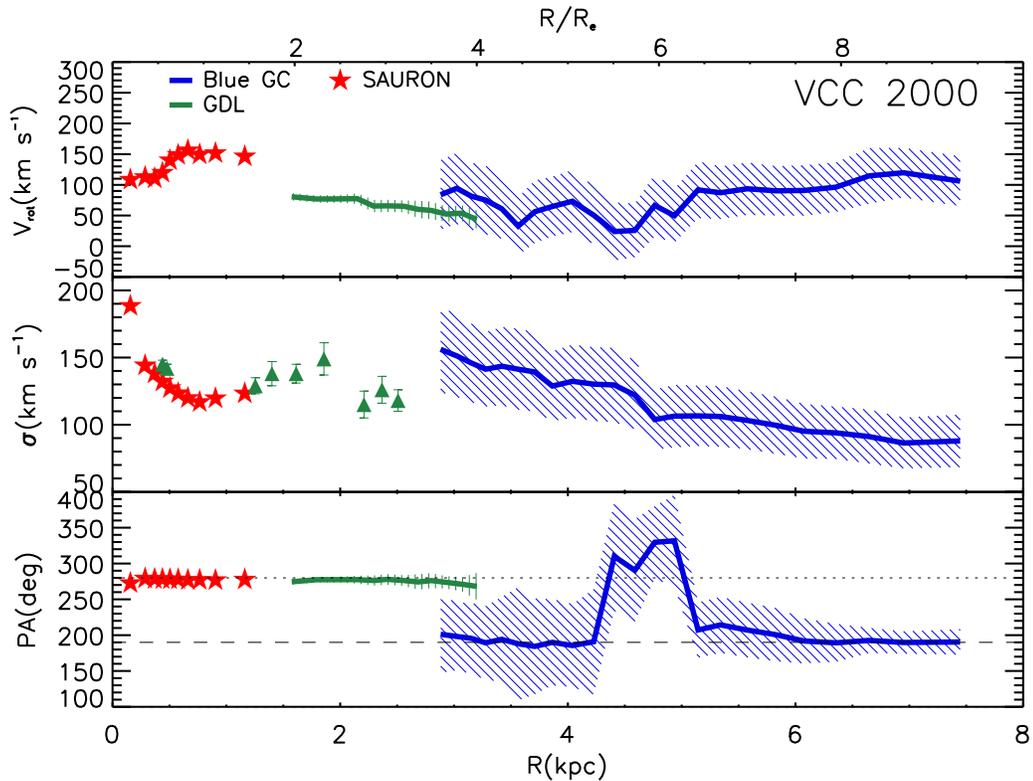}
\caption{The kinematics fitting result as a function of radius in VCC 2000. The symbols are shown in the upper left. The SAURON data and GDL show rotation around the same axis, but the GDL data (which is at larger radius) has a smaller rotation amplitude than the SAURON data. The velocity dispersion profile of the GCs decreases with radius. The $PA_{kin}$ of the blue GCs is largely not coincident with the $PA_{kin}$ of the GDL. }
\label{2000-1}}
\end{figure*}

\begin{figure}
\includegraphics[width=1.0\columnwidth]{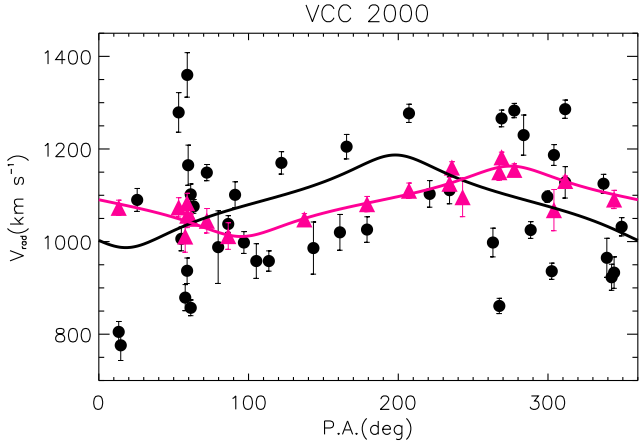}
\caption{Radial velocities versus position angle in VCC 2000. The GCs and the GDL both show rotation, but with different rotation axes.}
\label{2000-2}
\end{figure}

The right panel of Figure~\ref{fig:vfields-2000} shows the 2D velocity field of stars in VCC 2000. Both the SAURON and GMOS data show obvious rotation with the same $\rm PA_{kin}$, although the rotation velocity in the inner regions seems larger than in the outer regions. The globular clusters show rotation in the inner region $\lesssim$ 3$R_{e}$ and outer region. In the middle region, however, the rotation velocity dips to $\sim40{\rm km\ s}^{-1}$. 
There are 42 blue GCs and only 1 red GC in VCC 2000, so we only plot the blue GCs in Figure~\ref{2000-1}. 

Although there is little overlap between the SAURON and GMOS data for the stars, the stellar rotation axis ($\rm PA_{kin}$) is consistent across both data sets, with both showing rotation about the photometric minor axis. However, the amplitude of the rotation appears to decline with radius. The rotation amplitude of the GC system is consistent with that of the stars in their region of overlap, but the axis of rotation is off by 90 degrees, with the GCs rotating about the photometric major axis. The GCs and the stars appear kinematically decoupled in this galaxy. The velocity dispersion of the GCs monotonically declines. 


The best fitting $PA$ versus $V_{rad}$ curve is shown in Figure~\ref{2000-2}. The GC best-fitting amplitude is $100\pm42$ $\rm km\ s^{-1}$ with $\rm PA_{kin}$ = $198\pm32^{\circ}$. The GDL best-fitting amplitude is $76\pm6$ $\rm km\ s^{-1}$ with $\rm PA_{kin}$ = $275\pm4^{\circ}$. This again shows that the GCs and stars both show rotation, but with the different position angles for their rotation axes. The rotation velocity of the GDL is smaller than that derived from the SAURON data in the inner regions.

\subsection{VCC 1062}
\begin{figure*}[b]
\plottwo{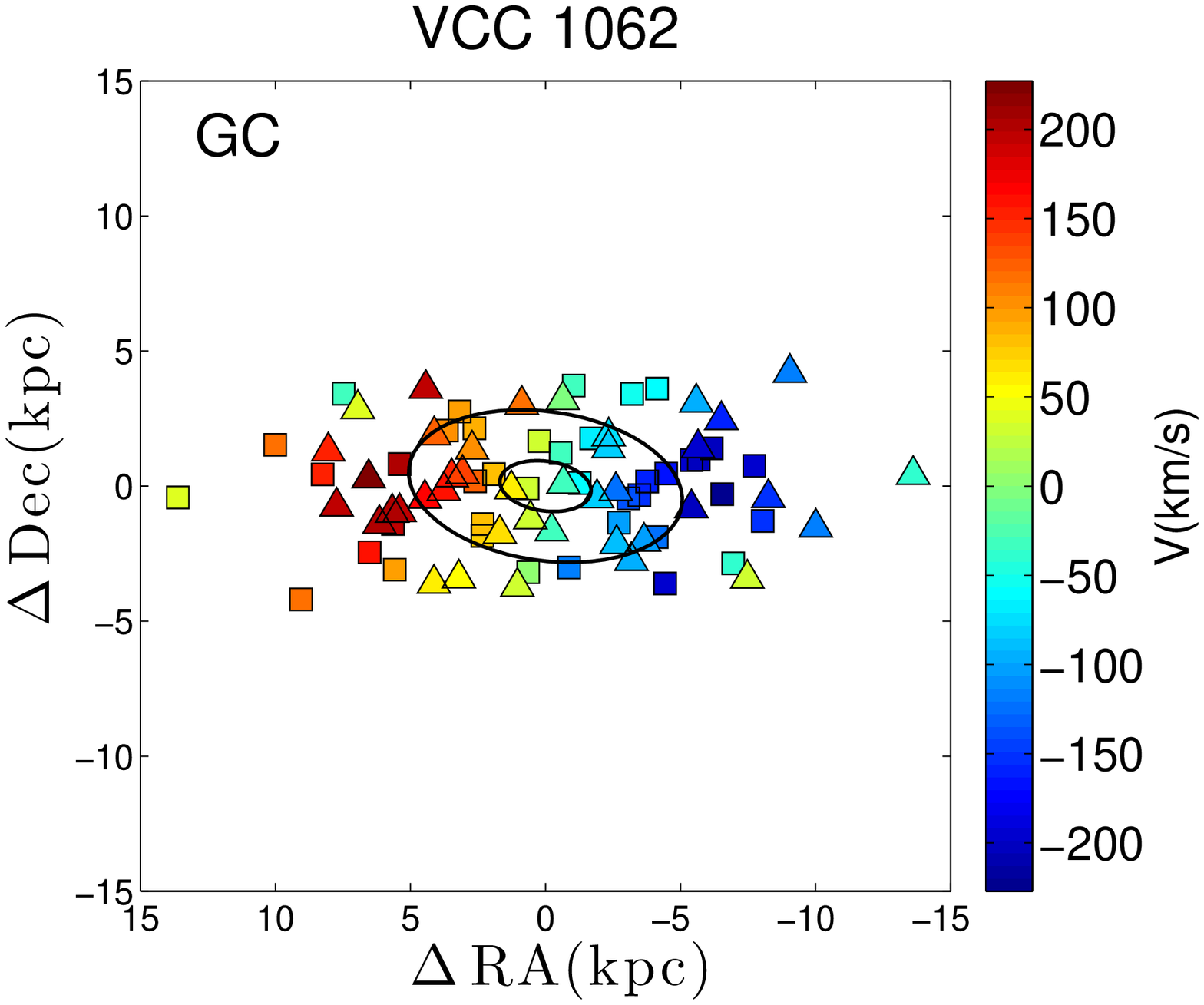}{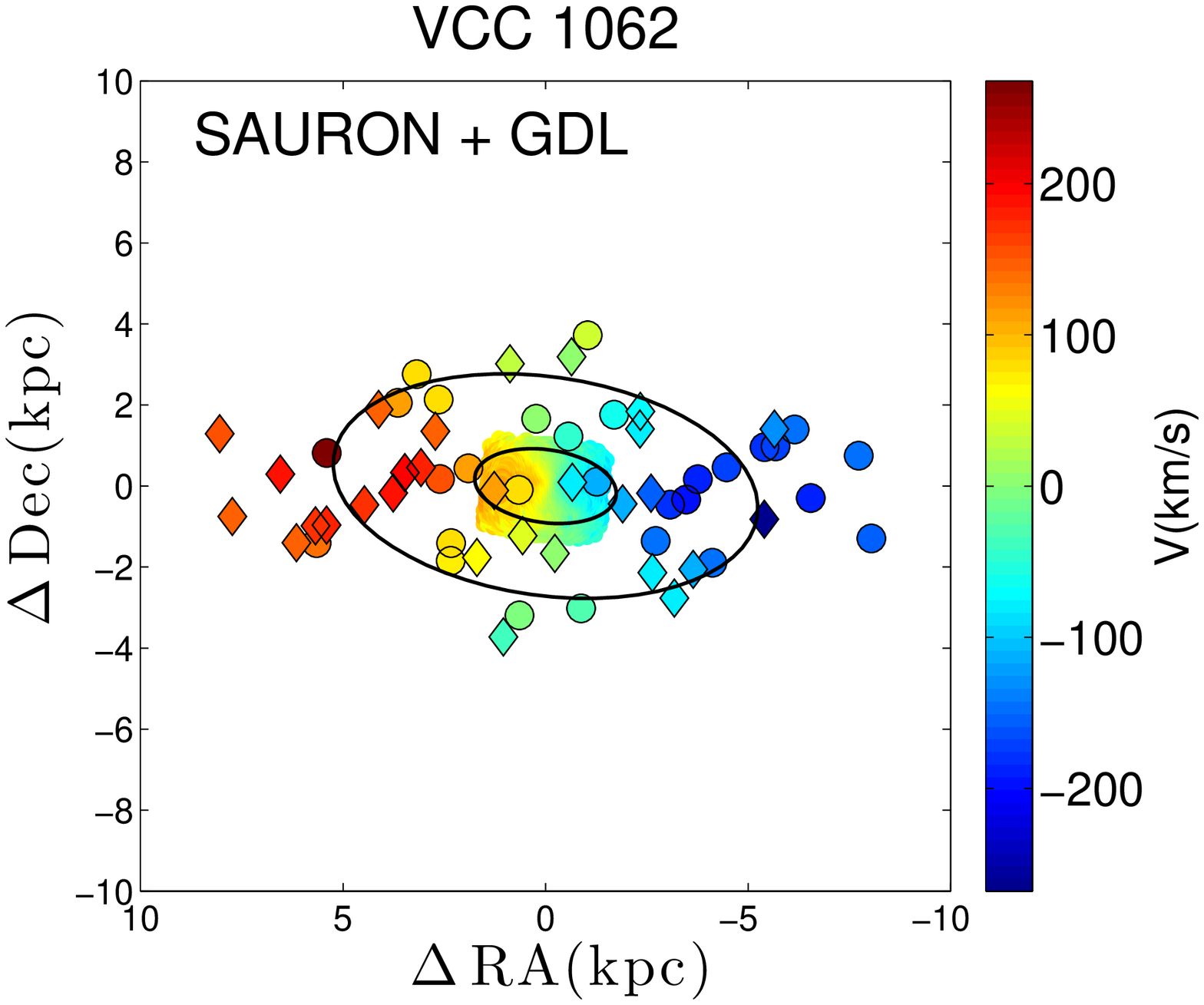}
\caption{Same as in Figure~\ref{fig:vfields-1231}, but for VCC 1062.}
\label{fig:vfields-1062}
\end{figure*}

The 2D velocity fields of VCC 1062 are shown in Figure~\ref{fig:vfields-1062}. In this galaxy, both the stars and GCs show strong rotation about the minor axis at all radii. The blue and red GCs also exhibit similar kinematics. In Figure~\ref{1062-1}, we can see that the rotation amplitude of the stars and GCs at large radii ($R>2$~kpc) stays flat at around 200 $\rm km\ s^{-1}$. The GCs also have a rotation axis similar to that fit to the stellar SAURON data in the inner region. 

The best fitting rotation curve of VCC 1062 is shown in Figure~\ref{1062-2}. We can see that the stars (GDL) and GC both show a significant rotation with the same $\rm PA_{kin}$. The GCs best-fitting amplitude is $153\pm28$ $\rm km\ s^{-1}$ with $\rm PA_{kin}$  =  $92\pm12^{\circ}$. The GDL best-fitting amplitude is $177\pm3$ $\rm km\ s^{-1}$ with $\rm PA_{kin}$ = $84\pm1^{\circ}$. The peak rotation is essentially along the photometric major axis (i.e., rotating about the minor axis). The rotation velocity of the GDL is larger than that for the SAURON data in the inner region.

\begin{figure*}
\includegraphics[scale=.90]{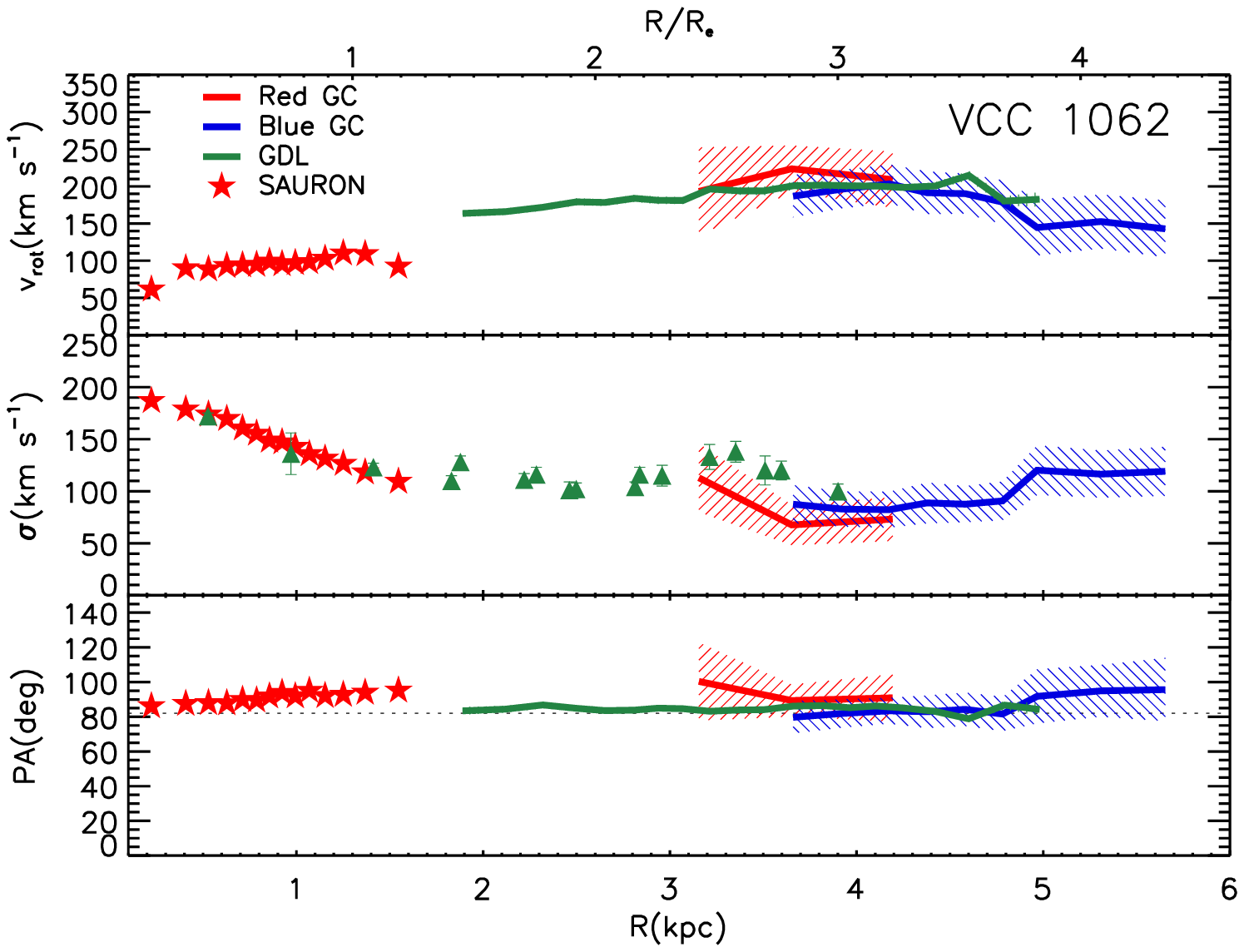}
\caption{The kinematic fitting results as a function of radius in VCC 1062. Red GCs, blue GCs, and the GDL all show significant rotation around the same rotation axis. The velocity dispersion profile of the red and blue GCs stays flat between 3 and 5 kpc, but the blue GC dispersion profile increases beyond 5 kpc.}
\label{1062-1}
\end{figure*}

\begin{figure}
\includegraphics[width=1.0\columnwidth]{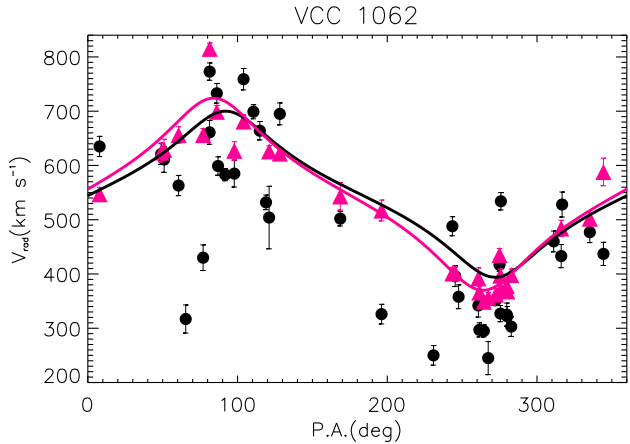}
\caption{Radial velocities versus position angle in VCC 1062. The GCs and GDL both show significant rotation about the same axis.}
\label{1062-2}
\end{figure}

\subsection{VCC 685}

\begin{figure*}
\plottwo {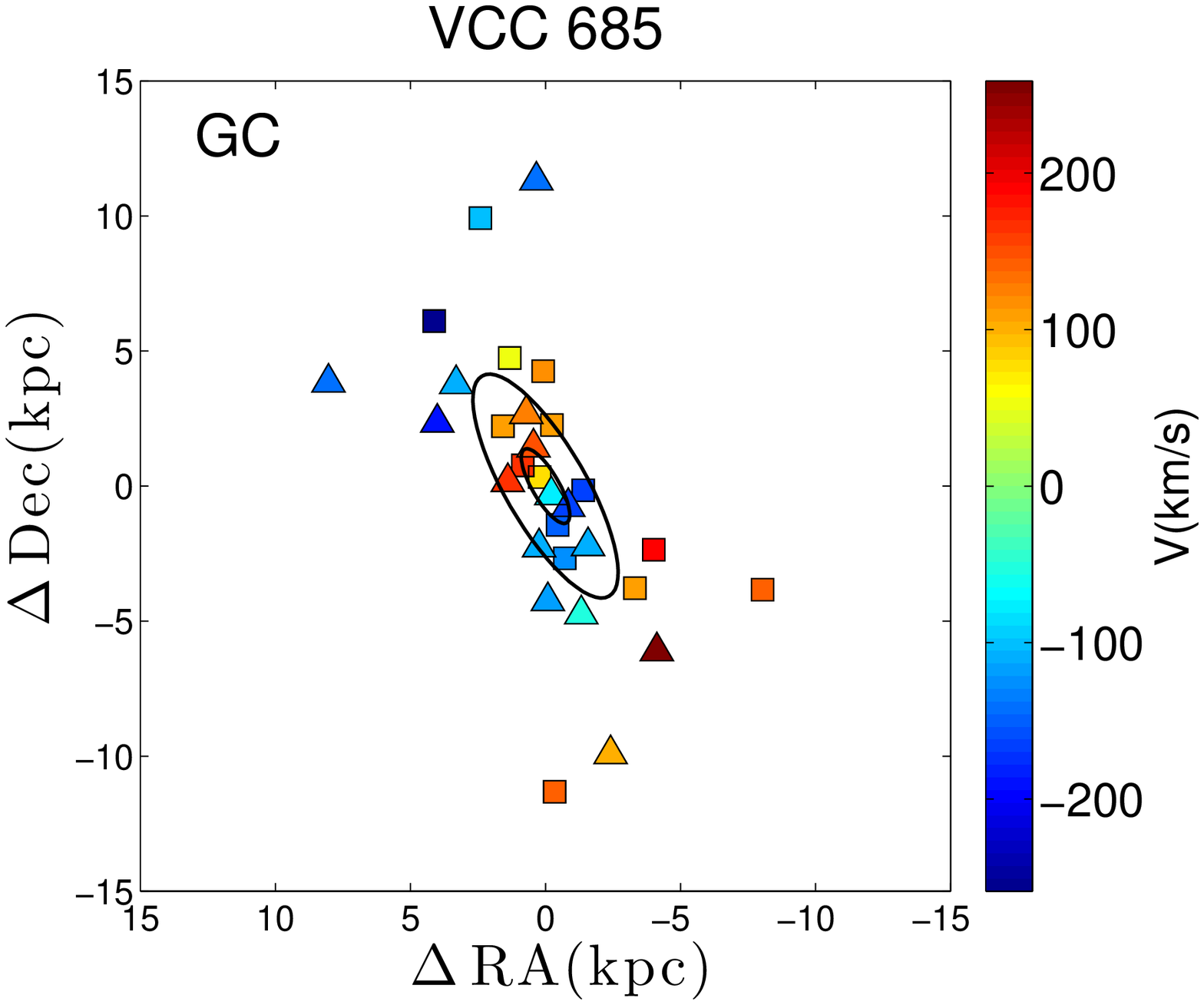}{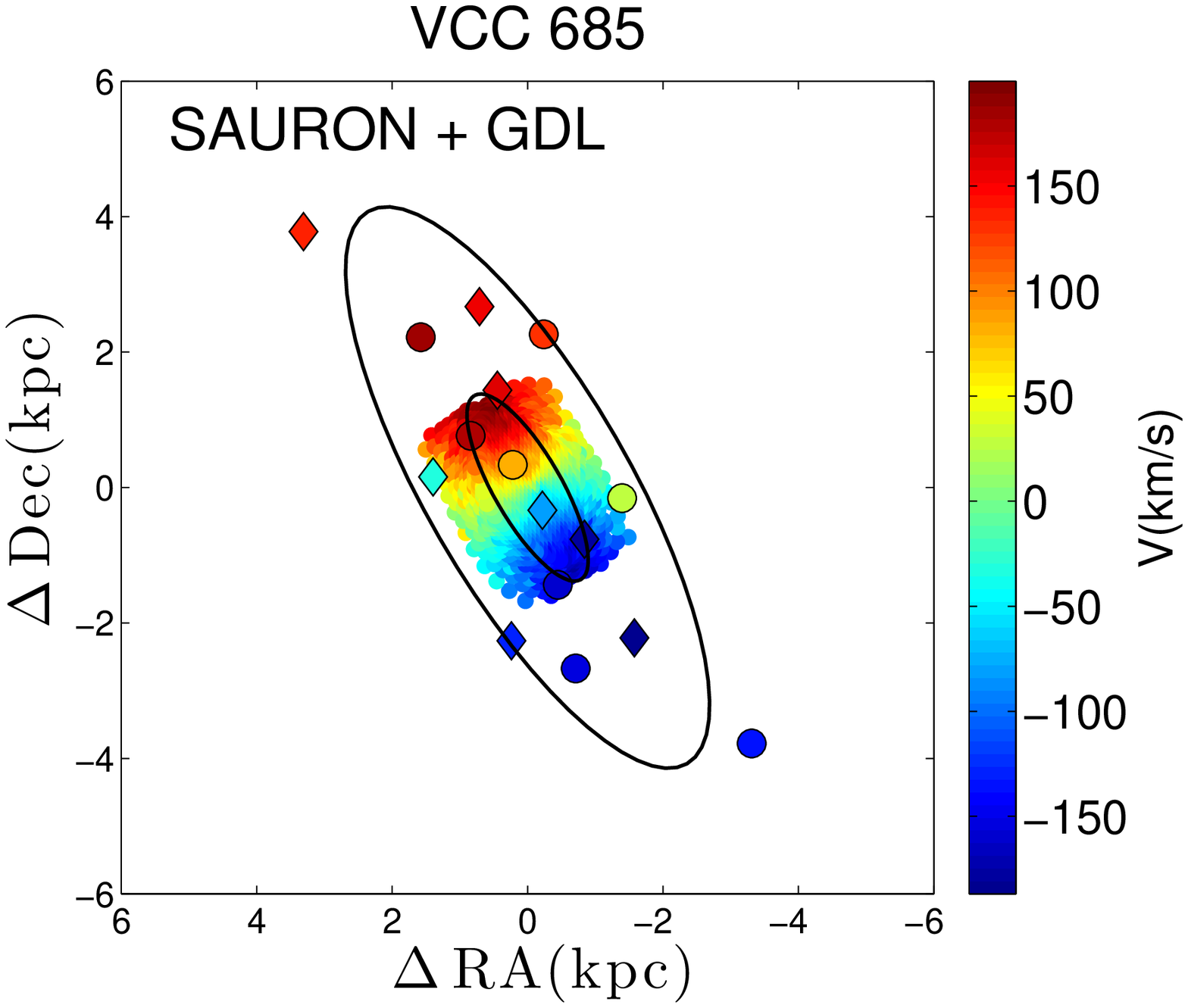}
\caption{Same as in Figure~\ref{fig:vfields-1231}, but for VCC 685}.
\label{fig:vfields-685}
\end{figure*}
 
VCC 685 is also classified as a fast rotator in the $\rm ATLAS^{3D}$ project. The stars clearly rotate about the photometric minor axis \citep{kr11}. Our GMOS GDL measurements are consistent with the SAURON data, and we can clearly see rotation in the 2D velocity field of the stars out to $3R_e$ (Figure~\ref{fig:vfields-685}, right panel, Figure~\ref{685-1}, top panel). In these inner regions, the GCs also show strong rotation about the photometric minor axis, but there is a remarkable shift at $R=3$--4 kpc. At these radii, the rotation axis of the GCs flips and exhibits counter-rotation. To better quantify this, we divide the GCs into two groups: 9 GCs in the inner region ($<3.5$~kpc), and 6 GCs in the outer region ( $>3.5$~kpc). The best fitting rotation curve for the inner and outer GC samples are shown in Figure~\ref{685-2}. The purple and green circles represent the inner and outer region GCs, respectively. The inner region GCs' best-fitting amplitude is $282\pm129$ $\rm km\ s^{-1}$ with $\rm PA_{kin}$  =  $64\pm38^{\circ}$. The outer region GCs best-fitting amplitude is $245\pm85$ $\rm km\ s^{-1}$ with $\rm PA_{kin}$  =  $224\pm13^{\circ}$.

Given that we only have 15 GCs in this galaxy, we wanted to see if this situation could occur randomly. To  test the significance of the counter rotation, we used the scrambling method discussed earlier where we fix the position angles of the GCs and randomly scramble the velocities and velocity errors. After 100,000 trials, we find that only 520 have kinematics similar to what we observe ($V_{rot,inner} \geq 282$, $V_{rot,outer} \geq 245$, and $121^\circ < \Delta PA_{kin}<239^\circ$). With only 0.52\% of trials randomly producing the observed kinematics, we conclude that this detection of counter-rotation in the GCs is significant.

In Figure~\ref{685-2}, we show that the best-fitting amplitude of GDL is $174\pm7$ $\rm km\ s^{-1}$ with the $\rm PA_{kin}$ is $28\pm5^{\circ}$. This is consistent with the results from fitting the SAURON data. These stellar kinematics are roughly (but not exactly) consistent with the kinematics of the inner GCs.

\begin{figure*}
\includegraphics[scale=.90]{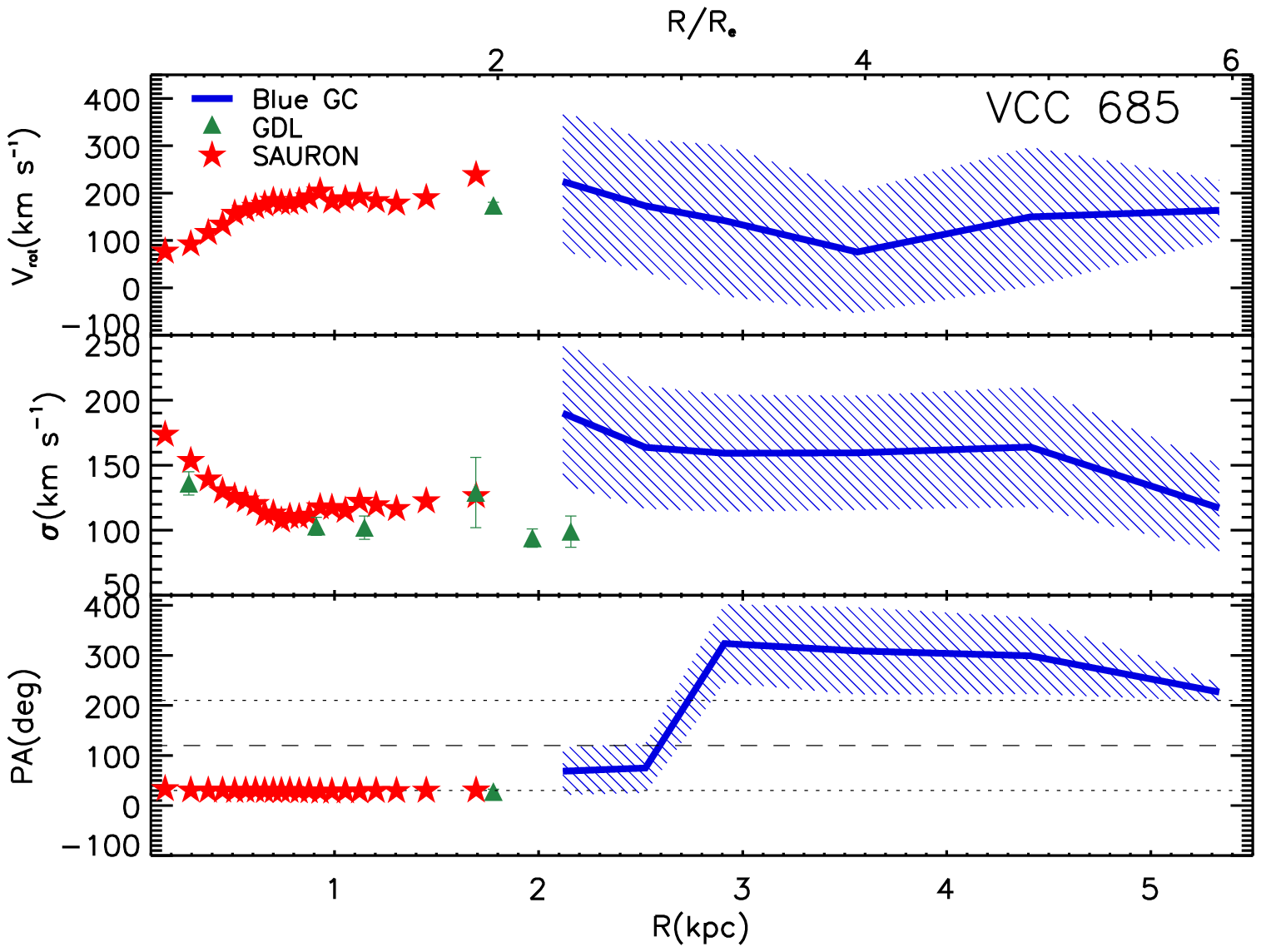}
\caption{The kinematics fitting result as a function of radius in VCC 685. The blue GCs exhibit counter-rotation, showing rotation in both the inner and outer regions, but around different rotation axes. The velocity dispersion profile of GC stays flat with radius. }
\label{685-1}
\end{figure*}

\begin{figure}
\includegraphics[width=1.0\columnwidth]{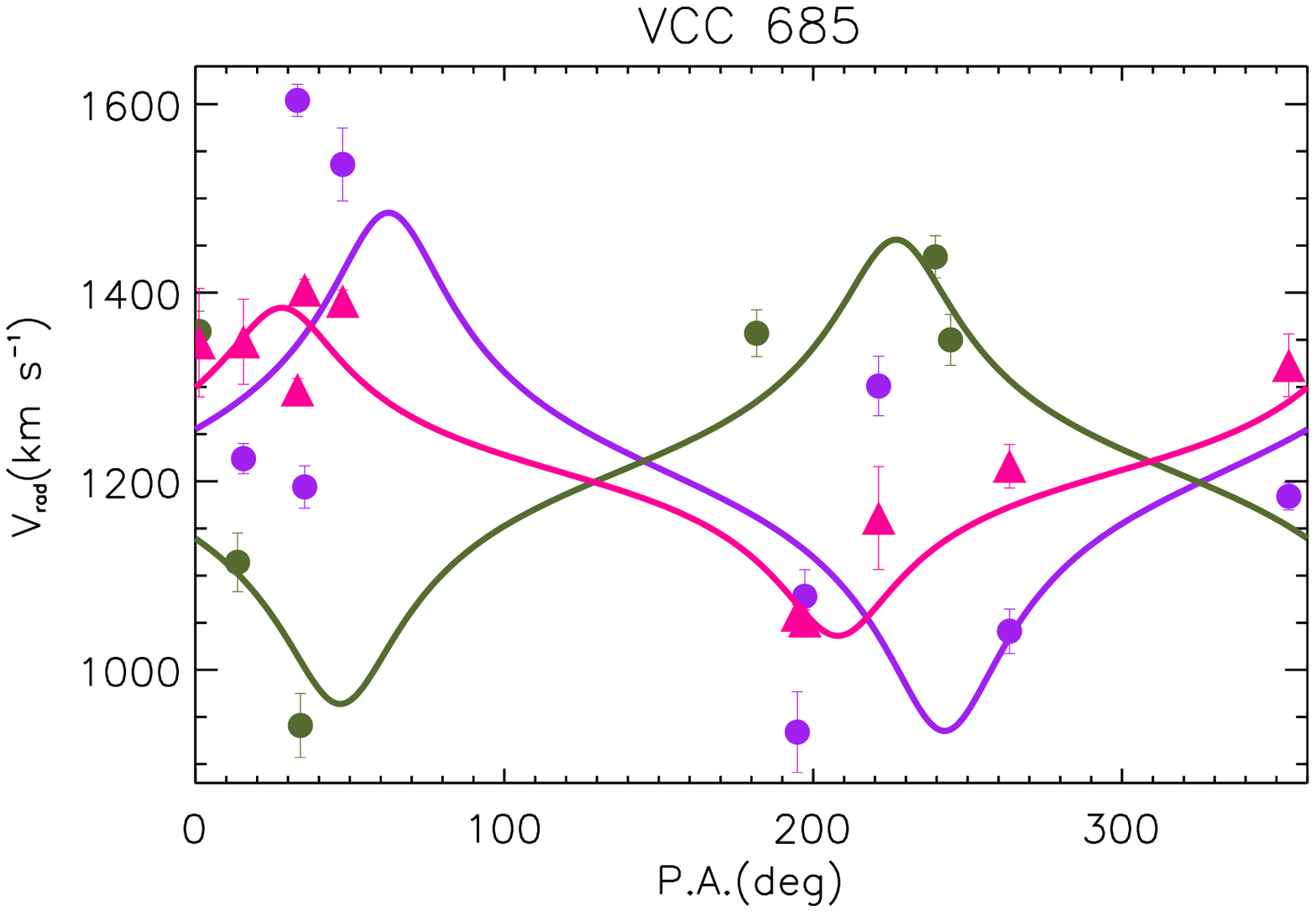}
\caption{Radial velocities versus position angle in VCC 685. The purple and green circles represent the inner and outer region GCs, respectively. The fits show strong counter-rotation. The magenta curve and points represents the rotation of the GDL, which lines up more with the rotation of the inner GCs. }
\label{685-2}
\end{figure} 

\section{Discussion}

\subsection{The Kinematics of GC Systems and Their Hosts}

The four galaxies in our sample are similar in many respects. All are classified as ``fast rotators'' by the $\text{ATLAS}^{\text{3D}}$ project, have similar morphological classifications, have nearly identical surface brightness profiles (S\'ersic $n\sim3$--4) and GC surface density profile shapes (S\'ersic $n\sim1.2$--1.8). Yet, their GC systems show different kinematics: VCC~1231's GCs show little or no rotation. VCC 2000 GCs have a rotation amplitude consistent with the stars, but with a different rotation axis ($\rm PA_{kin}$). VCC 1062's GCs show fast rotation consistent with the stars both in amplitude and position angle. VCC 685's GC system exhibits counter-rotation. This diversity suggests that these galaxies, despite their otherwise similar characteristics, had different assembly histories. Below, we discuss each of the galaxies in more detail.

\begin{itemize}
\item \emph{VCC 1231}: Although the GC system as a whole does not show significant rotation, the red and blue GC individually appear to have rotation around nearly opposite kinematic axes, and with the red GCs aligned with the rotation of the stars. This finding is currently of low significance due to the low number of GCs in each sample. Many observations suggest that the red GCs are more associated with the stellar main body \citep{pe04gc,fai11,str11,po13}. It will be interesting to observe more GCs to test if the metal-poor GCs are truly counter-rotating with respect to the metal-rich GCs and the stars. If true, and whether the metal-poor GCs are accreted from low-mass progenitors (e.g., \citealt{co98}), then this galaxy requires a formation scenario where the angular momenta of the dry merged components and the gaseous components are nearly opposite.

\item \emph{VCC 2000}: This galaxy's GC system is a more significant example of where the kinematics of blue GCs are different from those of the stars (see Figure \ref{2000-2}). In this case the rotation axis of the blue GCs is orthogonal to that of the stars. This galaxy GC system is consistent with a scenario where blue GCs are accreted from low-mass progenitor, and where the kinematics of these outer components are decoupled from the material that eventually settles to form the stellar disk. 

\item \emph{VCC 1062}: This galaxy's kinematics are the ``simplest'' in our sample. The red and blue GCs both have kinematics consistent with the stars. All three components of this galaxy show fast rotation about a common axis, the minor axis. In this case, the angular momenta of the progenitors that formed these components were all aligned.

\item \emph{VCC 685}: The kinematics of the GCs in VCC~685 are the most complex and tantalizing. Kinematically decoupled cores, or kinematically distinct components (KDCs), have been found in the stellar kinematics in the central regions of many galaxies. This designation refers to an abrupt kinematic change between then the inner and outer regions of a galaxy \citep{rix92,me98,we02,kr11,ar13,to14}. In VCC~685, we find evidence that the GCs in the inner and outer regions are counter-rotating with respect to one another, with the inner component roughly aligned with the stars. This suggests discrete merging events. Indeed, for stellar KDCs, gas-rich mergers are often invoked as an explanation. In this case, however, the kinematic transition happens at much larger radii than for the stellar KDCs that are traditionally discussed (although \citealt{ar13} find evidence of this behavior in some of their sample galaxies). More GC data, more stellar kinematic data at larger radii, and realistic simulations, are needed to investigate the counter-rotation in this galaxy. 

\end{itemize}

Lastly, we also analyze the kinematics of GCs subdivided by GC luminosity. Table~\ref{tbl-44} shows the rotation and velocity dispersion fitting results for subsamples of bright and faint GCs. We choose to divide the GC sample at the magnitude at which we have equal numbers of GCs in the two subsamples. The divided magnitude is shown in the last column of Table~\ref{tbl-44}. For VCC~1231, we find that the brighter GCs have significantly more rotation than the fainter ones, which are consistent with no rotation. This is possibly because the brightest GCs are more likely to be blue, and hence their kinematics mirror the blue GCs. The fainter sample contains both blue and red GCs, and so their kinematics are a combination of these apparently counter-rotating populations. The other two galaxies do not show very significant differences in the rotation properties between the bright and faint GCs. VCC 1062's bright GCs, however, have a noticeably lower velocity dispersion ($77\pm17$~km/s) compared to their fainter counterparts ($134\pm28$~km/s). These difference worth further exploration in the future with larger data sets.

\subsection{Specific Angular Momentum}
To provide a more quantitative look at the kinematics of early-type galaxies, we use the specific angular momentum.  An early study of the M87 GC system used a dimensionless spin parameter \citep{kisslerpatig98}. We use the dimensionless parameter defined by \cite{em07} to distinguish fast and slow rotators based on SAURON data within 1$R_{e}$. The expression for $\lambda$ is given by:

\begin{equation}
\lambda=\frac{\sum R_{i}F_{i}|V_{i}|}{\sum R_{i}F_{i}\sqrt{V_{i}^{2}+\sigma_{i}^{2}}}
\label{eqn:lambda}
\end{equation}

where $F_{i}$, $R_{i}$, $V_{i}$ and $\sigma_{i}$ are the flux, radius, velocity and velocity dispersion in the {\it i}th bin, respectively. Emsellem et al. (2011) define the dividing line between slow and fast rotators at 

\begin{equation}
\lambda=0.31\times\sqrt{\epsilon}
\label{eq8}
\end{equation}

where $\epsilon$ is the ellipticity of the system. However, \cite{em11}'s IFU study of ETG specific angular momentum was generally limited to stars within $\lesssim 1 R_e$. 

Unlike for filled IFU observations, our GC and GDL data only sparsely sample the outer halos of our sample galaxies. The density of our data, therefore, is not an exact representation of the true density of the GC or stars. Therefore, rather than sum over the full two-dimensional velocity field, which we do not have, we calculate $\lambda$ in one-dimension as follows: (1) Through our kinematic fitting, we obtain the rotation velocity and velocity dispersion profile with radius. This is done in elliptical annuli whose ellipticities are alternately set to two values: zero and the mean ellipticity of the stars. (2) For every bin in radius, we replace $F_i$ in Equation~\ref{eqn:lambda}  with the number of GCs expected at that radius, and use the rotation velocity at that radius. We call the $\lambda$ calculated in this way $\lambda_{1D}$. There is certainly a bias in using a one-dimensional method compared to $\lambda$ calculated with full two-dimensional data (which we now call $\lambda_{2D}$). This bias is due to the fact that the 1-D method assigns the peak velocity amplitude to every position at a given radius. In reality, the velocity along the minor axis will be smaller than along the major axis. We can calibrate this bias, however, by comparing $\lambda_{1D}$ and $\lambda_{2D}$ for galaxies where we have full IFU coverage. To do this, we apply our one-dimensional method to the 260 ETG galaxies from $\text{ATLAS}^{\text{3D}}$. The result is shown in Figure~\ref{com}. We can see that, as expected, $\lambda_{1D}$ is generally overestimated compared to $\lambda_{2D}$. However, they follow a well-defined relation described as:

\begin{equation}
\lambda_{2D} \approx 0.82\times\lambda_{1D}
\label{eq7}
\end{equation}

A similar relation was also found between the SAURON IFU data and long-slit ($V/\sigma$) measurements \citep{ca07}, and in the long-slit studies of $\lambda$ in early-type dwarf galaxies \citep{eli14}. Our scaling factor is larger than the value found by \cite{eli14} due to the fact that all of their long-slit data is along the major axis.

We apply this one-dimensional method to the GC data in our Gemini sample. We also include 6 galaxies with GC data from the literature (Table~\ref{tbl-3}). The uncertainties are determined from Monte Carlo simulations. Because the ellipticity of the GC system is not easily measured, we consider two cases, one where the ellipticity of the GC system is zero, and one where it is equal to that of the stars. The GC results are listed in Table~\ref{tbl-3} and Table~\ref{tbl-4}. In practice, the difference is small between these two cases, and our results are not dependent on the assumed ellipticity of the GC system. In Figure~\ref{lamda1}, we compare $\lambda_{1D}$ of GC (blue circles) and SAURON data (red circles) with ellipticity. In all cases, we use the ellipticity of the stars. Although this is potentially an issue, studies of the GC systems of ETGs in this mass range show that there is some correlation between the overall shape of GC systems and the ellipticity of the stellar distribution \citep{wang13}. The dashed lines link the same galaxy. The criterion of dividing slow and fast rotator with Equation~\ref{eq8} is rescaled by a factor 0.82 and plotted with a solid black line. 

\begin{deluxetable*}{cccccccccc}
\tabletypesize{\scriptsize}
\centering 
\tablecaption{General properties of literature galaxies\label{tbl-3}}
\tablewidth{0pt}
\tablehead{
\colhead{ID} & \colhead{$R_{e}$} &\colhead{$M_{K}$} & \colhead{D} &  \colhead{$\epsilon_{star}$} &
\colhead{$N_{GC}$} & \colhead{$\lambda_{1D,GC,\epsilon=0}$} & \colhead{$\lambda_{1D,GC,\epsilon=star}$} &  \colhead{Rotation} & \colhead{Ref.} \\
\colhead{} & \colhead{(kpc)} &\colhead{(mag)} & \colhead{(Mpc)} & \colhead{} & \colhead{} &
\colhead{} & \colhead{} & \colhead{$<1R_{e}$} &\colhead{} \\
\colhead{(1)} & \colhead{(2)} & \colhead{(3)} & \colhead{(4)} & \colhead{(5)} &
\colhead{(6)} & \colhead{(7)} & \colhead{(8)} & \colhead{(9)} & \colhead{(10)} 
}  
\startdata
NGC 4494 &  4.3 & $-24.2$ & 16.6 & 0.13& 117 &  $0.46\pm0.04$ & $0.48\pm0.04$& F & F+11\\
NGC 3379 &  2.0 & $-23.7$ & 10.2 & 0.15& 52 & $0.10\pm0.03$ & $0.13\pm0.03$ &F & P+04; P+06; B+06\\
NGC 4636 &  6.1 & $-24.3$ & 14.2 & 0.16& 459 & $0.21\pm0.02$ &$0.22\pm0.02$ &S & S+12\\
NGC 4472 (VCC 1226, M49)&  7.7 & $-25.6$ & 16.7 &0.19 & 263 & $0.17\pm0.04$ &$0.22\pm0.05$& S & C+03 \\
NGC 4649 (VCC 1978, M60)&  7.7 & $-25.3$ & 16.3 &0.19 & 121 & $0.53\pm0.02$ & $0.56\pm0.02$&F & H+08\\
NGC 4486 (VCC 1316, M87)&  13.1 & $-25.3$ & 16.5 &0.14& 1004 & $0.20\pm0.01$ &$0.19\pm0.01$& S & S+11; Z+15$^*$

\enddata
\tablecomments{The effective radii (2) are from \cite{Fe06} and \cite{em11}. The $K$-band absolute magnitude (3) and distance (4) are from \cite{po13}, Tables 1 and 7, where (3) is derived from 2MASS apparent magnitude and (4) are from \cite{to01} and \cite{mei07}.  (5) is the ellipticity with $1-(b/a)_{k}$, where the axis ratio are from 2MASS. (6) is the total number of GCs with measured radial velocities. (7) and (8) are the projected specific angular momentum with $\epsilon=0$ and $\epsilon=\epsilon_{star}$. (9) Whether a galaxy is classified as a fast or slow rotator in $\text{ATLAS}^{\text{3D}}$ \citep{em11} . (10) References: F+11\citep{fo11} ; P+04\citep{pu04}; P+06 \citep{pi06}; B+06\citep{be06}; S+12\citep{sc12}; C+03\citep{co03}; H+08\citep{hw08};  S+11\citep{str11}; Z+15\citep{hx15}.\\
$^{*}$ New GC data from our MMT observations, described by \cite{hx15}, are combined with previously published data. } 
\end{deluxetable*}

\begin{figure}
\includegraphics[scale=.50]{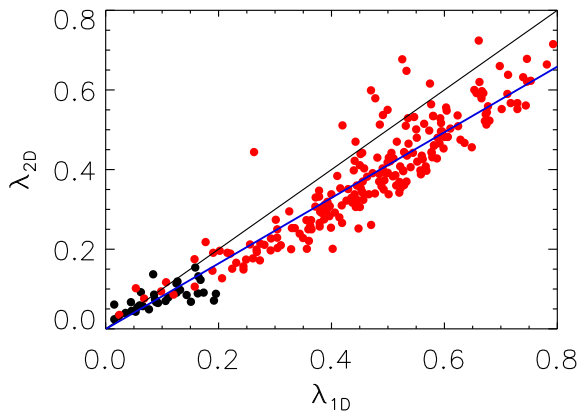}
\caption{The comparison of $\lambda_{1D}$ and $\lambda_{2D}$. The black points and the red points represent the slow and fast rotator. The black line is a 1:1 relation. The blue line show the best fitting linear relation to the data.}
\label{com}
\end{figure}

\begin{figure*}
\includegraphics[scale=.90]{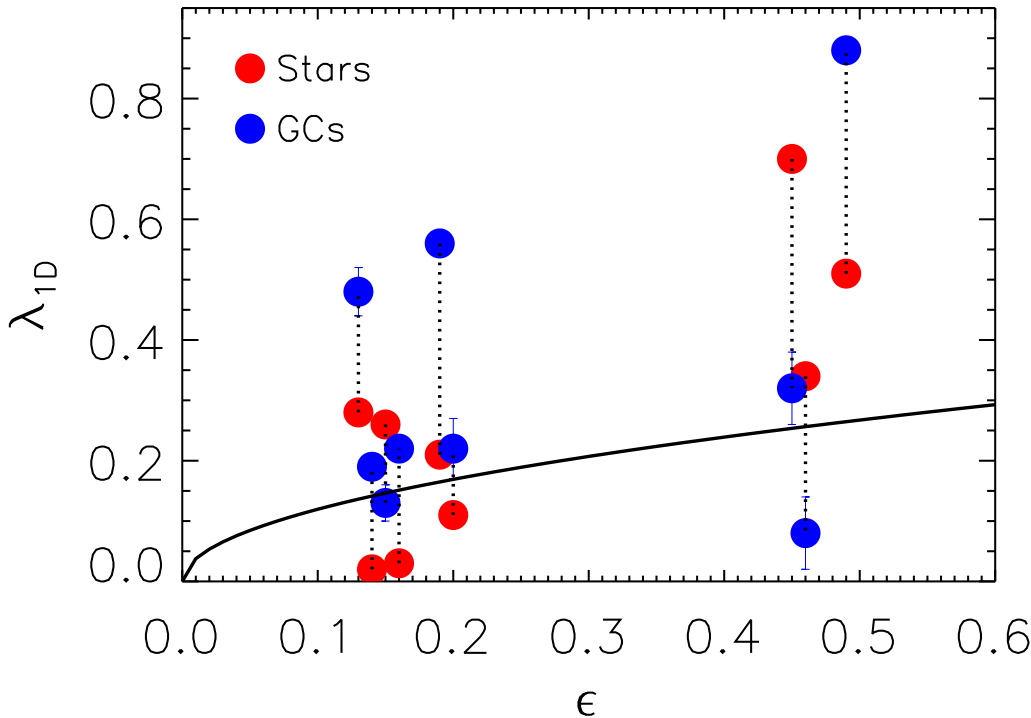}
\caption{The specific angular momentum of GCs and SAURON data as function of ellipticity . The red circles represent the $\lambda_{1D}$ of the GC. The blue circles represent the $\lambda_{1D}$ of the SAURON data. The dash lines link the same galaxy. The solid line is $\lambda_{1D}=0.38\times\sqrt{\epsilon}$. Most stellar slow rotators also show slow rotation in their GC systems, with $\lambda_{GC}\sim0.2$. Stellar fast rotators display a large range in $\lambda_{GC}$.}
\label{lamda1}
\end{figure*}

\begin{figure*}
\includegraphics[scale=.90]{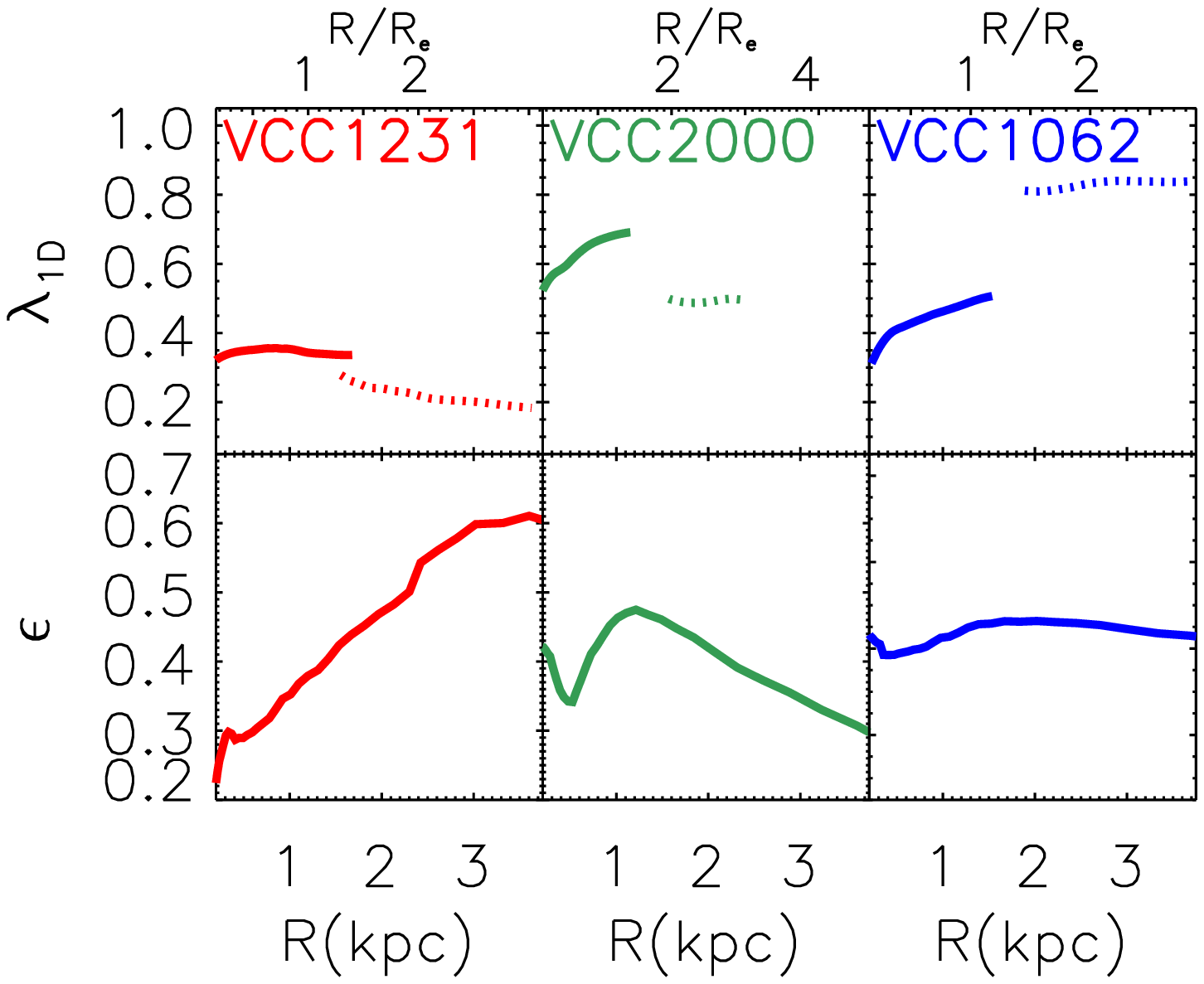}
\caption{The $\lambda_{1D}$ and the ellipticity profiles as radius for VCC 1231, 2000 and 1062. In the top panel, the solid curve represents the $\lambda_{1D}$ profile measurements from SAURON data. The dash curve represents the $\lambda_{1D}$ profile measurement from the GDL. The individual $\lambda_{1D}$ profiles (SAURON and GDL) are each cumulative with radius, but the two data sets are treated separately to allow for a comparison.}
\label{lamda2}
\end{figure*}

The three stellar slow rotators (red points below the line), are massive ETGs (NGC 4486, 4472 and 4636), and their GCs also are relatively slowly rotating, with $\lambda_{1D,GC}\sim0.2$. Massive ETGs may experience multiple mergers with different orbital configurations and mass ratio, and this leads to aggregate slow rotation in the halo. However, there are also cases of massive ETGs that have significant rotation in their outer regions  even though the galaxies are classified as slow rotators (e.g. NGC 1407, \citealt{ro09, po13} and NGC~5128 \citealt{pe04}). This case can be explained by angular momentum transfer to the outer regions by mergers.

For stellar fast rotators, however, the range in $\lambda_{1D,GC}$ is large. The most obvious examples are our three intermediate luminosity galaxies, where $\lambda_{1D,GC}$ of VCC 1062, 2000 and 1231 is 0.9, 0.3 and 0.1, respectively. Half of stellar fast rotators show the $\lambda_{1D}$ of GC are smaller than the $\lambda_{1D}$ of star.  Two stellar fast rotators (NGC 3379 and 4473) have $\lambda_{1D,GC}<0.2$, which means their outer regions are rotating very slowly. Similar results are also found by studying $\rm V_{rot}/\sigma$ (see the Figures 19 and 20 in \citealt{po13}). It seems that, at least for the less massive galaxies, GCs may be kinematically decoupled from the stars.    

\subsection{Stellar kinematic trends with galactocentric radius}

ETGs at $z\sim 2$ appear to be smaller and more dense than present-day ETGs of similar mass \citep{da05,tu07,ce09,van10,wi12} . A two phase model of galaxy assembly \citep{na09,os10,kh11} has been used to explain these observations. In the ``in situ'' phase, dissipative processes such as gas-rich major mergers and fragmenting turbulent disks \citep{de09,ce10} form the central region. In the ``ex situ'' phase, accreted stars from minor mergers expand a galaxy's outskirts. These two phases {\it may} create different kinematic components in ETGs, although it is not necessarily clear which component would have larger angular momentum. For example, the core region is likely to dissipate its angular momentum through violent relaxation, but a late binary merger can also imprint rotation in the inner regions \citep{co06, bo11,mo14}. The steady accretion processes in the outer region could create a dynamically hotter with low rotation, but could also imprint rotation if the accretion is anisotropic. However, if there are indeed two regimes of galaxy building, we may expect to see this expressed in the kinematics. Indeed, some observations have found different stellar kinematics between the inner and outer regions of early-type galaxies \citep{pro09, co09,ar13}.

One of the more comprehensive and recent studies of ETG stellar kinematics at large radii is by \cite{ar13}. They studied stellar kinematics in 22 nearby ETGs out to 2--4$R_{e}$ with multi-slit observations from the SLUGGS survey \citep{br14}. They found that galaxies classified as fast rotators based on data in their inner regions can show a range of $\lambda$ profiles in their outer regions, suggesting that ``the apparent unification of the elliptical and lenticular (S0) galaxy families in the $\text{ATLAS}^{\text{3D}}$ survey may be a consequence of a limited field of view''.  We measure $\lambda_{1D,GDL}$ in the range of $\sim1.5$--$3 R_{e}$, and compare this measurement to $\lambda_{1D,SAURON}$, which represents the specific angular momentum in the inner regions.  As in \cite{ar13}, we also find that $\lambda$ in the outer regions can differ from what is found in the inner regions. 

When trying to put into context the angular momentum profiles of galaxies, however, it is also important to consider ellipticity profiles. A changing ellipticity with radius leads to different expectations for a galaxy's $\lambda$ with radius. For example, it is not surprising that NGC 3377 shows a decrease in its $\lambda$ profile \citep{ar13}, as its stellar light also gets rounder with radius. In Figure~\ref{lamda2}, we plot our measurements of $\lambda_{1D}$ for the SAURON and GDL data versus radius, and also show the corresponding ellipticity profile ($\epsilon$), derived from NGVS imaging \citep{fe12}. For each galaxy's $\lambda_{1D}$ profile, the solid and dotted lines represent the SAURON and GDL data, respectively. Both lines are cumulative $\lambda_{1D}$ profiles with radius, but each data set is treated independently to allow for comparison (i.e., the GDL profile does not include the SAURON data). 

Each of the three galaxies exhibits different characteristics. In VCC~1231 (left, red), $\lambda_{1D}$ clearly decreases even though the ellipticity increases. This is not what one expects for an anisotropic rotator at constant inclination. In VCC~2000 (center, green), the rotation declines as the galaxy gets rounder. In VCC~1062 (right, blue), the ellipticity is nearly constant, but the rotation is increasing with radius. It is clear that while these galaxies may have been selected to be similar, their kinematics are quite different. 

Whether this supports a picture of ``two-phase'' formation is less clear. One expects continuous transitions from the inner to outer regions of galaxies that may not necessarily result from distinct mergers or formation episodes. \cite{ra14} have studied the stellar kinematics of 33 massive ETGs with IFU data out to 2--5$R_{e}$ and do not find a distinct transition between the inner and outer regions. Most of their sample galaxies are very massive, however, and it is possible that different kinematics between inner and outer regions are more easily seen in less massive systems, which are likely to have undergone fewer mergers than their massive counterparts. Individual merger events in massive galaxies are therefore more likely to be smoothed out by the larger number of merger and accretion events.

\begin{deluxetable*}{cccccc}
\tabletypesize{\scriptsize}
\centering 
\tablecaption{Specific angular momentum\label{tbl-4}}
\tablewidth{0pt}
\tablehead{
\colhead{ID} & \colhead{$\lambda_{2D}$} & \colhead{$\lambda_{1D}$} & \colhead{$\lambda_{1D}$} &
\colhead{$\lambda_{1D,GC,\epsilon=0}$} & \colhead{$\lambda_{1D,GC,\epsilon=star}$} \\
\colhead{} & \colhead{(SAURON)} & \colhead{(SAURON)} & \colhead{(GDL)} & \colhead{(GCs)} & \colhead{(GCs)}\\
\colhead{(1)} & \colhead{(2)} & \colhead{(3)} & \colhead{(4)} & \colhead{(5)} & \colhead{(6)}
}  
\startdata
VCC1231 & 0.23 & 0.34 & 0.17$\pm$0.02 &0.04$\pm$0.04 & $0.08\pm0.06$\\

VCC2000 & 0.55 & 0.70 & 0.49$\pm$0.02 &0.30$\pm$0.06 & $0.32\pm0.06$\\

VCC1062 & 0.36 & 0.51 & 0.84$\pm$0.01 &0.82$\pm$0.03 & $0.88\pm0.02$

\enddata
\tablecomments{(2) $\lambda_{2D}$ measurement from SAURON data, (3) $\lambda_{1D}$ measurement from SAURON data, (4) $\lambda_{1D}$ measurement from Gemini (GDL), (5) and (6) $\lambda_{1D,GC}$ for the GCs assuming different ellipticities. The uncertainties for the SAURON data are very small, and they are not listed in the table.}
\end{deluxetable*}

\section{Conclusions}
We have presented a kinematical analysis of halo stars and GCs in four intermediate luminosity galaxies in the Virgo cluster, based on Gemini/GMOS observations. These observations push these kinds of studies to lower mass ranges than previously done. Our galaxy diffuse light (GDL) and GC data extend out to 4$R_{e}$ and  8--14$R_{e}$, respectively. We compare the GCs with the stars, and also compare the stellar kinematics in the inner ($R\lesssim 1 R_e$) and outer regions. Although the four galaxies in our sample are all uniformly classified as ``fast rotators'' by the $\text{ATLAS}^{\text{3D}}$ project, we find that these galaxies show a diversity of GC and stellar kinematics in their outer regions.

\begin{enumerate}
 \item The GCs display a wide range of kinematic properties that in some galaxies follow the stars of their host galaxy, but in others do not. We see cases of nearly zero net rotation in the GCs (VCC~1231), rotation about a misaligned axis (VCC~2000), strong rotation (VCC~1062), and counter-rotation (VCC~685). 

\item We calculate the specific angular momentum, $\lambda$, of the GCs for three galaxies in our sample as well as in 6 galaxies from the literature. We find that massive ETGs that are classified as stellar slow rotators in their inner $R\lesssim 1R_e$ by $\text{ATLAS}^{\text{3D}}$ also tend to have slowly rotating GC systems (although at least one counter-example exists in the literature). The GC systems of stellar fast rotators, however, display a wide range of $\lambda$. Some galaxies whose stars are rapidly rotating have slowly or only moderately rotating GC systems, and vice versa. We interpret this diversity as the consequence of the stochastic nature of merging processes in less massive galaxies.

\item We compare the specific angular momentum for stars in the inner ($R\lesssim 1 R_e$) and outer ($1.5\lesssim R \lesssim 3.0$) regions of three of our sample galaxies. In conjunction with the ellipticity profiles derived from deep imaging, we find that one of our galaxies (VCC~1231) has a $\lambda$ profile that decreases despite increasing ellipticity with radius, which is not what is expected for a simple anisotropic rotator at fixed inclination. The other two galaxies show falling (VCC~2000) and rising (VCC~1062) $\lambda$ profiles. Even in this small sample, we can see that the stellar kinematics of otherwise morphologically similar galaxies can exhibit a diversity of properties at larger radii.

\end{enumerate}

The kinematic differences we and others have found between GCs and stars, and between the inner and outer regions, in intermediate luminosity ETGs are beginning to allow us to dissect the merging and accretion histories of normal galaxies. The complex array of properties we are finding shows that this kind of wide-field kinematic study of a larger sample of intermediate-mass galaxies will be needed to fully sample the parameter space of galaxy assembly.

\acknowledgments

We are grateful to Yiqing Liu for useful discussions. We thank Eric Emsellem for comments that helped shape the discussion. We also acknowledge the use of the MPFIT package written by Craig Markwardt. 

BL and EWP acknowledge support from the National Natural Science Foundation of China (NSFC) through Grant No.\ 11173003, and from the Strategic Priority Research Program, ``The Emergence of Cosmological Structures'', of the Chinese Academy of Sciences, Grant No.\ XDB09000105. AJ acknowledges support from BASAL CATA PFB-06. CL acknowledges support from the National Key Basic Research Program of China (2015CB857002) and NSFC Grant Nos.\ 11203017 and 11125313. SM acknowledges financial support from the Institut Universitaire de France (IUF), of which she is senior member, and the support of the French Agence Nationale de la Recherche (ANR) under the reference ANR10-BLANC-0506-01-Projet VIRAGE. This work was supported by the Sino-French LIA-Origins joint exchange program. T.H.P. acknowledges support through FONDECYT Regular Project Grant No. 1121005 and BASAL Center for Astrophysics and Associated Technologies (PFB-06).

EWP and THP gratefully acknowledge support from the Plaskett Fellowship, under which this work was started. We also thank the Gemini Observatory staff and contact scientists for their professional and friendly support of these observations.

Based on observations obtained at the Gemini Observatory (acquired through the Gemini Science Archive and processed using the Gemini IRAF package), which is operated by the 
Association of Universities for Research in Astronomy, Inc., under a cooperative agreement 
with the NSF on behalf of the Gemini partnership: the National Science Foundation 
(United States), the National Research Council (Canada), CONICYT (Chile), the Australian 
Research Council (Australia), Minist\'{e}rio da Ci\^{e}ncia, Tecnologia e Inova\c{c}\~{a}o 
(Brazil) and Ministerio de Ciencia, Tecnolog\'{i}a e Innovaci\'{o}n Productiva (Argentina). Data for this paper were obtained through program numbers: GN2006A-C-6, GN-2006A-Q-12, GN-2007A-Q-53, GN-2008A-Q-73, and GS-2008A-Q-8.

Observations reported here were obtained at the MMT Observatory, a joint facility of the University of Arizona and the Smithsonian Institution. MMT telescope time was granted, in part, by NOAO, through the Telescope System Instrumentation Program (TSIP). TSIP is funded by NSF. 

\bibliographystyle{apj}
\bibliography{gemini_kinematic}

\clearpage

\appendix

\section{UNSMOOTHED AND RAW VELOCITY FIELD}
The unsmoothed and raw GC and stellar (GDL) velocity and velocity dispersion fields are shown in Figure~\ref{app1} and Figure~\ref{app} . The rotation information we get from these figures is consistent with the smoothed and point symmetric velocity fields in Figures~\ref{fig:vfields-1231}, \ref{fig:vfields-2000}, \ref{fig:vfields-1062}, and \ref{fig:vfields-685} . Our quantitative analysis of specific angular momentum is based on the unsmoothed and non-symmetrized data.

The 2D velocity dispersion field of VCC 1231 shows good agreement with SAURON in the overlap region. Our result is also consistent with \cite{fo13} and \cite{ar13}, who find two velocity dispersion peaks at 3 kpc along the major axis. VCC 1231 is also called as ``2$\sigma$'' galaxy.

\begin{figure}
\centering 
\plottwo{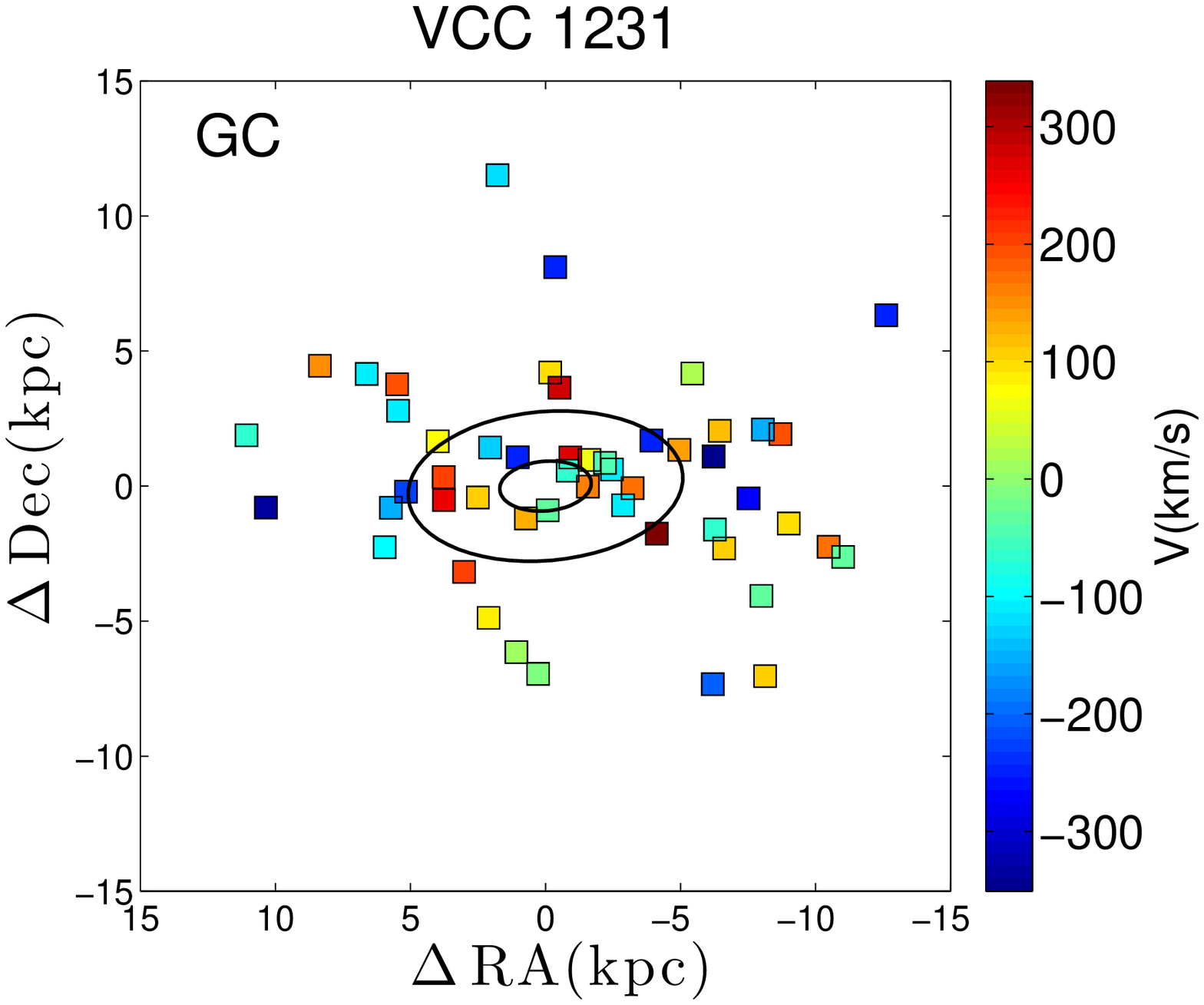}{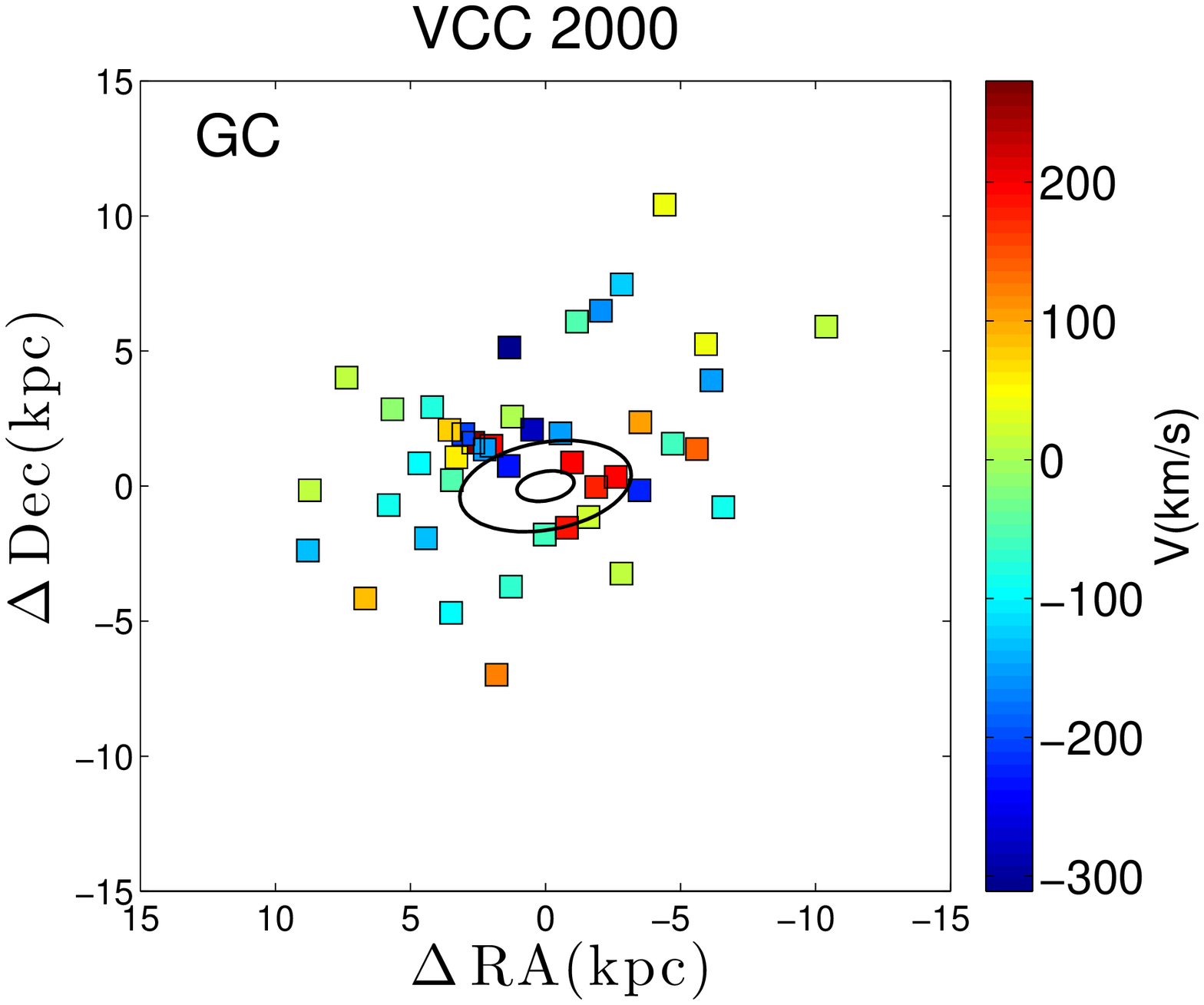}
\plottwo{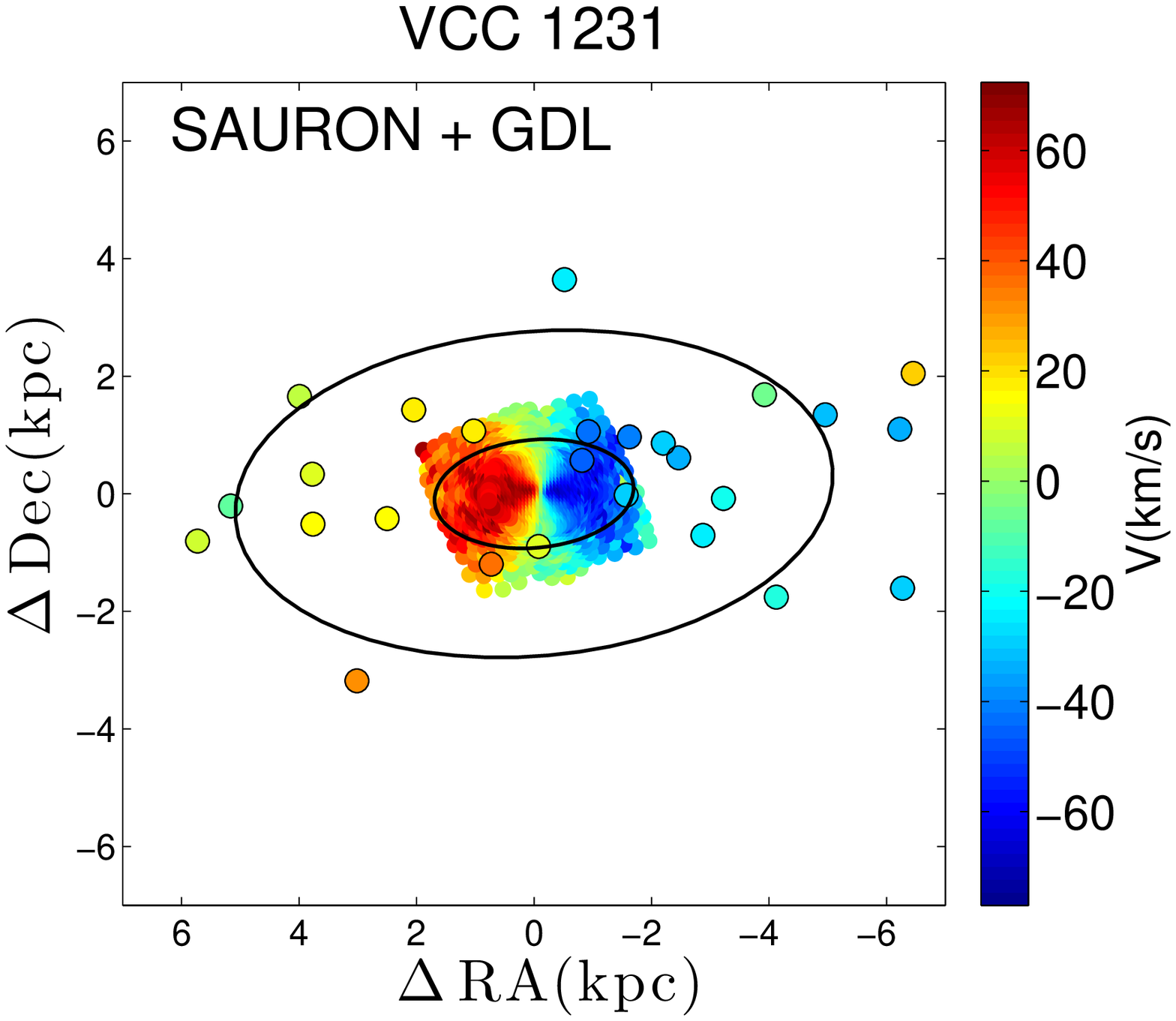}{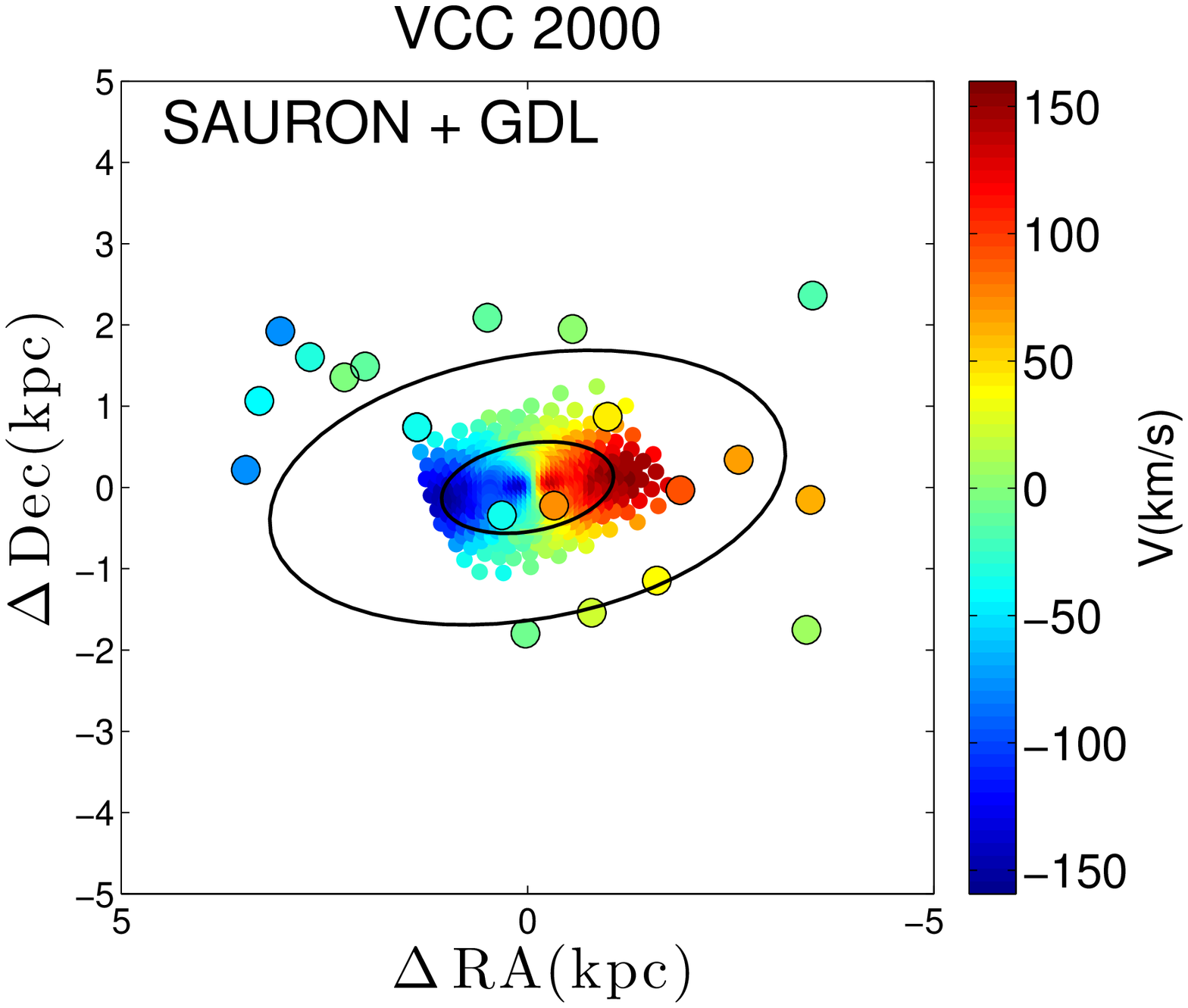}
\plottwo{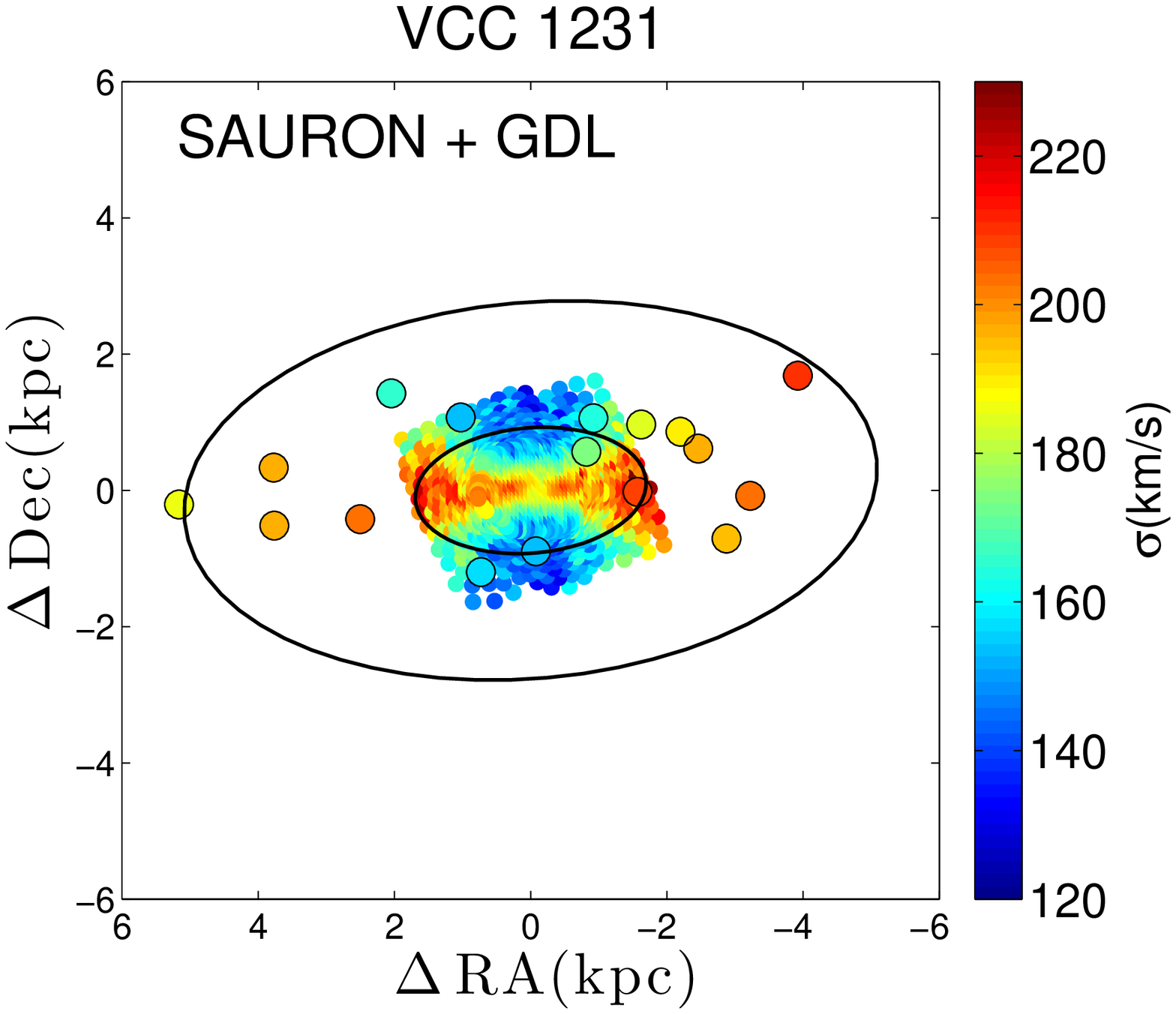}{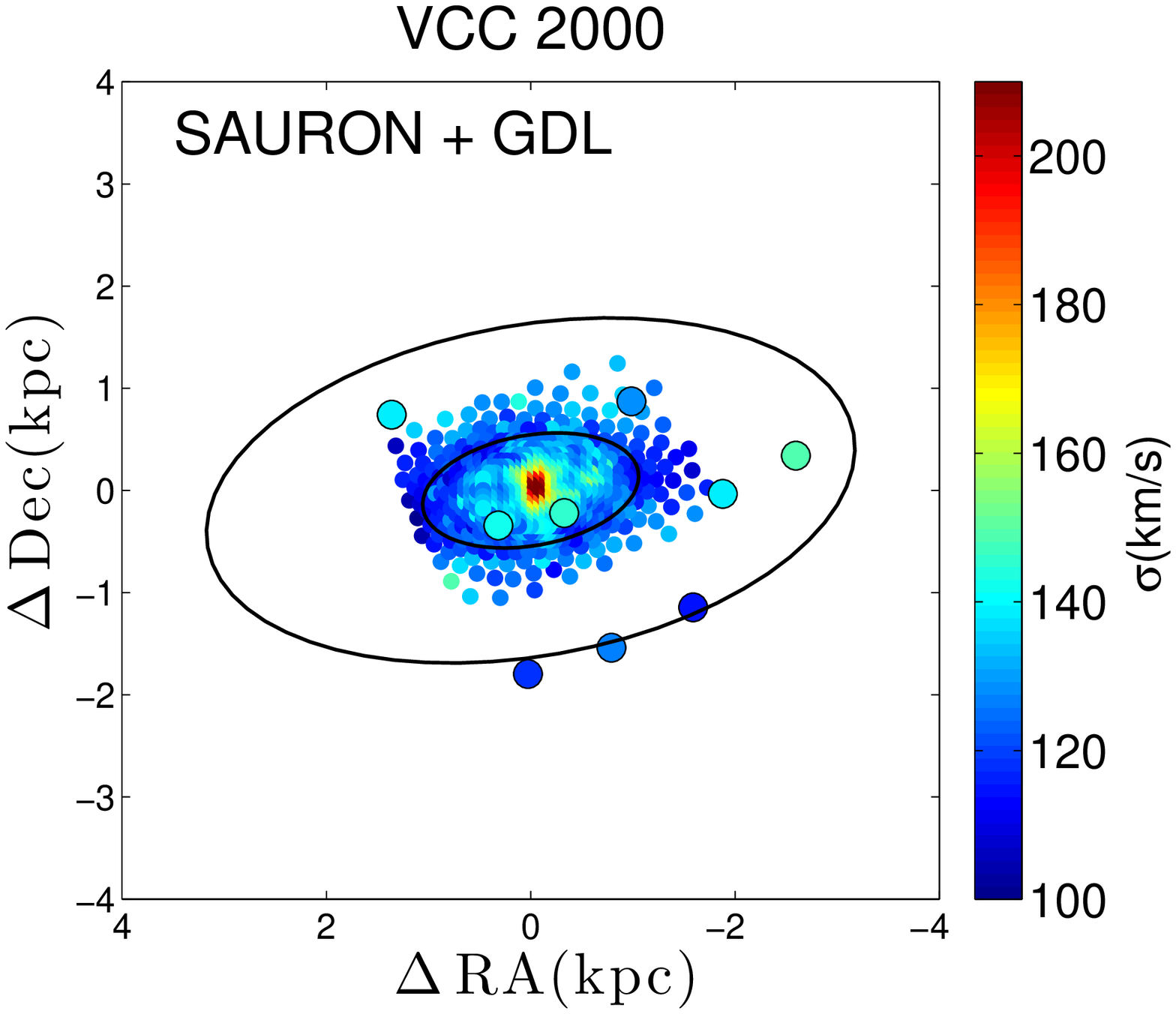}
\caption{The 2D velocity and velocity dispersion fiedls of VCC 1231 and 2000, without smooth and double data. } 
\label{app1}
\end{figure}

\begin{figure}
\centering 
\plottwo{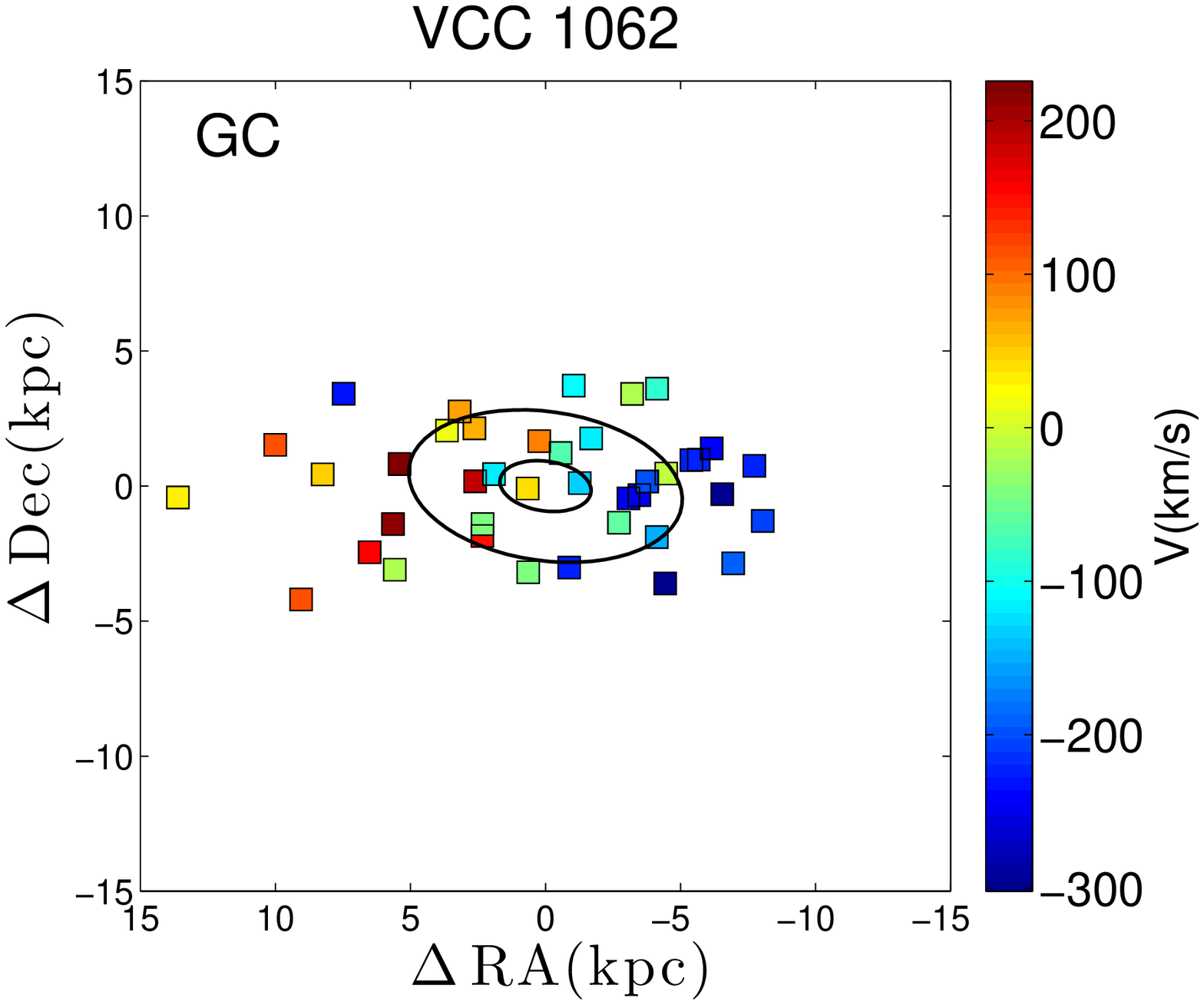}{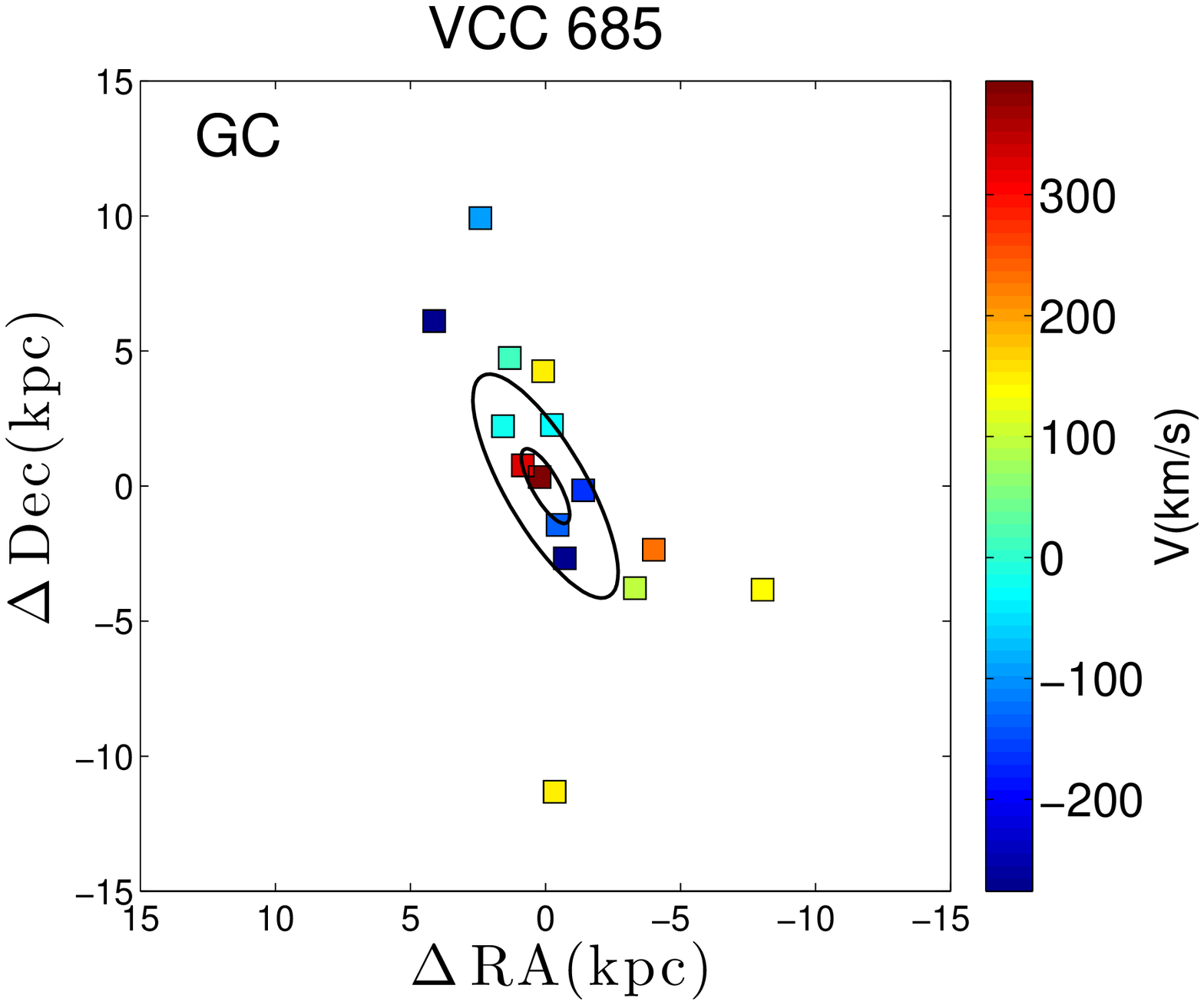}
\plottwo{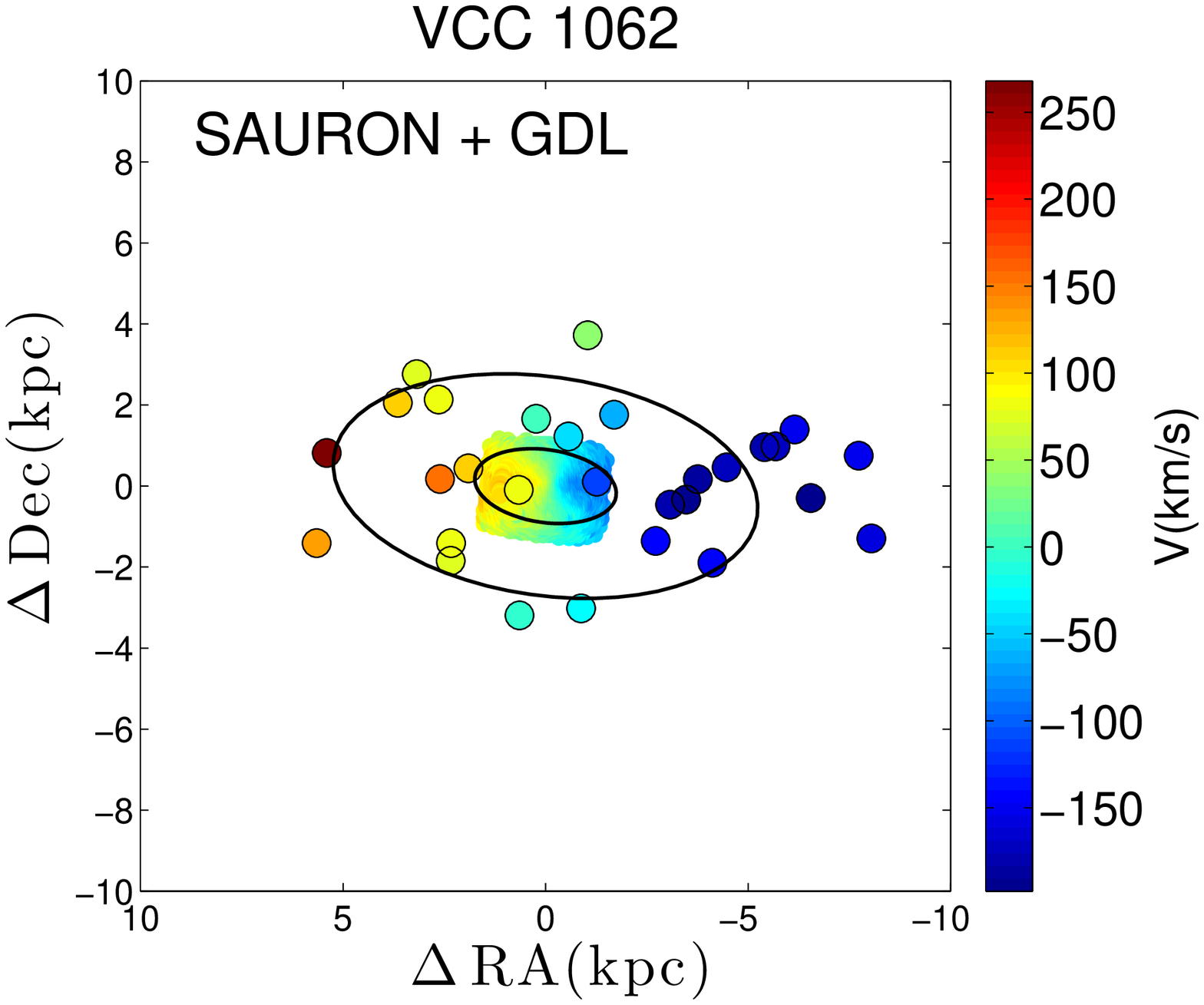}{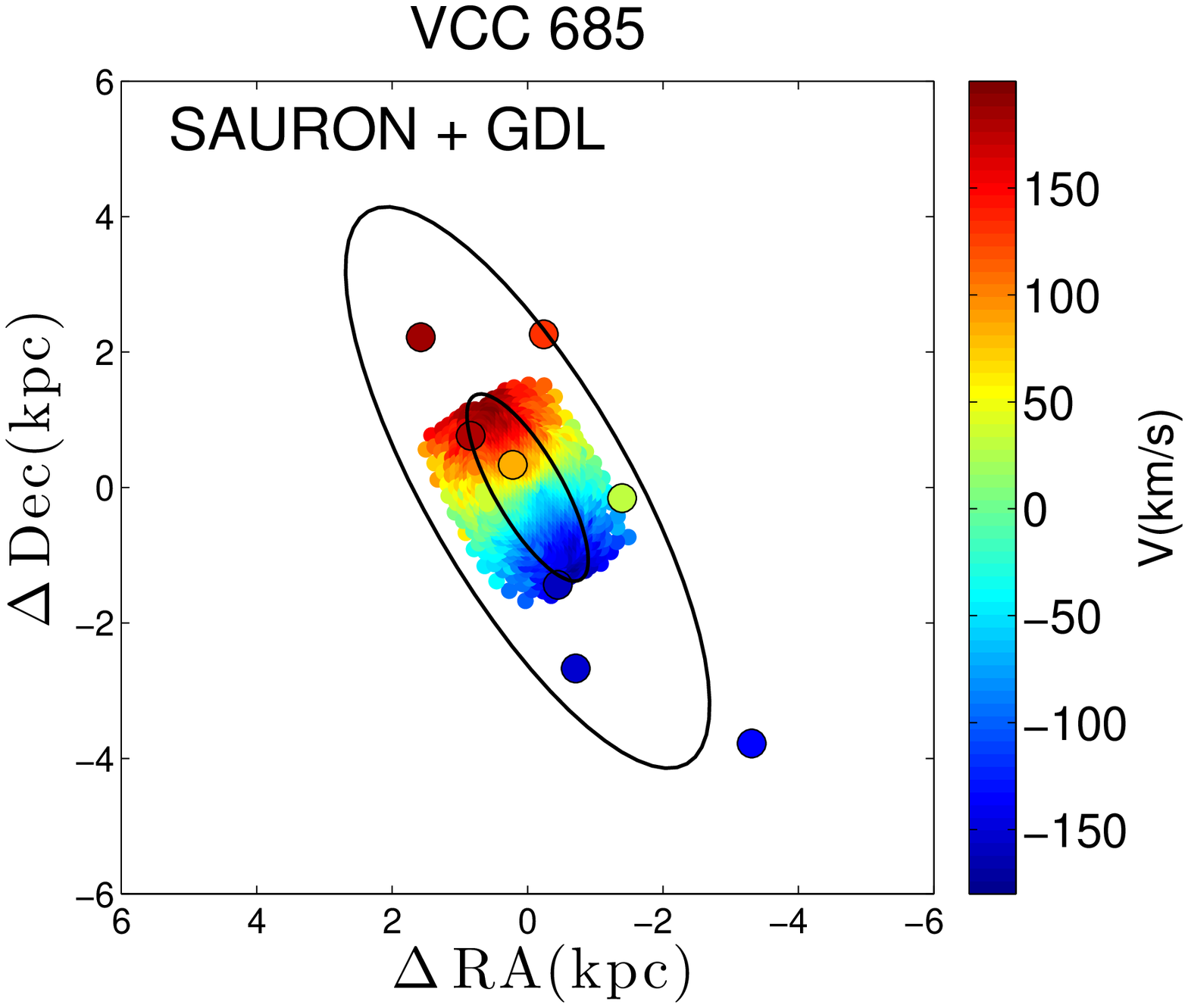}
\plottwo{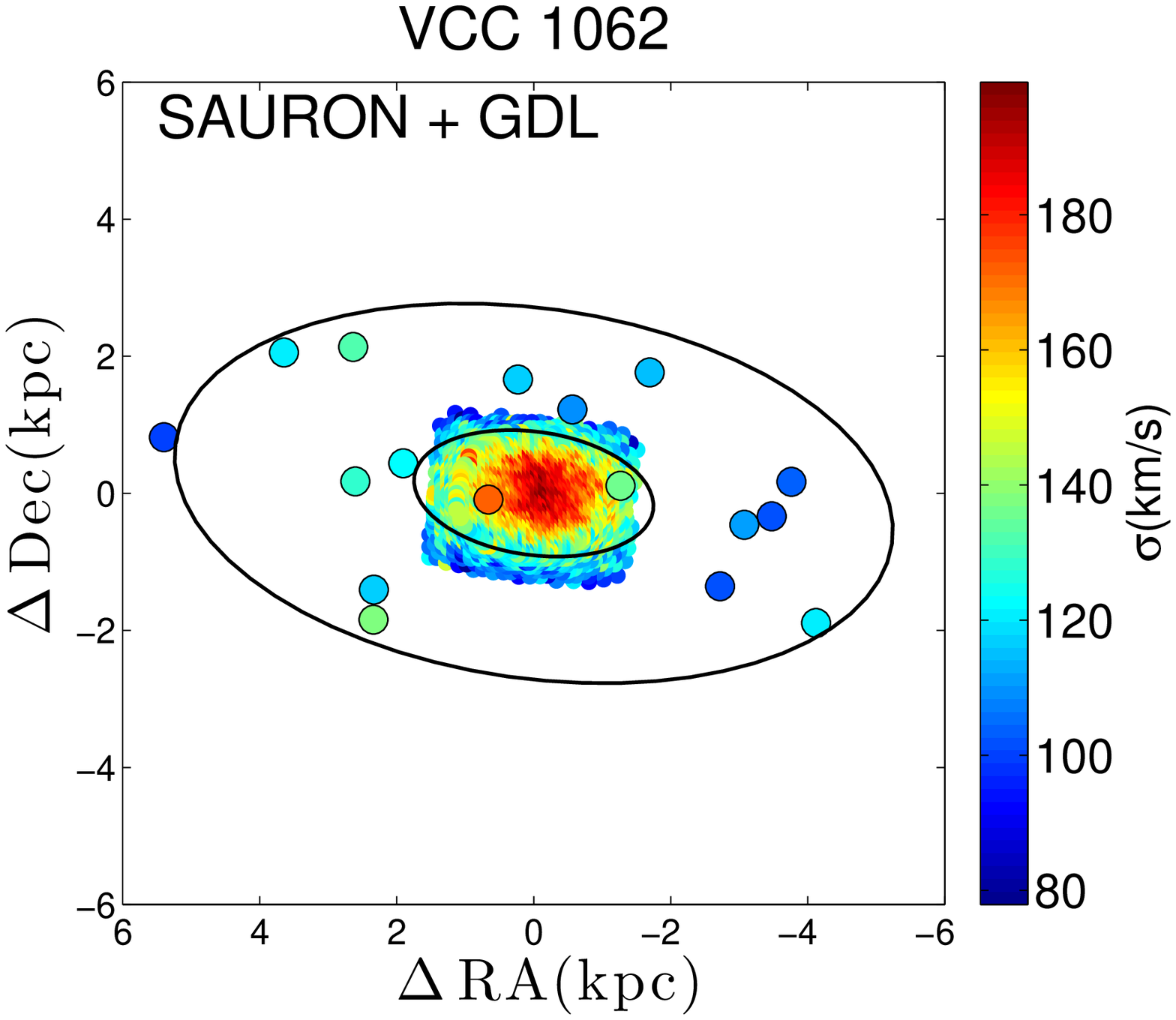}{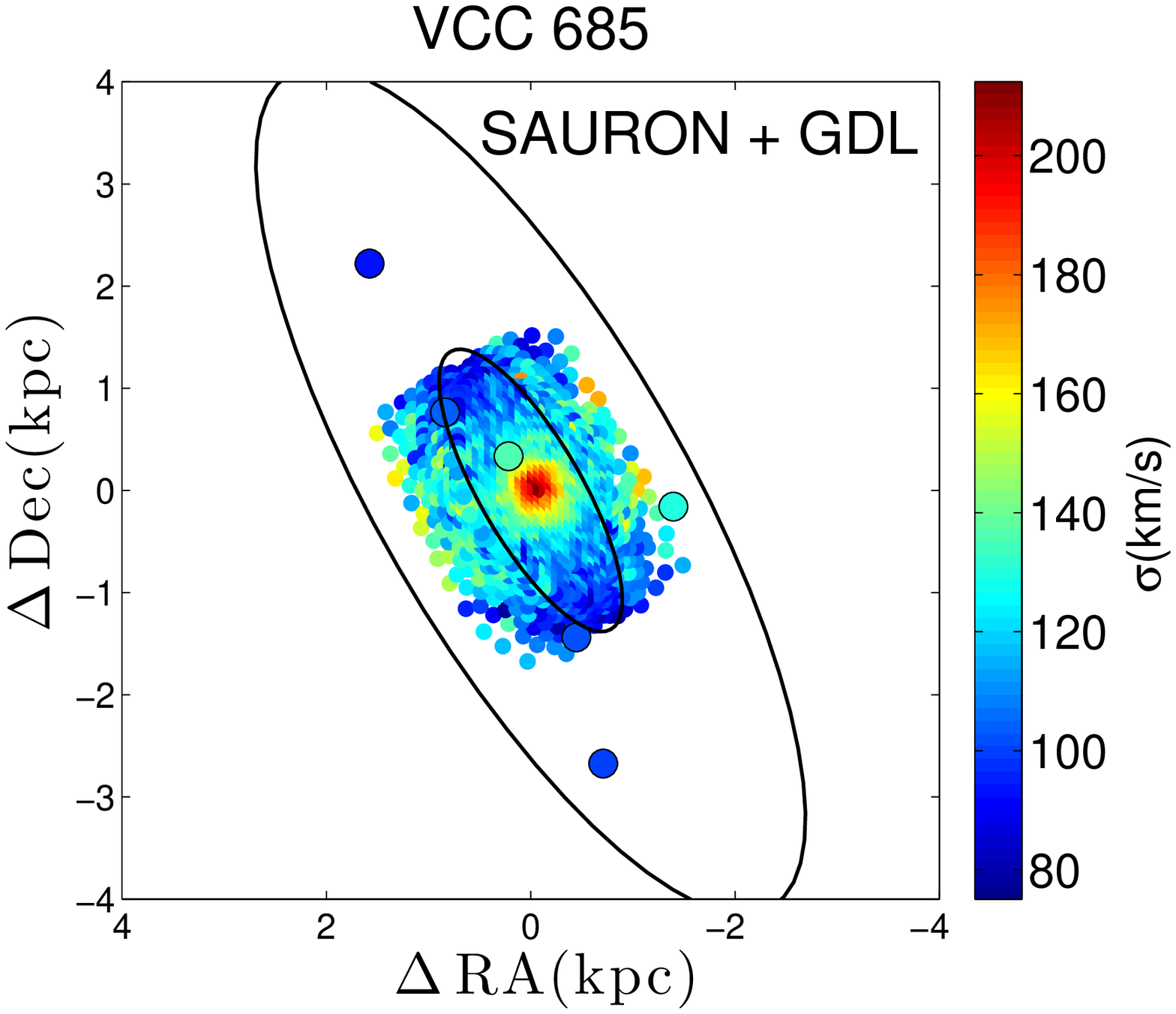}
\caption{Same as in Figure \ref{app1}, but for VCC 1062 and 685. } 
\label{app}
\end{figure}

\clearpage
\LongTables

\begin{deluxetable*}{llllllll}
\tabletypesize{\scriptsize}
\centering 
\tablewidth{0pt}
\tablecaption{Globular Cluster\label{gc_table}}
\tablehead{
\colhead{ID}
& \colhead{R.A.(J2000)}
& \colhead{Decl.(J2000)}
& \colhead{\it v$_{\rm los}$}
& \colhead{\it g}
& \colhead{\it z}
& \colhead{\it g-z}
& \colhead{\it E(B-V)}\\
\colhead{}
& \colhead{(deg)}
& \colhead{(deg)}
& \colhead{\rm (km s$^{-1}$)}
& \colhead{(mag)}
& \colhead{(mag)}
& \colhead{(mag)}
& \colhead{(mag)}\\
\colhead{(1)}
& \colhead{(2)}
& \colhead{(3)}
& \colhead{(4)}
& \colhead{(5)}
& \colhead{(6)}
& \colhead{(7)}
& \colhead{(8)}
}
\startdata

\multicolumn{8}{c}{VCC 1231}\\
\hline
VCC1231GC-01&187.420170& 13.436530&2450$\pm$19& 21.280$\pm$0.016& 20.229$\pm$0.024&  0.991$\pm$0.007&0.028\\
VCC1231GC-02&187.422675& 13.437170&2107$\pm$22& 21.960$\pm$0.026& 20.922$\pm$0.029&  0.973$\pm$0.033&0.028\\
VCC1231GC-03&187.428810& 13.436997&2377$\pm$20& 22.192$\pm$0.022& 20.893$\pm$0.020&  1.253$\pm$0.028&0.028\\
VCC1231GC-04&187.413315& 13.420913&2429$\pm$15& 22.233$\pm$0.030& 20.788$\pm$0.019&  1.374$\pm$0.024&0.028\\
VCC1231GC-05&187.418970& 13.424147&2357$\pm$17& 23.033$\pm$0.032& 22.196$\pm$0.031&  0.817$\pm$0.022&0.028\\
VCC1231GC-06&187.411215& 13.419499&2229$\pm$21& 23.147$\pm$0.038& 22.185$\pm$0.031&  0.907$\pm$0.025&0.028\\
VCC1231GC-07&187.405065& 13.453014&2007$\pm$18& 22.096$\pm$0.007& 21.359$\pm$0.016&  0.703$\pm$0.012&0.028\\
VCC1231GC-08&187.429815& 13.401846&2058$\pm$49& 22.933$\pm$0.013& 22.271$\pm$0.037&  0.551$\pm$0.030&0.028\\
VCC1231GC-09&187.432710& 13.444921&2282$\pm$20& 22.855$\pm$0.022& 21.790$\pm$0.021&  1.018$\pm$0.025&0.028\\
VCC1231GC-10&187.441245& 13.429025&2435$\pm$28& 21.842$\pm$0.037& 20.977$\pm$0.018&  0.799$\pm$0.039&0.028\\
VCC1231GC-11&187.437780& 13.422738&2599$\pm$24& 21.821$\pm$0.012& 20.903$\pm$0.020&  0.832$\pm$0.019&0.028\\
VCC1231GC-12&187.447410& 13.432945&2324$\pm$20& 21.209$\pm$0.029& 20.154$\pm$0.020&  0.992$\pm$0.028&0.028\\
VCC1231GC-13&187.450110& 13.433290&2533$\pm$21& 21.813$\pm$0.019& 20.817$\pm$0.012&  0.946$\pm$0.020&0.028\\
VCC1231GC-14&187.454700& 13.403273&2250$\pm$24& 22.485$\pm$0.018& 21.570$\pm$0.016&  0.871$\pm$0.022&0.028\\
VCC1231GC-15&187.452990& 13.445077&2356$\pm$20& 21.861$\pm$0.017& 20.343$\pm$0.013&  1.436$\pm$0.016&0.028\\
VCC1231GC-16&187.456470& 13.424834&2389$\pm$27& 22.005$\pm$0.020& 20.962$\pm$0.015&  0.990$\pm$0.021&0.028\\
VCC1231GC-17&187.463295& 13.427741&2367$\pm$17& 21.306$\pm$0.014& 20.320$\pm$0.017&  0.920$\pm$0.019&0.028\\
VCC1231GC-18&187.474695& 13.439762&2151$\pm$26& 22.359$\pm$0.022& 21.412$\pm$0.016&  0.885$\pm$0.022&0.028\\
VCC1231GC-19&187.468155& 13.427391&2523$\pm$17& 21.806$\pm$0.032& 20.469$\pm$0.011&  1.274$\pm$0.028&0.028\\
VCC1231GC-20&187.475715& 13.426306&2107$\pm$29& 22.550$\pm$0.020& 21.521$\pm$0.019&  0.967$\pm$0.022&0.028\\
VCC1231GC-21&187.469025& 13.435520&2337$\pm$20& 22.932$\pm$0.025& 21.607$\pm$0.014&  1.278$\pm$0.024&0.028\\
VCC1231GC-22&187.479150& 13.444892&2148$\pm$23& 22.673$\pm$0.023& 21.639$\pm$0.015&  1.014$\pm$0.026&0.028\\
VCC1231GC-23&187.485810& 13.446020&2412$\pm$15& 21.166$\pm$0.032& 20.194$\pm$0.022&  0.880$\pm$0.022&0.028\\
VCC1231GC-24&187.496295& 13.436360&2200$\pm$11& 21.183$\pm$0.004& 20.467$\pm$0.008&  0.647$\pm$0.007&0.028\\
VCC1231GC-25&187.429695& 13.433433&1909$\pm$15& 22.667$\pm$0.017& 21.218$\pm$0.019&  1.384$\pm$0.019&0.028\\
VCC1231GC-26&187.424715& 13.427611&1988$\pm$23& 22.616$\pm$0.017& 21.261$\pm$0.018&  1.294$\pm$0.019&0.028\\
VCC1231GC-27&187.422900& 13.414083&2226$\pm$26& 22.785$\pm$0.056& 21.737$\pm$0.062&  0.909$\pm$0.021&0.028\\
VCC1231GC-28&187.422360& 13.402969&2364$\pm$14& 21.274$\pm$0.003& 20.445$\pm$0.007&  0.763$\pm$0.006&0.028\\
VCC1231GC-29&187.450515& 13.431453&2198$\pm$31& 21.206$\pm$0.014& 20.287$\pm$0.018&  0.849$\pm$0.014&0.028\\
VCC1231GC-30&187.447620& 13.429241&2418$\pm$23& 21.843$\pm$0.025& 20.514$\pm$0.016&  1.262$\pm$0.027&0.028\\
VCC1231GC-31&187.442595& 13.426672&2151$\pm$20& 21.411$\pm$0.016& 20.467$\pm$0.013&  0.876$\pm$0.015&0.028\\
VCC1231GC-32&187.461765& 13.411035&2351$\pm$16& 22.139$\pm$0.080& 21.053$\pm$0.022&  0.999$\pm$0.074&0.028\\
VCC1231GC-33&187.434585& 13.434356&2397$\pm$16& 22.953$\pm$0.027& 21.884$\pm$0.025&  0.990$\pm$0.025&0.028\\
VCC1231GC-34&187.444185& 13.431620&2155$\pm$69& 22.671$\pm$0.023& 21.758$\pm$0.023&  0.826$\pm$0.023&0.028\\
VCC1231GC-35&187.457610& 13.433330&2013$\pm$24& 22.274$\pm$0.020& 21.276$\pm$0.020&  0.931$\pm$0.023&0.028\\
VCC1231GC-36&187.452285& 13.459685&2012$\pm$38& 22.952$\pm$0.034& 21.852$\pm$0.026&  1.031$\pm$0.031&0.028\\
VCC1231GC-37&187.460490& 13.472438&2140$\pm$21& 21.361$\pm$0.005& 20.300$\pm$0.007&  1.046$\pm$0.006&0.028\\
VCC1231GC-38&187.473540& 13.428547&2033$\pm$35& 22.964$\pm$0.053& 22.082$\pm$0.042&  0.821$\pm$0.064&0.028\\
VCC1231GC-39&187.493505& 13.426348&1929$\pm$34& 20.809$\pm$0.002& 19.955$\pm$0.005&  0.755$\pm$0.005&0.028\\
VCC1231GC-40&187.428210& 13.420699&2371$\pm$32& 23.131$\pm$0.017& 21.801$\pm$0.012&  1.271$\pm$0.019&0.028\\
VCC1231GC-41&187.429515& 13.423301&2199$\pm$48& 23.347$\pm$0.025& 22.404$\pm$0.022&  0.885$\pm$0.029&0.028\\
VCC1231GC-42&187.451670& 13.442974&2538$\pm$16& 22.347$\pm$0.016& 21.181$\pm$0.025&  1.114$\pm$0.029&0.028\\
VCC1231GC-43&187.457820& 13.406286&2272$\pm$15& 21.564$\pm$0.016& 20.385$\pm$0.018&  1.075$\pm$0.014&0.028\\
VCC1231GC-44&187.465260& 13.417413&2467$\pm$11& 20.955$\pm$0.024& 19.776$\pm$0.012&  1.126$\pm$0.023&0.028\\
VCC1231GC-45&187.453365& 13.425970&2230$\pm$14& 21.304$\pm$0.021& 20.272$\pm$0.015&  0.957$\pm$0.017&0.028\\
VCC1231GC-46&187.438560& 13.435633&2014$\pm$17& 22.761$\pm$0.015& 21.614$\pm$0.020&  1.070$\pm$0.020&0.028\\
VCC1231GC-47&187.445190& 13.432550&2213$\pm$36& 23.151$\pm$0.026& 21.727$\pm$0.022&  1.358$\pm$0.031&0.028\\
VCC1231GC-48&187.461540& 13.434665&2125$\pm$27& 22.784$\pm$0.024& 21.886$\pm$0.016&  0.837$\pm$0.026&0.028\\
VCC1231GC-49&187.474830& 13.443451&2450$\pm$20& 22.730$\pm$0.021& 21.425$\pm$0.020&  1.236$\pm$0.026&0.028\\
VCC1231GC-50&187.468185& 13.430573&2459$\pm$38& 23.187$\pm$0.015& 21.905$\pm$0.026&  1.206$\pm$0.021&0.028\\
VCC1231GC-51&187.476593& 13.420838&2161$\pm$24& 22.740$\pm$0.012& 21.886$\pm$0.028&  0.678$\pm$0.021&0.028\\
\cutinhead{VCC 2000}
VCC2000GC-01&191.092740& 11.213082&1081$\pm$12& 20.572$\pm$0.002& 19.893$\pm$0.005&  0.611$\pm$0.004&0.034\\
VCC2000GC-02&191.122215& 11.219080& 949$\pm$42& 22.741$\pm$0.027& 21.912$\pm$0.027&  0.778$\pm$0.025&0.034\\
VCC2000GC-03&191.125230& 11.215344& 907$\pm$28& 22.478$\pm$0.029& 21.556$\pm$0.029&  0.906$\pm$0.022&0.034\\
VCC2000GC-04&191.128680& 11.213797&1016$\pm$19& 22.727$\pm$0.066& 21.802$\pm$0.052&  0.889$\pm$0.028&0.034\\
VCC2000GC-05&191.109285& 11.205482& 920$\pm$17& 21.151$\pm$0.015& 20.243$\pm$0.014&  0.889$\pm$0.022&0.034\\
VCC2000GC-06&191.138010& 11.200358&1074$\pm$24& 22.548$\pm$0.069& 21.258$\pm$0.023&  1.181$\pm$0.070&0.034\\
VCC2000GC-07&191.135130& 11.198518& 789$\pm$22& 21.885$\pm$0.032& 20.882$\pm$0.063&  0.863$\pm$0.032&0.034\\
VCC2000GC-08&191.131050& 11.197987& 917$\pm$33& 22.523$\pm$0.032& 21.546$\pm$0.013&  0.895$\pm$0.035&0.034\\
VCC2000GC-09&191.114865& 11.196526&1009$\pm$18& 22.778$\pm$0.027& 21.642$\pm$0.014&  1.071$\pm$0.028&0.034\\
VCC2000GC-10&191.149545& 11.201708& 990$\pm$25& 22.250$\pm$0.023& 21.305$\pm$0.020&  0.888$\pm$0.025&0.034\\
VCC2000GC-11&191.143635& 11.196664&1344$\pm$48& 22.456$\pm$0.036& 21.622$\pm$0.029&  0.760$\pm$0.031&0.034\\
VCC2000GC-12&191.140995& 11.196223&1263$\pm$42& 22.247$\pm$0.019& 21.427$\pm$0.014&  0.752$\pm$0.020&0.034\\
VCC2000GC-13&191.146065& 11.194601&1133$\pm$17& 22.156$\pm$0.019& 20.993$\pm$0.015&  1.116$\pm$0.021&0.034\\
VCC2000GC-14&191.151405& 11.193774& 972$\pm$78& 22.561$\pm$0.020& 21.807$\pm$0.013&  0.705$\pm$0.022&0.034\\
VCC2000GC-15&191.167200& 11.189954&1085$\pm$28& 21.389$\pm$0.004& 20.758$\pm$0.011&  0.576$\pm$0.007&0.034\\
VCC2000GC-16&191.119545& 11.199561&1171$\pm$22& 22.462$\pm$0.017& 21.455$\pm$0.015&  0.959$\pm$0.019&0.034\\
VCC2000GC-17&191.125890& 11.190406&1250$\pm$18& 22.211$\pm$0.033& 20.846$\pm$0.015&  1.279$\pm$0.025&0.034\\
VCC2000GC-18&191.107575& 11.187544& 982$\pm$31& 22.356$\pm$0.015& 21.531$\pm$0.021&  0.742$\pm$0.021&0.034\\
VCC2000GC-19&191.123100& 11.191837&1267$\pm$15& 21.029$\pm$0.024& 20.032$\pm$0.022&  0.907$\pm$0.021&0.034\\
VCC2000GC-20&191.129370& 11.193860&1270$\pm$19& 21.611$\pm$0.027& 20.632$\pm$0.019&  0.903$\pm$0.011&0.034\\
VCC2000GC-21&191.111385& 11.195764&1214$\pm$43& 23.032$\pm$0.029& 22.234$\pm$0.030&  0.719$\pm$0.034&0.034\\
VCC2000GC-22&191.141985& 11.195713& 921$\pm$27& 22.467$\pm$0.032& 21.466$\pm$0.018&  0.942$\pm$0.033&0.034\\
VCC2000GC-23&191.138505& 11.193366& 841$\pm$17& 21.266$\pm$0.019& 20.263$\pm$0.013&  0.933$\pm$0.017&0.034\\
VCC2000GC-24&191.146710& 11.191358&1022$\pm$18& 20.553$\pm$0.058& 19.691$\pm$0.052&  0.787$\pm$0.023&0.034\\
VCC2000GC-25&191.155815& 11.187847& 982$\pm$23& 21.848$\pm$0.029& 20.955$\pm$0.023&  0.870$\pm$0.031&0.034\\
VCC2000GC-26&191.150445& 11.183163& 942$\pm$21& 21.543$\pm$0.031& 20.743$\pm$0.034&  0.731$\pm$0.011&0.034\\
VCC2000GC-27&191.145060& 11.197882& 863$\pm$28& 22.877$\pm$0.024& 22.085$\pm$0.020&  0.729$\pm$0.026&0.034\\
VCC2000GC-28&191.159205& 11.174632&1154$\pm$24& 21.152$\pm$0.003& 20.439$\pm$0.008&  0.635$\pm$0.006&0.034\\
VCC2000GC-29&191.138430& 11.210131& 760$\pm$32& 22.579$\pm$0.030& 21.743$\pm$0.021&  0.768$\pm$0.026&0.034\\
VCC2000GC-30&191.127015& 11.186158&1095$\pm$29& 22.510$\pm$0.032& 21.358$\pm$0.017&  1.079$\pm$0.024&0.034\\
VCC2000GC-31&191.119665& 11.189951& 845$\pm$16& 20.777$\pm$0.022& 19.771$\pm$0.019&  0.937$\pm$0.022&0.034\\
VCC2000GC-32&191.130135& 11.184649&1261$\pm$19& 21.572$\pm$0.020& 20.567$\pm$0.008&  0.900$\pm$0.019&0.034\\
VCC2000GC-33&191.133315& 11.183667&1010$\pm$27& 22.594$\pm$0.032& 21.563$\pm$0.024&  0.970$\pm$0.036&0.034\\
VCC2000GC-34&191.110065& 11.210609&1112$\pm$33& 22.839$\pm$0.022& 22.013$\pm$0.020&  0.825$\pm$0.023&0.034\\
VCC2000GC-35&191.122260& 11.178163&1087$\pm$28& 22.925$\pm$0.038& 21.779$\pm$0.016&  1.083$\pm$0.035&0.034\\
VCC2000GC-36&191.116050& 11.230333&1109$\pm$20& 20.599$\pm$0.002& 19.878$\pm$0.006&  0.629$\pm$0.004&0.034\\
VCC2000GC-37&191.140260& 11.163837&1189$\pm$26& 22.380$\pm$0.012& 21.439$\pm$0.014&  0.902$\pm$0.016&0.034\\
VCC2000GC-38&191.138190& 11.176321&1004$\pm$38& 21.193$\pm$0.030& 20.322$\pm$0.026&  0.792$\pm$0.017&0.034\\
VCC2000GC-39&191.147055& 11.198483&1149$\pm$43& 23.000$\pm$0.026& 22.169$\pm$0.025&  0.777$\pm$0.023&0.034\\
VCC2000GC-40&191.146830& 11.172589& 970$\pm$56& 23.030$\pm$0.033& 22.174$\pm$0.033&  0.809$\pm$0.024&0.034\\
VCC2000GC-41&191.155275& 11.201397&1060$\pm$14& 21.365$\pm$0.015& 20.465$\pm$0.011&  0.851$\pm$0.013&0.034\\
VCC2000GC-42&191.161890& 11.205892&1086$\pm$22& 22.710$\pm$0.037& 21.815$\pm$0.026&  0.843$\pm$0.031&0.034\\
VCC2000GC-43&191.167455& 11.181426& 942$\pm$37& 22.561$\pm$0.010& 21.812$\pm$0.028&  0.658$\pm$0.019&0.034\\
\cutinhead{VCC 1062}
VCC1062GC-01&187.040970&  9.794455& 683$\pm$12& 22.890$\pm$0.020& 21.534$\pm$0.020&  1.325$\pm$0.023&0.022\\
VCC1062GC-02&187.000530&  9.817132& 444$\pm$19& 21.895$\pm$0.013& 20.964$\pm$0.013&  0.886$\pm$0.014&0.022\\
VCC1062GC-03&187.004025&  9.816408& 512$\pm$23& 22.575$\pm$0.045& 21.757$\pm$0.062&  0.729$\pm$0.060&0.022\\
VCC1062GC-04&186.995685&  9.807220& 306$\pm$18& 24.395$\pm$0.038& 23.127$\pm$0.057&  1.231$\pm$0.059&0.022\\
VCC1062GC-05&186.999240&  9.805346& 518$\pm$15& 21.978$\pm$0.024& 21.020$\pm$0.018&  0.902$\pm$0.022&0.022\\
VCC1062GC-06&186.989805&  9.792907& 342$\pm$21& 22.503$\pm$0.029& 21.568$\pm$0.029&  0.859$\pm$0.023&0.022\\
VCC1062GC-07&186.991335&  9.802511& 229$\pm$30& 23.130$\pm$0.020& 22.134$\pm$0.022&  0.980$\pm$0.023&0.022\\
VCC1062GC-08&186.985650&  9.798775& 326$\pm$21& 22.977$\pm$0.016& 21.823$\pm$0.026&  1.109$\pm$0.020&0.022\\
VCC1062GC-09&187.012260&  9.817571& 421$\pm$21& 22.418$\pm$0.012& 21.570$\pm$0.025&  0.791$\pm$0.017&0.022\\
VCC1062GC-10&187.009785&  9.810227& 417$\pm$21& 22.582$\pm$0.021& 21.196$\pm$0.017&  1.305$\pm$0.023&0.022\\
VCC1062GC-11&187.017090&  9.809849& 619$\pm$18& 22.279$\pm$0.023& 21.295$\pm$0.017&  0.908$\pm$0.020&0.022\\
VCC1062GC-12&187.026240&  9.811627& 595$\pm$23& 21.870$\pm$0.032& 20.928$\pm$0.051&  0.805$\pm$0.021&0.022\\
VCC1062GC-13&187.025115&  9.796717& 679$\pm$20& 21.588$\pm$0.025& 20.716$\pm$0.038&  0.778$\pm$0.017&0.022\\
VCC1062GC-14&187.037430&  9.792029& 516$\pm$13& 21.699$\pm$0.017& 20.740$\pm$0.012&  0.914$\pm$0.020&0.022\\
VCC1062GC-15&187.030080&  9.811315& 547$\pm$18& 22.869$\pm$0.026& 21.944$\pm$0.024&  0.891$\pm$0.021&0.022\\
VCC1062GC-16&187.047585&  9.805256& 583$\pm$17& 22.437$\pm$0.021& 21.309$\pm$0.016&  1.088$\pm$0.021&0.022\\
VCC1062GC-17&187.044615&  9.816442& 301$\pm$25& 22.926$\pm$0.020& 21.979$\pm$0.015&  0.900$\pm$0.022&0.022\\
VCC1062GC-18&187.054260&  9.809342& 645$\pm$21& 23.102$\pm$0.038& 21.681$\pm$0.027&  1.413$\pm$0.025&0.022\\
VCC1062GC-19&187.000560&  9.796535& 380$\pm$19& 22.458$\pm$0.039& 20.965$\pm$0.037&  1.487$\pm$0.020&0.022\\
VCC1062GC-20&187.005885&  9.798541& 472$\pm$17& 22.254$\pm$0.049& 20.662$\pm$0.061&  1.482$\pm$0.033&0.022\\
VCC1062GC-21&186.994635&  9.807293& 309$\pm$21& 22.574$\pm$0.020& 21.113$\pm$0.018&  1.414$\pm$0.022&0.022\\
VCC1062GC-22&187.003020&  9.802365& 279$\pm$11& 21.228$\pm$0.017& 20.009$\pm$0.020&  1.165$\pm$0.019&0.022\\
VCC1062GC-23&186.999375&  9.790121& 234$\pm$18& 23.134$\pm$0.033& 22.105$\pm$0.017&  1.038$\pm$0.030&0.022\\
VCC1062GC-24&186.986835&  9.806428& 311$\pm$14& 22.979$\pm$0.016& 21.892$\pm$0.028&  1.094$\pm$0.019&0.022\\
VCC1062GC-25&187.014075&  9.808210& 461$\pm$19& 22.665$\pm$0.033& 21.234$\pm$0.024&  1.293$\pm$0.017&0.022\\
VCC1062GC-26&187.018665&  9.791675& 486$\pm$13& 21.268$\pm$0.019& 20.328$\pm$0.015&  0.859$\pm$0.019&0.022\\
VCC1062GC-27&187.012890&  9.792312& 310$\pm$18& 22.946$\pm$0.019& 21.573$\pm$0.021&  1.289$\pm$0.020&0.022\\
VCC1062GC-28&187.028310&  9.813976& 605$\pm$19& 22.592$\pm$0.020& 21.547$\pm$0.017&  0.991$\pm$0.018&0.022\\
VCC1062GC-29&187.037685&  9.798342& 743$\pm$19& 21.771$\pm$0.058& 20.911$\pm$0.051&  0.866$\pm$0.016&0.022\\
VCC1062GC-30&187.025085&  9.798357& 488$\pm$57& 23.113$\pm$0.030& 22.145$\pm$0.051&  0.866$\pm$0.037&0.022\\
VCC1062GC-31&187.050615&  9.787901& 648$\pm$17& 22.409$\pm$0.009& 21.621$\pm$0.021&  0.718$\pm$0.015&0.022\\
VCC1062GC-32&187.067940&  9.802051& 567$\pm$11& 21.506$\pm$0.004& 20.749$\pm$0.010&  0.720$\pm$0.007&0.022\\
VCC1062GC-33&187.004535&  9.801900& 281$\pm$12& 24.777$\pm$0.062& 23.460$\pm$0.066&  1.271$\pm$0.068&0.022\\
VCC1062GC-34&187.011405&  9.804032& 401$\pm$11& 21.189$\pm$0.016& 20.091$\pm$0.020&  1.012$\pm$0.016&0.022\\
VCC1062GC-35&187.018740&  9.803278& 569$\pm$24& 21.804$\pm$0.017& 20.602$\pm$0.019&  1.157$\pm$0.021&0.022\\
VCC1062GC-36&187.026120&  9.804276& 717$\pm$18& 22.399$\pm$0.021& 21.325$\pm$0.021&  1.002$\pm$0.024&0.022\\
VCC1062GC-37&187.023465&  9.805268& 414$\pm$23& 22.974$\pm$0.022& 21.496$\pm$0.045&  1.366$\pm$0.030&0.022\\
VCC1062GC-38&187.036740&  9.806686& 757$\pm$15& 22.842$\pm$0.027& 21.257$\pm$0.025&  1.534$\pm$0.033&0.022\\
VCC1062GC-39&187.001925&  9.804249& 336$\pm$10& 20.526$\pm$0.020& 19.357$\pm$0.023&  1.111$\pm$0.023&0.022\\
VCC1062GC-40&186.992850&  9.808851& 287$\pm$18& 22.179$\pm$0.008& 21.139$\pm$0.014&  0.991$\pm$0.011&0.022\\
\cutinhead{VCC 685}
VCC685GC-01&185.989590& 16.651251&1341$\pm$24& 22.262$\pm$0.008& 21.616$\pm$0.026&  0.609$\pm$0.017&0.028\\
VCC685GC-02&185.959680& 16.679079&1334$\pm$27& 22.130$\pm$0.021& 21.205$\pm$0.014&  0.868$\pm$0.019&0.028\\
VCC685GC-03&185.975325& 16.684580&1422$\pm$22& 21.954$\pm$0.014& 20.998$\pm$0.011&  0.897$\pm$0.016&0.028\\
VCC685GC-04&185.978040& 16.679291&1285$\pm$31& 22.256$\pm$0.014& 21.422$\pm$0.017&  0.754$\pm$0.017&0.028\\
VCC685GC-05&185.988135& 16.683411& 918$\pm$42& 22.367$\pm$0.021& 21.238$\pm$0.022&  1.021$\pm$0.016&0.028\\
VCC685GC-06&185.985465& 16.692770&1025$\pm$23& 21.566$\pm$0.019& 20.658$\pm$0.017&  0.844$\pm$0.010&0.028\\
VCC685GC-07&185.989155& 16.688009&1062$\pm$28& 21.987$\pm$0.026& 21.125$\pm$0.018&  0.811$\pm$0.024&0.028\\
VCC685GC-08&185.989965& 16.701780&1168$\pm$14& 21.484$\pm$0.021& 20.311$\pm$0.013&  1.115$\pm$0.023&0.028\\
VCC685GC-09&185.991735& 16.694599&1588$\pm$17& 21.529$\pm$0.012& 20.532$\pm$0.018&  0.903$\pm$0.017&0.028\\
VCC685GC-10&185.994150& 16.696199&1520$\pm$38& 21.916$\pm$0.023& 20.950$\pm$0.028&  0.910$\pm$0.018&0.028\\
VCC685GC-11&185.991225& 16.709190&1343$\pm$21& 20.597$\pm$0.018& 19.737$\pm$0.012&  0.811$\pm$0.006&0.028\\
VCC685GC-12&185.997015& 16.701620&1178$\pm$22& 21.662$\pm$0.016& 20.724$\pm$0.012&  0.878$\pm$0.020&0.028\\
VCC685GC-13&185.996025& 16.711010&1208$\pm$15& 21.145$\pm$0.015& 19.992$\pm$0.012&  1.084$\pm$0.017&0.028\\
VCC685GC-14&186.006900& 16.716101& 925$\pm$33& 22.649$\pm$0.019& 21.725$\pm$0.015&  0.871$\pm$0.021&0.028\\
VCC685GC-15&186.000240& 16.730289&1098$\pm$31& 21.078$\pm$0.003& 20.493$\pm$0.012&  0.508$\pm$0.008&0.028\\
\enddata
\tablecomments{
Col.(1):\ Object ID;
Col.(2):\ Right ascension in decimal degrees (J2000);
Col.(3):\ Declination in decimal degrees (J2000);
Col.(4):\ Heliocentric radial velocity and uncertainty;
Cols.(5,6):\ $g$ and $z$ band magnitudes from the ACSVCS and NGVS (not corrected for Galactic extinction);
Col.(7):\ $g-z$ color;
Col.(8):\ The Galactic reddening determined by Schlegel et al.\
(1998). The corrections for foreground reddening were taken to be $A_g
= 3.634E(B-V )$ and $A_z = 1.485E(B-V )$ in the g and z bands, respectively \citep{jo04}.
}
\end{deluxetable*}

\begin{deluxetable*}{lllll}
\tablecolumns{5}
\tablewidth{0pt}
\tablecaption{Galaxy Diffuse Light}
\tablehead{
\colhead{ID}
& \colhead{R.A.(J2000)}
& \colhead{Decl.(J2000)}
& \colhead{\it v$_{\rm los}$}
& \colhead{\it $\sigma_{\rm los}$}\\
\colhead{}
& \colhead{(deg)}
& \colhead{(deg)}
& \colhead{\rm (km s$^{-1}$)}
& \colhead{\rm (km s$^{-1}$)}\\
\colhead{(1)}
& \colhead{(2)}
& \colhead{(3)}
& \colhead{(4)}
& \colhead{(5)}
}
\startdata

\multicolumn{5}{c}{VCC 1231}\\
\hline
VCC1231GDL-01&187.450515& 13.431453&2188$\pm$14& 174$\pm$ 4\\
VCC1231GDL-02&187.447620& 13.429241&2205$\pm$18& 208$\pm$ 4\\
VCC1231GDL-03&187.453365& 13.425970&2244$\pm$12& 153$\pm$ 4\\
VCC1231GDL-04&187.456470& 13.424834&2270$\pm$14& 157$\pm$ 5\\
VCC1231GDL-05&187.457610& 13.433330&2252$\pm$14& 154$\pm$ 5\\
VCC1231GDL-06&187.447410& 13.432945&2193$\pm$14& 185$\pm$ 5\\
VCC1231GDL-07&187.450110& 13.433290&2191$\pm$14& 163$\pm$ 3\\
VCC1231GDL-08&187.445190& 13.432550&2206$\pm$14& 189$\pm$ 2\\
VCC1231GDL-09&187.444185& 13.431620&2200$\pm$18& 197$\pm$ 5\\
VCC1231GDL-10&187.441245& 13.429025&2215$\pm$18& 204$\pm$ 5\\
VCC1231GDL-11&187.442595& 13.426672&2209$\pm$16& 195$\pm$ 5\\
VCC1231GDL-12&187.461540& 13.434665&2251$\pm$16& 165$\pm$ 6\\
VCC1231GDL-13&187.463295& 13.427741&2250$\pm$15& 203$\pm$ 3\\
VCC1231GDL-14&187.468185& 13.430573&2244$\pm$17& 197$\pm$ 6\\
VCC1231GDL-15&187.468155& 13.427391&2249$\pm$17& 196$\pm$ 5\\
VCC1231GDL-16&187.473540& 13.428547&2225$\pm$25& 186$\pm$11\\
VCC1231GDL-17&187.438560& 13.435633&2229$\pm$22& 211$\pm$10\\
VCC1231GDL-18&187.475715& 13.426306&2243$\pm$29&   \nodata\\
VCC1231GDL-19&187.469025& 13.435520&2239$\pm$25&   \nodata\\
VCC1231GDL-20&187.434585& 13.434356&2203$\pm$17&   \nodata\\
VCC1231GDL-21&187.437780& 13.422738&2217$\pm$24&   \nodata\\
VCC1231GDL-22&187.428810& 13.436997&2257$\pm$29&   \nodata\\
VCC1231GDL-23&187.429695& 13.433433&2200$\pm$20&   \nodata\\
VCC1231GDL-24&187.429515& 13.423301&2206$\pm$24&   \nodata\\
VCC1231GDL-25&187.451670& 13.442974&2210$\pm$26&   \nodata\\
VCC1231GDL-26&187.465260& 13.417413&2265$\pm$32&   \nodata\\
\cutinhead{VCC 2000}
VCC2000GDL-01&191.134440& 11.189226&1032$\pm$12& 142$\pm$ 3\\
VCC2000GDL-02&191.131935& 11.189683&1144$\pm$12& 145$\pm$ 3\\
VCC2000GDL-03&191.138505& 11.193366&1034$\pm$10& 138$\pm$ 7\\
VCC2000GDL-04&191.129370& 11.193860&1115$\pm$10& 129$\pm$ 6\\
VCC2000GDL-05&191.125890& 11.190406&1166$\pm$11& 138$\pm$ 9\\
VCC2000GDL-06&191.123100& 11.191837&1139$\pm$12& 149$\pm$12\\
VCC2000GDL-07&191.127015& 11.186158&1109$\pm$13& 115$\pm$10\\
VCC2000GDL-08&191.130135& 11.184649&1094$\pm$16& 126$\pm$10\\
VCC2000GDL-09&191.133315& 11.183667&1065$\pm$16& 118$\pm$ 8\\
VCC2000GDL-10&191.138505& 11.193366&1034$\pm$ 9&   \nodata\\
VCC2000GDL-11&191.119665& 11.189951&1134$\pm$17&   \nodata\\
VCC2000GDL-12&191.146710& 11.191358& 996$\pm$28&   \nodata\\
VCC2000GDL-13&191.141985& 11.195713&1067$\pm$21&   \nodata\\
VCC2000GDL-14&191.140995& 11.196223&1058$\pm$20&   \nodata\\
VCC2000GDL-15&191.135130& 11.198518&1057$\pm$16&   \nodata\\
VCC2000GDL-16&191.131050& 11.197987&1075$\pm$19&   \nodata\\
VCC2000GDL-17&191.143635& 11.196664&1040$\pm$25&   \nodata\\
VCC2000GDL-18&191.146065& 11.194601&1029$\pm$26&   \nodata\\
VCC2000GDL-19&191.145060& 11.197882& 995$\pm$33&   \nodata\\
VCC2000GDL-20&191.119545& 11.199561&1052$\pm$44&   \nodata\\
VCC2000GDL-21&191.119845& 11.183841&1080$\pm$42&   \nodata\\
\cutinhead{VCC 1062}
VCC1062GDL-01&187.018740&  9.803278& 611$\pm$17& 172$\pm$ 3\\
VCC1062GDL-02&187.026120&  9.804276& 683$\pm$11& 128$\pm$ 6\\
VCC1062GDL-03&187.023465&  9.805268& 640$\pm$12& 123$\pm$ 4\\
VCC1062GDL-04&187.014075&  9.808210& 486$\pm$11& 110$\pm$ 5\\
VCC1062GDL-05&187.011405&  9.804032& 419$\pm$12& 136$\pm$20\\
VCC1062GDL-06&187.017090&  9.809849& 531$\pm$12& 116$\pm$ 7\\
VCC1062GDL-07&187.009785&  9.810227& 468$\pm$14& 115$\pm$10\\
VCC1062GDL-08&187.026240&  9.811627& 613$\pm$19& 133$\pm$12\\
VCC1062GDL-09&187.030080&  9.811315& 640$\pm$15& 120$\pm$14\\
VCC1062GDL-10&187.036740&  9.806686& 799$\pm$10& 100$\pm$ 7\\
VCC1062GDL-11&187.025115&  9.796717& 606$\pm$11& 138$\pm$10\\
VCC1062GDL-12&187.025085&  9.798357& 610$\pm$10& 116$\pm$ 7\\
VCC1062GDL-13&187.000560&  9.796535& 386$\pm$10& 120$\pm$ 9\\
VCC1062GDL-14&187.005885&  9.798541& 385$\pm$11& 101$\pm$ 8\\
VCC1062GDL-15&187.004535&  9.801900& 350$\pm$11& 111$\pm$ 6\\
VCC1062GDL-16&187.001925&  9.804249& 346$\pm$10& 104$\pm$ 5\\
VCC1062GDL-17&187.003020&  9.802365& 333$\pm$ 9& 102$\pm$ 6\\
VCC1062GDL-18&186.999240&  9.805346& 356$\pm$12&   \nodata\\
VCC1062GDL-19&186.995685&  9.807220& 352$\pm$10&   \nodata\\
VCC1062GDL-20&186.994635&  9.807293& 362$\pm$10&   \nodata\\
VCC1062GDL-21&187.037685&  9.798342& 665$\pm$12&   \nodata\\
VCC1062GDL-22&186.992850&  9.808851& 382$\pm$11&   \nodata\\
VCC1062GDL-23&186.991335&  9.802511& 340$\pm$10&   \nodata\\
VCC1062GDL-24&186.986835&  9.806428& 381$\pm$12&   \nodata\\
VCC1062GDL-25&186.985650&  9.798775& 376$\pm$19&   \nodata\\
VCC1062GDL-26&187.012890&  9.792312& 501$\pm$19&   \nodata\\
VCC1062GDL-27&187.028310&  9.813976& 608$\pm$17&   \nodata\\
VCC1062GDL-28&187.018665&  9.791675& 527$\pm$25&   \nodata\\
VCC1062GDL-29&187.012260&  9.817571& 572$\pm$25&   \nodata\\
\cutinhead{VCC 685}
VCC685GDL-01&185.985465& 16.692770&1224$\pm$34& 129$\pm$27\\
VCC685GDL-02&185.988135& 16.683411&1042$\pm$11&  99$\pm$12\\
VCC685GDL-03&185.997015& 16.701620&1381$\pm$12&  94$\pm$ 7\\
VCC685GDL-04&185.994150& 16.696199&1372$\pm$12& 103$\pm$ 7\\
VCC685GDL-05&185.989155& 16.688009&1034$\pm$12& 102$\pm$ 9\\
VCC685GDL-06&185.991735& 16.694599&1278$\pm$12& 136$\pm$ 9\\
VCC685GDL-07&185.978040& 16.679291&1060$\pm$90&   \nodata\\
VCC685GDL-08&185.989965& 16.701780&1323$\pm$71&   \nodata
\enddata
\tablecomments{
Col.(1):\ Slit ID;
Col.(2):\ Right ascension in decimal degrees (J2000);
Col.(3):\ Declination in decimal degrees (J2000);
Col.(4):\ Heliocentric radial velocity and uncertainty;
Col.(5):\ Velocity dispersion and uncertainty.
}
\label{gdl_table}
\end{deluxetable*}

\end{document}